\documentclass[12pt]{article}
\usepackage{amsfonts}
\usepackage[fleqn]{amsmath}
\usepackage{amssymb} 
\usepackage{hhline}
\usepackage{dsfont}
\usepackage{nicefrac}
\usepackage{upgreek}
\usepackage{mathrsfs}
\usepackage[scale={.75,.75}]{geometry}
\usepackage[T1]{fontenc}
\usepackage{bbm} 
\usepackage{fleqn} 
\usepackage{mathrsfs} 
\usepackage{cite} 
\usepackage{caption}
\usepackage{cite}
\usepackage{psfrag}
\usepackage{graphicx}
\usepackage{verbatim}
\usepackage{hyperref}

\numberwithin{equation}{section}
\numberwithin{table}{section}
\allowdisplaybreaks

\newcommand{\Slash}[1]{\displaystyle{\not}{#1}}

\newcommand{\be}{\begin{equation}}
\newcommand{\ee}{\end{equation}}
\newcommand{\bea}{\begin{array}}
\newcommand{\eea}{\end{array}}

\newcommand{\bes}{\begin{split}}
\newcommand{\ees}{\end{split}}

\newcommand{\tm}{y}
\newcommand{\e}{\epsilon}

\newcommand{\beq}{\begin{eqnarray}}
\newcommand{\eeq}{\end{eqnarray}}

\newcommand{\C}{\mathcal{C}}

\def\lsim{\mathrel{\raise.3ex\hbox{$<$\kern-.75em\lower1ex\hbox{$\sim$}}}}
\def\gsim{\mathrel{\raise.3ex\hbox{$>$\kern-.75em\lower1ex\hbox{$\sim$}}}}

\numberwithin{equation}{section}
\numberwithin{table}{section}
\allowdisplaybreaks

\begin{document}

\title{ 
{\normalsize     
DESY 10-218\hfill\mbox{}\\
December 2010\hfill\mbox{}\\}
\vspace{1cm}
\textbf{Quantum Leptogenesis I} \\[8mm]}
%
\author{{ A.~Anisimov,$^1$
W.~Buchm\"uller,$^2$
M.~Drewes,$^3$
S.~Mendizabal$^4$}\\[0.5cm]
{\it\normalsize
$^1$Fakult\"at f\"ur Physik,
Universit\"at Bielefeld, 33615 Bielefeld, Germany}\\
{\it\normalsize
$^2$Deutsches Elektronen-Synchrotron DESY, 22603 Hamburg, Germany}\\
{\it\normalsize
$^3$Institute de Th\'eorie des Ph\'enom\`enes Physiques
EPFL, 1015 Lausanne, Switzerland}\\
{\it\normalsize
$^4$Institut f\"ur Theoretische Physik, Goethe-Universit\"at, 
60438 Frankfurt am Main, Germany}\\[0.15cm]
}
\date{\mbox{}}
\maketitle

\thispagestyle{empty}


\begin{abstract}
\noindent
Thermal leptogenesis explains the observed matter-antimatter asymmetry of
the universe in terms of neutrino masses, consistent with neutrino oscillation
experiments. We present a full quantum mechanical calculation of the generated
lepton asymmetry based on Kadanoff-Baym equations. Origin of the asymmetry
is the departure from equilibrium of the statistical propagator of the
heavy Majorana neutrino, together with CP violating couplings. The lepton
asymmetry is calculated directly in terms of Green's functions without 
referring to ``number densities''. Compared to Boltzmann and quantum
Boltzmann equations, the crucial difference are memory effects, rapid
oscillations much faster than the heavy neutrino equilibration time. These
oscillations strongly suppress the generated lepton asymmetry, unless 
the standard model gauge interactions, which cause thermal damping, 
are properly taken into account. We find
that these damping effects essentially compensate the enhancement due to
quantum statistical factors, so that finally the conventional Boltzmann
equations again provide rather accurate predictions for the lepton asymmetry. 
\end{abstract}

\newpage

\section{Introduction}

Standard thermal leptogenesis \cite{fy86} provides a simple and elegant 
explanation of the origin of matter in the universe. Baryogenesis via
leptogenesis 
naturally emerges in grand unified extensions of the Standard Model, 
which incorporate right-handed neutrinos and the seesaw mechanism,
and the predicted connection between the cosmological matter-antimatter 
asymmetry and neutrino properties is in remarkable agreement with the present 
evidence for neutrino masses \cite{Buchmuller:2005eh}.

Leptogenesis is an out-of-equilibrium process in the high-temperature symmetric
phase of the Standard Model. It makes use of nonperturbative properties of
the Standard Model, the sphaleron processes which change baryon and 
lepton number \cite{Kuzmin:1985mm}, and it requires CP violation in the lepton
sector and quantum interference in the thermal bath. Almost all quantitative
studies of leptogenesis to date are based on Boltzmann's classical kinetic 
equations for the description of the nonequilibrium process 
\cite{Buchmuller:2005eh}. 

In this article, we discuss a full quantum mechanical calculation of the 
generated lepton asymmetry based on Kadanoff-Baym equations \cite{kb62}
and the Schwinger-Keldysh formalism \cite{sch61,kel64,bak63}. The main result
has previously been reported in \cite{Anisimov:2010aq}. Here we give a 
detailed derivation of the result, discuss its interpretation and set the 
stage for future computations. Further work is still needed to obtain a 
`quantum theory of leptogenesis' that can predict the cosmological 
matter-antimatter asymmetry in terms of neutrino properties without 
uncontrolable assumptions.

Conventional leptogenesis calculations based on kinetic equations suffer 
from a basic conceptual problem: the Boltzmann equations are classical 
equations for the time evolution of phase space distribution functions; 
the involved collision terms, however, are obtained from zero-temperature 
S-matrix elements which involve quantum interferences. This is in contrast 
to other successful applications of the Boltzmann equations in cosmology, like 
primordial nucleosynthesis, decoupling of photons or freeze-out of weakly 
interacting dark matter particles, where the collision terms arise from
tree-level S-matrix elements. In the case of leptogenesis, clearly a full
quantum mechanical treatment is necessary to understand the range of validity
of the Boltzmann equations and to determine the size of possible corrections
\cite{Buchmuller:2000nd}. 

In recent years, various attempts have been made to go beyond Boltzmann
equations. In \cite{Buchmuller:2000nd}, a solution of Kadanoff-Baym equations 
for leptogenesis has been found to leading order in a derivative expansion in
terms of distribution functions satisfying the Boltzmann equations. Various
thermal corrections, in particular quantum statistical factors and thermal 
masses, have been included \cite{Covi:1997dr,Giudice:2003jh,Kiessig:2010pr,
Anisimov:2010gy}. 
Quantum Boltzmann equations have been derived from Kadanoff-Baym equations
for scalar and Yukawa theories \cite{Lindner:2005kv,Lindner:2007am} and
for leptogenesis \cite{DeSimone:2007rw,Garny:2009rv,Garny:2009qn,
Beneke:2010wd}. Except for \cite{DeSimone:2007rw}, they do not contain memory
effects, but they yield the correct statistical factors which go beyond
the Boltzmann equations \cite{Anisimov:2010aq,DeSimone:2007rw,Garny:2009rv,
Garny:2010nj,Beneke:2010wd}. Quantum Boltzmann equations have important
applications for resonant leptogenesis \cite{DeSimone:2007rw},
flavoured leptogenesis \cite{Cirigliano:2009yt,Beneke:2010dz} and
$N_2$-leptogenesis \cite{Garbrecht:2010sz}.
Similar techniques have been developed for
electroweak baryogenesis \cite{Prokopec:2003pj,Prokopec:2004ic,
Konstandin:2004gy, Konstandin:2005cd} and for coherent baryogenesis 
\cite{Herranen:2010mh}. 

The quantum treatment of leptogenesis discussed in this paper is entirely based
on Green's functions, thus avoiding all approximations needed to arrive at
Boltzmann equations. Our work is based on \cite{Anisimov:2008dz}, where the
approach to thermal equilibrium has been discussed in terms of Green's 
functions for a toy model, a scalar field coupled to a large thermal bath.
In leptogenesis it is the heavy neutrino which is weakly coupled to the
standard model plasma containing many degrees of freedom. The nonequilibrium 
propagator of the heavy neutrino is obtained by solving the Kadanoff-Baym 
equations. The induced quantum corrections of the lepton (and Higgs) 
propagators then yield the wanted lepton asymmetry.

\begin{figure}[t]
\begin{center}
\includegraphics[width=12cm]{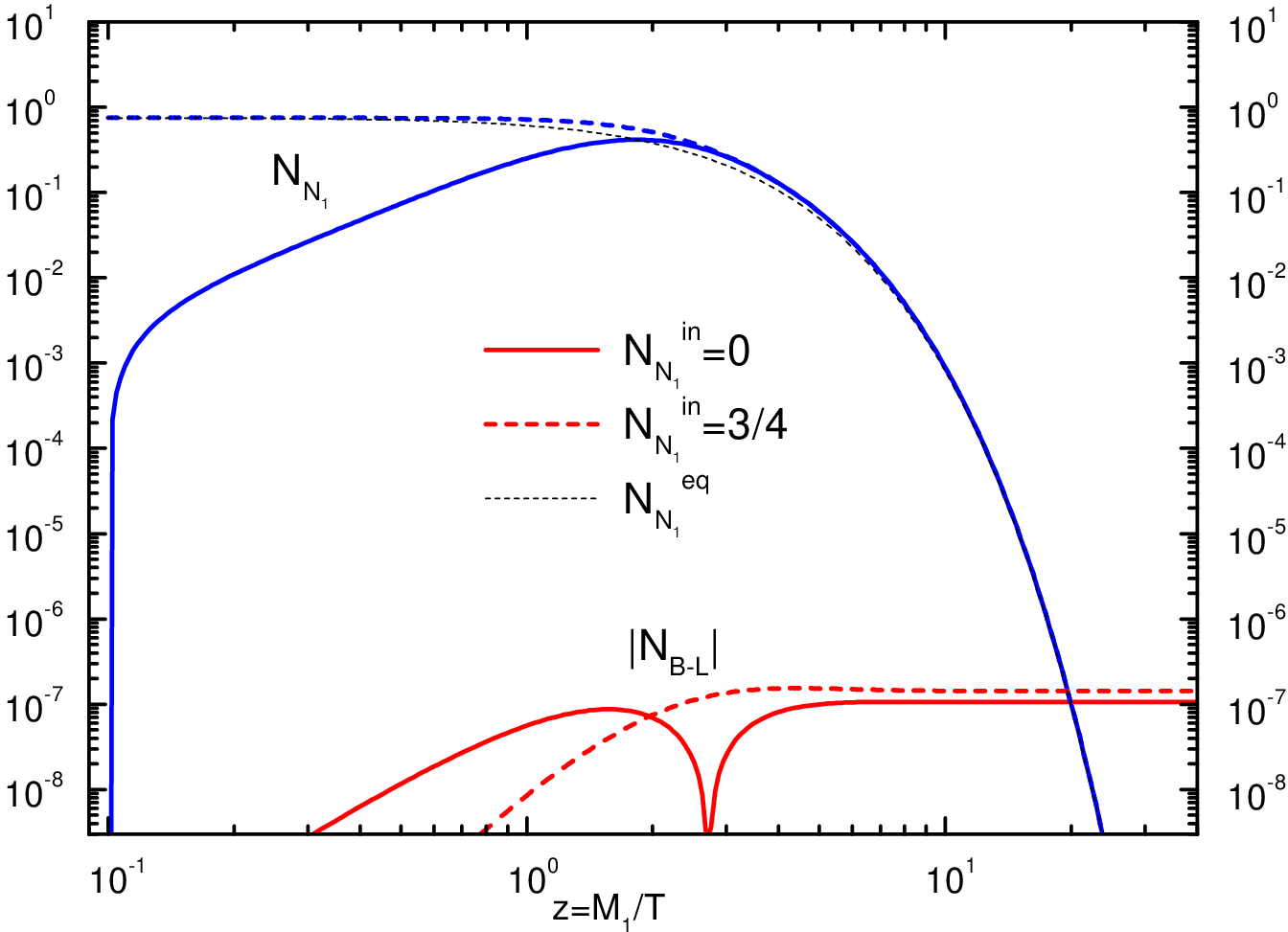}
\end{center}
\caption{Evolution of heavy neutrino abundance ${\rm N_{N_1}}$ and lepton 
asymmetry ${\rm N_{B-L}}$ for typical leptogenesis parameter:
$M_1 = 10^{10}~\mathrm{GeV}$, $\widetilde{m}_1 = 8\pi \Gamma_1 
(v_{\rm ew}/M_1)^2 = 10^{-3}~\mathrm{eV}$, $\epsilon = 10^{-6}$; the
inverse temperature $z=M_1/T$ is the time variable.  The
dashed (full) lines correspond to thermal (vacuum) initial conditions 
for the heavy neutrino abundance; the dotted line represents 
the equilibrium abundance. From \cite{Buchmuller:2002rq}.}
\label{boltzmannleptogenesis}
\end{figure}

In general baryogenesis requires departure from thermal equilibrium. For
the cosmological baryon asymmetry, this is provided by the Hubble expansion
of the universe and, possibly, also by initial conditions. This can be seen
in Fig.~\ref{boltzmannleptogenesis} where the time evolution of heavy
neutrino abundance and lepton asymmetry, as predicted by the Boltzmann
equations, are shown for two different initial conditions: thermal and zero
heavy neutrino abundance. In the first case, the Hubble expansion leads to
an excess of the neutrino abundance at $T \simeq 0.3~M_1$; shortly afterwards,
washout processes are no longer in equilibrium and the lepton asymmetry is
`frozen in'. This is the standard out-of-equilibrium decay scenario of 
baryogenesis. In the second case, interactions with the thermal bath first
bring the heavy neutrino into thermal equilibrium; due to the departure
from thermal equilibrium during this time, an initial lepton asymmetry is 
generated.
Around $T \simeq 0.3~M_1$, this asymmetry is washed out and, as in the first
case, the final lepton asymmetry is generated. Remarkably, the initial and
the final asymmetry have about the same size. For the generation of the
initial asymmetry the change of temperature due to the Hubble expansion is
not important. This allows us to make a significant technical simplification
in our analysis. Since our goal is the comparison of Boltzmann and 
Kadanoff-Baym equations, we concentrate on the computation of the initial
asymmetry at constant temperature. We expect differences between the classical 
and the quantum approach to be of similar size in the generation of the final 
asymmetry. In our numerical analysis we shall consider temperatures $T\lsim M$,
where the heavy neutrino production rate is not strongly affected by the
effect of thermal masses of lepton and Higgs fields \cite{Giudice:2003jh,
Kiessig:2010pr,Anisimov:2010gy}.

We consider an extension of the Standard Model with additional gauge singlet
fermions, i.e., right-handed neutrinos, whose masses and couplings are
described by the Lagrangian (sum over $i,j$),
\begin{align}\label{LNR1}
\mathcal{L}=\mathcal{L}_{SM} + \overline{\nu_R}_i i\Slash{\partial}\nu_{R i} 
+ \overline{l_L}_i\tilde{\phi}\lambda_{ij}^*\nu_{R j} 
+ \overline{\nu_R}_j \lambda_{ij} l_{L i} \phi
- \frac{1}{2} M_{ij}\left(\overline{\nu_R^c}_i \nu_{R j} + 
\overline{\nu_R}_j \nu_{R i}^c \right) \ .
\end{align}
Here $\nu_{R}^{c}=C\bar{\nu}_{R}^{T}$, $C$ is the charge conjugation matrix 
and $\widetilde{\phi} = i\sigma_2 \phi^*$; SU(2) isospin indices have been 
omitted.
For simplicity, we consider the case of hierarchical Majorana masses, 
$M_{k>1}\gg M_{1}\equiv M$, and small Yukawa couplings of the lightest heavy
neutrino $N_1 \equiv N$, $\lambda_{i1} \ll 1$, such that the decay width is 
much smaller than the mass. Leptogenesis is then dominated by decays and 
inverse decays of $N$, and it is convenient to integrate out the heavier
neutrinos. From Eq.~(\ref{LNR1}) one then obtains the effective Lagrangian 
\begin{align}\label{LRN2}
\mathcal{L} =  &\mathcal{L}_{SM} 
+ \frac{1}{2}\overline{N}i\Slash{\partial}N 
+ \overline{l_L}_i \widetilde{\phi}\lambda_{i1}^* N + 
      N^T \lambda_{i1} C l_{L i}\phi - \frac{1}{2}M N^T C N \nonumber\\
      &\;  +\frac{1}{2}\eta_{ij} l_{L i}^T \phi Cl_{L j}\phi
  + \frac{1}{2} \eta_{ij}^{*} \overline{l_L}_i \widetilde{\phi} C 
    \overline{l_L}_j^T \widetilde{\phi}\ ,
\end{align}
with $N = \nu_{R1} + \nu_{R1}^c$, and the familiar dimension-5 coupling
\begin{align}
\eta_{ij}=\sum_{k>1}\lambda_{ik}\frac{1}{M_k} \lambda_{kj}^T\ .
\end{align}
Using this effective Lagrangian has the advantage that vertex- and
self-energy contributions to the CP asymmetry in the heavy neutrino decay
\cite{Covi:1996wh,Flanz:1994yx,Buchmuller:1997yu} are
obtained from a single graph \cite{Buchmuller:2000nd}.

The paper is organized as follows. In Section~2 we present solutions of the 
Boltzmann equations for the heavy neutrino distribution function and the lepton
asymmetry, which are useful for later comparison with the Kadanoff-Baym 
equations. Some results from nonequilibrium quantum field theory (QFT), 
in particular equilibrium correlation functions and Kadanoff-Baym equations, 
are recalled in Section~3. Section~4 contains some of the main results
of this paper: analytic solutions of spectral function and statistical 
propagator for 
the heavy neutrino. These are needed for the computation of the lepton
asymmetry, which is carried out in Section~5. A detailed comparison of
the Boltzmann result and the Kadanoff-Baym result is given in Section~6,
and numerical results for the generated lepton asymmetries are compared
in Section~7. Summary and conclusions are given in Section~8, and various
details, including equilibrium correlation functions, Feynman rules, 
a discussion of the zero-width limit and the computation of some integrals 
are contained in Appendices A~-~D.

\section{Boltzmann equations}
The Boltzmann equations for the time evolution of the distribution functions
of heavy neutrinos, lepton and Higgs doublets are well known 
\cite{HahnWoernle:2009qn}. As discussed in the previous section, we focus
on the generation of the `initial asymmetry' 
(cf.~Fig.~\ref{boltzmannleptogenesis}), which allows us to neglect Hubble
expansion and washout terms and to work at constant temperature $T$. 
The distribution function of the heavy neutrinos is then determined
by the first-order differential equation\footnote{To simplify notation,
we use the same symbol for the modulus of 3-momentum and 4-momentum, e.g.,
$k= |{\bf k}|$ and $k = (|{\bf k}|,{\bf k})$.}
\begin{align}\label{bN}
\frac{\partial}{\partial t}f_N(t,\omega_{\bf p}) =
&-\frac{2}{\omega_{\bf p}}\int_{\bf k,q}(2\pi)^4\delta^4(k+q-p) 
\left(\lambda^{\dagger}\lambda\right)_{11} p\cdot k \nonumber\\
&\times[f_N(t,\omega_{\bf p})(1-f_l(k))(1+f_{\phi}(q))
-f_l(k)f_{\phi}(q)(1-f_N(t,\omega_{\bf p}))]\ ,
\end{align}
with vacuum initial condition,
\begin{align}
f_N(0,\omega_{\bf p}) = 0 \ ;
\end{align}
here $\omega_{\bf p} =\sqrt{M^2+{\bf p}^2}$, $k$ and $q$ are the 
energies of $N$, $l$ and $\phi$ 
with equilibrium distribution functions $f_l$ and $f_{\phi}$, respectively; 
the averaged decay matrix element is
$|M(N(p)\rightarrow l(k)\phi(q))|^2 = 
2\left(\lambda^{\dag}\lambda\right)_{11} p\cdot k$ 
(cf.~\cite{Buchmuller:2000nd}). 
For the momentum integrations we use the notation
\begin{equation}\label{notation}
\int_{\bf p}\ldots = \int \frac{d^3 p}{(2\pi)^3 2\omega}\ldots \ .
\end{equation}
In most leptogenesis calculations one directly computes the number density,
\begin{align}\label{numberN}
n_N(t) = \int \frac{d^3p}{(2\pi)^3}\ f_N(t,\omega_{\bf p}) \ ,
\end{align}
assuming kinetic equilibrium.

The sum of decay and inverse decay widths, whose inverse is the time 
needed to reach thermal equilibrium \cite{Weldon:1983jn}, is given by 
\begin{align}\label{gammaN}
\Gamma_{\bf p} = \left(\lambda^{\dagger}\lambda\right)_{11}
\frac{2}{\omega_{\bf p}} \int_{\bf k,q} &(2\pi)^4\delta^4(k+q-p)\ p\cdot k\ 
f_{l\phi}(k,q)\ ,
\end{align}
where we have introduced the statistical factor (cf.~\cite{Weldon:1983jn})
\begin{align}\label{leptonhiggs}
f_{l\phi}(k,q) & = f_l(k)f_{\phi}(q)+(1-f_l(k))(1+f_{\phi}(q)) \nonumber\\
& = 1 - f_l(k) + f_{\phi}(q) \ .
\end{align}
Neglecting the momentum dependence of the heavy neutrino width
($\Gamma_{\bf p} \equiv \Gamma$), one easily
obtains the solution of the Boltzmann equation (\ref{bN}) with vacuum initial
condition,
\begin{equation}\label{solN}
f_N(t,\omega_{\bf p})=f_N^{eq}(\omega_{\bf p})
\left(1-e^{-\Gamma t}\right) \ ,
\end{equation}
where the equilibrium distribution is
\begin{align}
f_N^{eq}(\omega_{\bf p}) = \frac{1}{e^{\beta\omega_{\bf p}}+1} \ ,
\end{align}
and $\beta=1/T$ is the inverse temperature.

To compute the lepton asymmetry, we need
the Boltzmann equation for the lepton distribution function,
\begin{align}\label{bl}
\frac{\partial}{\partial t}f_{l}(t,k)=-\frac{1}{2k}&\int_{\bf q,p} 
(2\pi)^4\delta^4(k+q-p) \nonumber\\
\times&\left[|M(l\phi\rightarrow N)|^2f_{l}(k)f_{\phi}(q)
(1-f_N(t,\omega_{\bf p}))\right. \nonumber \\
-&\left.\ |M(N\rightarrow l\phi)|^2f_N(t,\omega_{\bf p})(1-f_{l}(k))
(1+f_{\phi}(q))\right]\ , 
\end{align}
where now $\mathcal{O}(\lambda^4)$ corrections to the matrix elements have to
be kept. Using Eq.~(\ref{solN}) one obtains for the lepton asymmetry 
\begin{align}
f_{Li}(t,k) = f_{li}(t,k) - f_{\bar{l}i}(t,k) \ ,
\end{align}
with initial condition $f_{Li}(0,k) = 0$, 
\begin{align}\label{soll1}
f_{Li}(t,k) = - &\epsilon_{ii}\frac{1}{k}
\int_{\bf q,p}(2\pi)^4\delta^4(k+q-p)\ p\cdot k\  f_{l\phi}(k,q) 
f_N^{eq}(\omega_{\bf p}) 
\frac{1}{\Gamma}\left(1-e^{-\Gamma t}\right)\ ,
\end{align}
where we have defined
\begin{align}\label{cpeps}
\epsilon_{ij} = 
\frac{3}{16\pi}{\rm Im}\{\lambda^*_{i1}(\eta\lambda^*)_{j1}\} M \ .
\end{align}
Summing over all lepton flavours, the generated lepton asymmetry is 
proportional to the familiar CP asymmetry \cite{Buchmuller:2000nd},
\begin{align}
\epsilon = \sum_i\frac{\epsilon_{ii}}{\left(\lambda^{\dag}\lambda\right)_{11}} 
= \frac{3}{16\pi}\frac{{\rm Im}\left(\lambda^{\dag}\eta\lambda^*\right)_{11}M}
{\left(\lambda^{\dag}\lambda\right)_{11}} \ .
\end{align}

For later comparison with solutions of the Kadanoff-Baym equations, 
it is convenient to rewrite Eq.~(\ref{soll1}) as a 4-fold integral,
\begin{align}\label{soll2}
f_{Li}(t,k) = - \epsilon_{ii}\frac{16\pi}{k}
&\int_{\bf q,p,q',k'}  k\cdot k'\ (2\pi)^4\delta^4(k+q-p)
(2\pi)^4\delta^4(k'+q'-p) \nonumber\\
&\times f_{l\phi}(k,q)
f_N^{eq}(\omega_{\bf p})
\frac{1}{\Gamma}\left(1-e^{-\Gamma t}\right)\ .
\end{align}
The integrand is now proportional to the averaged matrix element
$|M(l\phi\rightarrow \bar{l}\bar{\phi})|^2 = 2 k\cdot k' 
(\lambda^{\dagger}\lambda)_{11}/M^2$ (cf.~\cite{Buchmuller:2000nd}), 
which involves the product of the 4-vectors $k$ and $k'$.
At low temperatures, $T \ll M$, the integrand falls off like 
$e^{-\beta\omega_{\bf p}} < e^{-\beta M}$, i.e., the generated asymmetry 
is strongly suppressed. In standard leptogenesis calculations one
considers the integrated lepton asymmetry,
\begin{align}\label{number}
n_L = \sum_i \int \frac{d^3k}{(2\pi)^3}\ f_{Li}(t,k) \ .
\end{align}
The number densities $n_N$ (\ref{numberN}) and $n_L$ correspond to the comoving
number densities ${\rm N_{N_1}}$ and ${\rm |N_{B-L}|}$ shown in 
Fig.~\ref{boltzmannleptogenesis}, in the initial phase of the time evolution,
i.e., for $T \gsim 0.3\ M$. 

\section{Nonequilibrium QFT and Kadanoff-Baym equations}
		\label{KBESec}
In the following, we briefly introduce concepts and quantities from 
nonequilibrium quantum field theory that are necessary for our computation
(cf.~\cite{Berges:2004yj,Chou:1984es}). A thermodynamical system is 
represented by a 
statistical ensemble described by a density matrix $\varrho$. The expectation 
value for an operator $\mathcal{A}$ is then given by 
\begin{align}
\langle\mathcal{A}\rangle={\rm Tr}\left(\varrho\mathcal{A}\right) \ ,
\end{align}
where we have adopted the usual normalisation ${\rm Tr}\varrho=1$. Solving the 
initial value problem for $\varrho$ allows to compute all observables for all 
times. Direct computation of the time evolution of $\varrho$ is difficult. 
Generically, the von Neumann (or quantum Liouville) equation of motion for 
$\varrho$ can only be solved perturbatively for a reduced density matrix with 
an effective Hamiltonian. In most practical applications to date, 
a number of additional assumptions are made that lead to 
effective Boltzmann equations, which can take account of coherent 
oscillations\footnote{See \cite{Raffelt:1992uj,Sigl:1992fn} 
for an application to neutrino oscillations.},
or quantum corrected Boltzmann equations (cf.~Section~6)\footnote{In 
\cite{Gagnon:2009zza,Gagnon:2010kt} an approach based on first principles 
has been suggested 
that is applicable if the occupation numbers for the out-of-equilibrium
fields are small.}.

Instead of the time evolution of the density matrix, one can also directly
study the equations of motion of the correlation functions of the theory. 
The infinitely many degrees of freedom of the initial density matrix are 
then mapped onto their infinitely many initial conditions. Though a full 
characterisation of the system in principle involves all $n$-point functions, 
it is often sufficient to study the one- and two-point function. This applies 
to the problem considered in this work.

\subsection{Correlation functions for lepton and Higgs fields}
Leptogenesis occurs at temperatures above the electroweak scale where sphaleron
processes are active and transfer the generated lepton asymmetry to a 
baryon asymmetry. Hence, the Standard Model is in the symmetric phase and the 
four real degrees of freedom of the Higgs doublet correspond to four massless 
real scalar fields.

The spectral function and statistical propagator of a real scalar field $\phi$,
 $\Delta^{-}$ and $\Delta^{+}$, respectively, are defined as
\begin{align}
\Delta^{-}(x_{1},x_{2}) &= i\langle [\phi(x_{1}),\phi(x_{2})]\rangle \ ,
\label{dminus}\\
\Delta^{+}(x_{1},x_{2}) 
&= \frac{1}{2}\langle\{\phi(x_{1}),\phi(x_{2})\}\rangle \ .\label{dplus}
\end{align}
Here only contributions from connected diagrams are to be included to compute 
the dressed correlation functions. These fulfill the symmetry relations
\begin{align}
\Delta^{-}(x_1,x_2) &=-\Delta^{-}(x_2,x_1)\ ,\label{dminussymm}\\
\Delta^{+}(x_1,x_2) &= \Delta^{+}(x_2,x_1)\ ,\label{dplussymm}
\end{align}
which follow directly from the definitions.

The functions $\Delta^{\pm}$ have an intuitive physical interpretation.
The spectral function $\Delta^{-}$ is the Fourier transform of the spectral 
density, 
\begin{align}
\rho_{\textbf{q}}(t,\omega) = 
-i\int\frac{dy}{2\pi}e^{i\omega y}\Delta^{-}(t+\frac{y}{2},t-\frac{y}{2})\ ,
\end{align}
where we have used the relative and total time coordinates, $y=t_{1}-t_{2}$ 
and $t=(t_{1}+t_{2})/2$, respectively. 

The spectral density $\rho_{\textbf{q}}(t,\omega)$ characterises the density 
of quantum mechanical states in phase space. Propagating states, or 
resonances, appear as peaks in the spectral function.
The statistical propagator contains the information about the occupation 
number of each state.

In the following we shall also need the Wightman functions
\begin{align}
\Delta^>(x_1,x_2) &= \langle \phi(x_1)\phi(x_2)\rangle \ , \label{forw}\\ 
\Delta^<(x_1,x_2) &= \langle \phi(x_2)\phi(x_1)\rangle \ , \label{back} 
\end{align}
which are related to $\Delta^{\pm}$ by
\begin{align}
\Delta^{-}(x_1,x_2) &= i\left(\Delta^>(x_1,x_2)-\Delta^<(x_1,x_2)\right)\ ,\\
\Delta^{+}(x_1,x_2) &= \frac{1}{2}\left(\Delta^>(x_1,x_2) 
+ \Delta^<(x_1,x_2)\right) \ . 
\end{align}
Using microcausality and the condition for canonical quantization, 
\begin{align}
[\phi(x_1),\phi(x_2)]|_{t_{1}=t_{2}} 
&= [\dot\phi(x_1),\dot\phi(x_2)]|_{t_{1}=t_{2}} = 0\ , \\
[\phi(x_1),\dot\phi(x_2)]|_{t_{1}=t_{2}} &= i\delta({\bf x}_1-{\bf x}_2)\ ,
\end{align}
one obtains boundary conditions in $y=t_1-t_2$ for $\Delta^{-}$,
\begin{align}
\Delta^{-}(x_{1},x_{2})|_{t_{1}=t_{2}} &= 0\ , \label{con1}\\
\partial_{t_{1}}\Delta^{-}(x_{1},x_{2})|_{t_{1}=t_{2}}
&=-\partial_{t_{2}}\Delta^{-}(x_{1},x_{2})|_{t_{1}=t_{2}}
 = \delta({\bf x}_1-{\bf x}_2)\ , \label{con2}\\
\partial_{t_{1}}\partial_{t_{2}}
\Delta^{-}(x_{1},x_{2})|_{t_{1}=t_{2}} &= 0\ .
\label{con3}
\end{align}
Note that these conditions do not depend on the physical initial conditions 
of the system encoded in the initial density matrix. These enter via the 
initial conditions for the statistical propagator.

Analogous to $\Delta^{\pm}$,
one can define the spectral functions and statistical propagators 
for fermions. The fermionic fields in the Lagrangian (\ref{LRN2}) are 
massless left-handed leptons (Weyl fields $l_{Li}$) and a massive neutrino 
(Majorana field $N$). 
For the massless leptons, spectral function and statistical propagator are 
defined as
\begin{align}
(S^-_{Lij})_{\alpha\beta}(x_1,x_2) 
&= i\langle \{l_{Li\alpha}(x_1),\bar{l}_{Lj\beta}(x_2)\}\rangle\ ,\\
(S^+_{Lij})_{\alpha\beta}(x_1,x_2)
&={1\over 2}\langle [l_{Li\alpha}(x_1),\bar{l}_{Lj\beta}(x_2)]\rangle\ ,
\end{align}
where $\alpha$ and $\beta$ are spinor indices, and SU(2) indices were
omitted for notational simplicity. The subscript $L$ denotes the projection 
to left-handed fields, i.e., 
$S^{\pm}_L = P_L S^{\pm}$, where $P_{L}=(1-\gamma^{5})/2$ and $S^{\pm}$ are
the propagators for Dirac fermions. As for bosons, we shall need the
functions
\begin{align}
(S^>_{Lij})_{\alpha\beta}(x_1,x_2) &= 
\langle l_{Li\alpha}(x_1)\bar{l}_{Lj\beta}(x_2)\rangle \ ,\\
(S^<_{Lij})_{\alpha\beta}(x_1,x_2) &=
-\langle\bar{l}_{Lj\beta}(x_2) l_{Li\alpha}(x_1)\rangle \ ,
\end{align}
which are related to spectral function and statistical propagator by
\begin{align}
S^{-}_{Lij}(x_1,x_2) &= i\left(S^>_{Lij}(x_1,x_2)-S^<_{Lij}(x_1,x_2)\right)\ ,
\label{relation1}\\
S^{+}_{Lij}(x_1,x_2) &= \frac{1}{2}\left(S^>_{Lij}(x_1,x_2) 
+ S^<_{Lij}(x_1,x_2)\right) \ . \label{relation2}
\end{align}
The propagators $S^{\pm}$ have the symmetry properties
\begin{align}
\gamma_0 \left[S^-_{Lij}(x_1,x_2)\right]^{\dagger}\gamma_0 &= 
-S^-_{Lji}(x_2,x_1) \ , \\
\gamma_0 \left[S^+_{Lij}(x_1,x_2)\right]^{\dagger}\gamma_0 &= 
S^+_{Lji}(x_2,x_1) \ .
\end{align}
The canonical quantization condition,
\begin{align}
\{l_{Li\alpha}(x_1),l^{\dagger}_{Lj\beta}(x_2)\}
 = P_{L\alpha\beta}\delta_{ij}\delta({\bf x}_1-{\bf x}_2) \ ,
\end{align}
implies the boundary condition for the spectral function
\begin{align}
S_{Lij}^{-}(x_{1},x_{2})|_{t_{1}=t_{2}} = 
iP_{L}\delta_{ij}\delta({\bf x}_1-{\bf x}_2) \ .
\end{align}

Finally, spectral function and statistical propagator for the Majorana field 
$N$ read
\begin{align}
G^-_{\alpha\beta}(x_1,x_2) 
&= i\langle \{N_{\alpha}(x_1),N_{\beta}(x_2)\}\rangle\ ,\\
G^+_{\alpha\beta}(x_1,x_2)
&={1\over 2}\langle [N_{\alpha}(x_1),N_{\beta}(x_2)]\rangle \ .
\end{align}
They have the symmetries
\begin{align}
G^-(x_1,x_2) &= G^-(x_2,x_1)^T\ ,\label{GminusSymmetry}\\
G^+(x_1,x_2) &= -G^+(x_2,x_1)^T \ .\label{GplusSymmetry}
\end{align}
The canonical quantization condition, together with the Majorana property 
$N = C\bar{N}^T$, implies the boundary condition
\begin{align}\label{Gboundary}
G^{-}(x_{1},x_{2})|_{t_{1}=t_{2}} = 
i\gamma^{0}\delta({\bf x}_1-{\bf x}_2) C^{-1} \ .
\end{align}
As for scalars, the physical initial conditions enter as boundary conditions 
for the statistical propagator.
In the following, we will consider two types of initial conditions, 
thermal equilibrium and Gaussian initial correlations,
for which we solve the equations of motion in the following section. 
Analogous to real scalars, the functions $G^{\gtrless}$ are 
defined as
\begin{align}
G^{>}_{\alpha\beta}(x_1,x_2) 
&= \langle N_{\alpha}(x_1)N_{\beta}(x_2)\rangle\ ,\label{Glarger}\\
G^{<}_{\alpha\beta}(x_1,x_2)
&= -\langle N_{\beta}(x_2)N_{\alpha}(x_1)\rangle ,\label{Gsmaller}
\end{align}
with the usual relations to spectral function and statistical propagator,
\begin{align}
G^{-}(x_1,x_2) &= i\left(G^>(x_1,x_2)-G^<(x_1,x_2)\right)\ ,
\label{Grel1}\\
G^{+}(x_1,x_2) &= \frac{1}{2}\left(G^>(x_1,x_2) 
+ G^<(x_1,x_2)\right) \ . \label{Grel2}
\end{align}

\subsection{Equations of motion}\label{KBESection}
\begin{figure}[t]
\centering
\includegraphics[width=12cm]{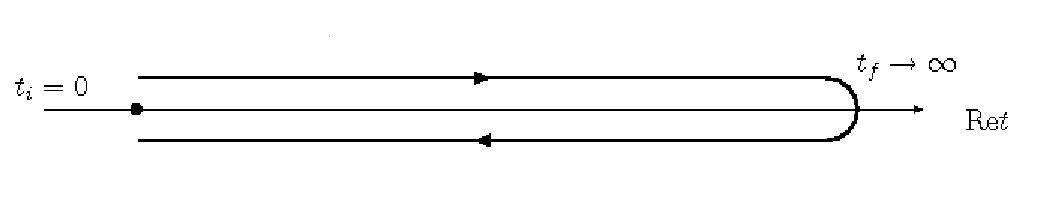}
\caption{Path in the complex time plane for nonequilibrium Green's 
functions. The contour runs from some initial time $x^{0}=t_{i}+i\epsilon$ 
($t_{i} = 0)$ parallel to the real axis ($x^{0}=t+i\epsilon$) up to some final 
time $t_{f}+i\epsilon$ and returns to $t_{i}-i\epsilon$. To compute physical 
correlation functions for arbitrary times $t>t_{i}$, one takes the limits 
$t_{f}\rightarrow\infty$ and $\epsilon \rightarrow 0$.\label{contour1}}
\end{figure}

In thermal leptogenesis, the deviation from thermal equilibrium that is 
necessary to create a matter-antimatter asymmetry is due to the heavy Majorana 
neutrinos which are out of equilibrium. The equations of motion for their
correlation functions $G^{\pm}$ can be obtained via the Schwinger-Keldysh 
formalism \cite{kel64}. The basic quantity is the Green's function with 
time arguments defined on a contour $\C$ in the complex $x^0$-plane, known as 
the Keldysh contour (cf.~Fig.\ref{contour1}), 
\begin{equation}\label{causal}
G_\C(x_1,x_2)
=\theta_\C(x^0_1,x^0_2)G^>(x_1,x_2) 
+ \theta_\C(x^0_2,x^0_1)G^<(x_1,x_2)\ .
\end{equation}
Here the $\theta$-functions enforce path ordering along the contour $\C$.
The necessity of considering Green's functions with time arguments on the 
Keldysh contour (rather than the real axis) is a consequence of the fact that 
nonequilibrium processes are initial value problems. The system is prepared at 
initial time $t_{i}$, its state at later times is unknown. Hence, the usual 
approach to define a S-matrix by projection onto asymptotic `in' and `out' 
states, sending initial and final time to infinity, cannot be applied. When 
using the Keldysh contour which starts and ends at the same time 
$t_{i}$\footnote{Due to this fact this formalism is sometimes called  
`in-in' formalism, in contrast to the `in-out' formalism used to compute the 
S-matrix.}, no knowledge of the system's state at $t=\pm\infty$ is needed to 
define a generating functional for correlation functions. 

The Green's function $G_\C$ satisfies the Schwinger-Dyson equation
\begin{align}\label{sde}
C(i\Slash{\partial}_{1}-M)G_\C(x_{1},x_{2})  
- i\int_{\C}d^{4}x' C\Sigma_{\C}(x_{1},x')G_{\C}(x',x_{2})
=i\delta_{\C}(x_{1}-x_{2})\ ,
\end{align}
where $C\Sigma_{\C}(x_{1},x')$ is the self-energy\footnote{An explicit factor 
$C$ is factorized for later convenience.} on the contour and 
$\Slash{\partial}_{1} = \gamma^{\mu}\partial/\partial x_1^{\mu}$. 
Like the Green's function, also the self-energy can be decomposed as
\begin{align}
\Sigma_\C(x_1,x_2) =\theta_\C(x^0_1,x^0_2)\Sigma^>(x_1,x_2) 
+ \theta_\C(x^0_2,x^0_1)\Sigma^<(x_1,x_2)\ .
\end{align}
In the Schwinger-Dyson equation (\ref{sde}) the time coordinates of $G_\C$ 
and $\Sigma_\C$ can lie on the upper or the lower branch of the contour.

The familar time-ordered Feynman propagator is obtained from 
$G_{\C}(x_{1},x_{2})$ when both time arguments lie on the upper branch,
and therefore denoted by $G^{11}$. Correspondingly, $G_{\C}(x_{1},x_{2})$ 
with both time arguments on the lower part of the contour corresponds to 
an anti-time-ordered propagator, denoted as $G^{22}$. For correlators with 
one time argument on the upper and one on the lower part of the contour, 
referred to as $G^{12}$ and $G^{21}$, the order of field operators is
fixed by the path ordering: operators on the upper branch are always `earlier' 
than those on the lower branch (cf.~\ref{causal}). Altogether, one has
\begin{align}
G^{12}(x_1,x_2) &= G^<(x_1,x_2)\ , \label{uplow} \\
G^{21}(x_1,x_2) &= G^>(x_1,x_2)\ , \label{lowup} \\
G^{11}(x_1,x_2)   &= G^{+}(x_{1},x_{2}) 
- \frac{i}{2}{\rm sign}(x^0_1-x^0_2)G^{-}(x_{1},x_{2}) \ ,\label{upup}\\
G^{22}(x_1,x_2) &= G^{+}(x_{1},x_{2}) 
+ \frac{i}{2}{\rm sign}(x^0_1-x^0_2)G^{-}(x_{1},x_{2}) \ ;\label{lowlow}
\end{align}
the last two relations are easily verified by inserting the definitions
of $G^{\pm}$.

In a perturbative expansion of the Schwinger-Dyson equation (\ref{sde}) in 
terms of Feynman diagrams, time arguments of internal vertices can lie on 
either branch. Hence, the number of contributing graphs 
doubles with each internal vertex since this can lie on the upper 
or the lower branch\footnote{This fact is sometimes referred to as 
`doubling of degrees of freedom'.}. Two upper 
vertices are connected by $G^{11}$, two lower vertices by $G^{22}$ and 
vertices of different type by $G^{12}$ and $G^{21}$. Each lower vertex 
leads to an additional factor $-1$. 

Like the Green's function, also the self-energy $\Sigma_{\C}$, the sum of
all one-particle irreducible graphs, can be dissected into components 
$\Sigma^{kl}$, with $k$ and $l$ being `contour indices' as defined above.
Analogous to (\ref{uplow}) and (\ref{lowup}) one then defines self-energies 
$\Sigma^{\gtrless}$ and, following (\ref{Grel1}) and (\ref{Grel2}),
self-energies $\Sigma^{\pm}$ via the equations
\begin{align}
\Sigma^{-}(x_1,x_2) &= i\left(\Sigma^>(x_1,x_2) - 
\Sigma^<(x_1,x_2)\right) \ , \\
\Sigma^{+}(x_1,x_2) &= \frac{1}{2}\left(\Sigma^>(x_1,x_2) + 
\Sigma^<(x_1,x_2)\right) \ .
\end{align}
Since the self-energies $\Sigma^{kl}$ are directly related to the full
Green's functions $G^{kl}$, they also satisfy the relations
(\ref{uplow}) - (\ref{lowlow}).

Using the above relations for $G^{kl}$ and $\Sigma^{kl}$, one obtains, after 
a straightforward calculation, from the Schwinger-Dyson equation (\ref{sde}) 
a system of two coupled differential equations for $G^{\pm}_{\bf p}$, 
the Kadanoff-Baym equations. Due to spatial homogeneity, we can consider
the equations for each Fourier mode separately,
\begin{align}
C(i\gamma^0\partial_{t_1}-{\bf p}\pmb{\gamma}-M)G_{\bf p}^-(t_1,t_2)=
&-\int_{t_1}^{t_2} dt' C\Sigma_{\bf p}^{-}(t_1,t')G_{\bf p}^-(t',t_2)\ ,
\label{kb1}\\
C(i\gamma^0\partial_{t_1}-{\bf p}\pmb{\gamma}-M)G_{\bf p}^+(t_1,t_2) =
&-\int^{t_2}_{t_i}dt' C\Sigma_{\bf p}^{+}(t_1,t')G_{\bf p}^-(t',t_2)\nonumber\\
&+\int^{t_1}_{t_i}dt' C\Sigma^-_{\bf p}(t_1,t')G_{\bf p}^{+}(t',t_2)\ .
\label{kb2}
\end{align}
For the lepton propagators $S^{\pm}_{L{\bf k}}$ one obtains the same equations,
with $C\Sigma_{\bf p}^{\pm}$ replaced by the lepton self-energies  
$\Pi_{\bf k}^{\pm}$ and no charge conjugation matrix C multiplying the kinetic
term.

The Kadanoff-Baym equations (\ref{kb1}) and (\ref{kb2}) are exact. 
They contain all quantum 
and non-Markovian effects including the dependence on the initial time $t_i$.
Furthermore, in contrast to usual linear response techniques, 
they do not rely on any assumption regarding the size of the initial deviation 
from equilibrium. The equations in this form are valid for arbitrary 
nonequilibrium initial states which can be parameterized by Gaussian initial 
correlations. This covers the case considered in this work since the
generated lepton asymmetry involves to leading order in the Yukawa coupling
only the 2-point functions of the heavy neutrino. When higher order initial 
correlations play a significant role, the Kadanoff-Baym formalism is still 
applicable, but the equation for the statistical 
propagator contains extra terms at $t_{i}$ \cite{Garny:2009ni}.
In \cite{Berges:2004yj}, thermalization has been studied for a scalar field
theory using the equation of motion for the statistical propagator. 

In nonequilibrium quantum field theory, instead of distribution functions, 
quantum mechanical correlation functions $G^{\pm}$ characterise the state 
of the system. The interactions enter via the self-energies $\Sigma^{\pm}$ 
which, via the generalized cutting rules,
contain all possible processes. Encoding this information in the self-energies 
avoids potential problems related to the definition of asymptotic states for 
unstable particles as well as the substraction of real intermediate state
contributions in Boltzmann equations. Note, finally,  that 
the integro-differential equations (\ref{kb1}), (\ref{kb2}) do not suffer 
from the late time uncertainties or secular terms that perturbative expansions 
of Boltzmann equations are often plagued with when applied to multiscale 
problems (cf.~\cite{Berges:2004yj}).

\subsection{Weak coupling to a thermal bath}
The Kadanoff-Baym equations provide a tool to study the dynamics of arbitrary 
nonequilibrium systems. Unfortunately, in most cases they can only be solved 
numerically. As discussed in the introduction, in this work we consider a 
rather simple system: one field that is out of equilibrium ($N$) is weakly 
coupled to a large thermal bath of Standard Model fields. This leads to
a number of simplifications compared to the general case that allow to find 
analytic solutions. We have previously studied scalar field models of this
type \cite{Anisimov:2008dz,Thesis}. Here we extend the methods developed 
therein to the case of thermal leptogenesis.

\begin{figure}
\begin{center}
\includegraphics[width=14cm]{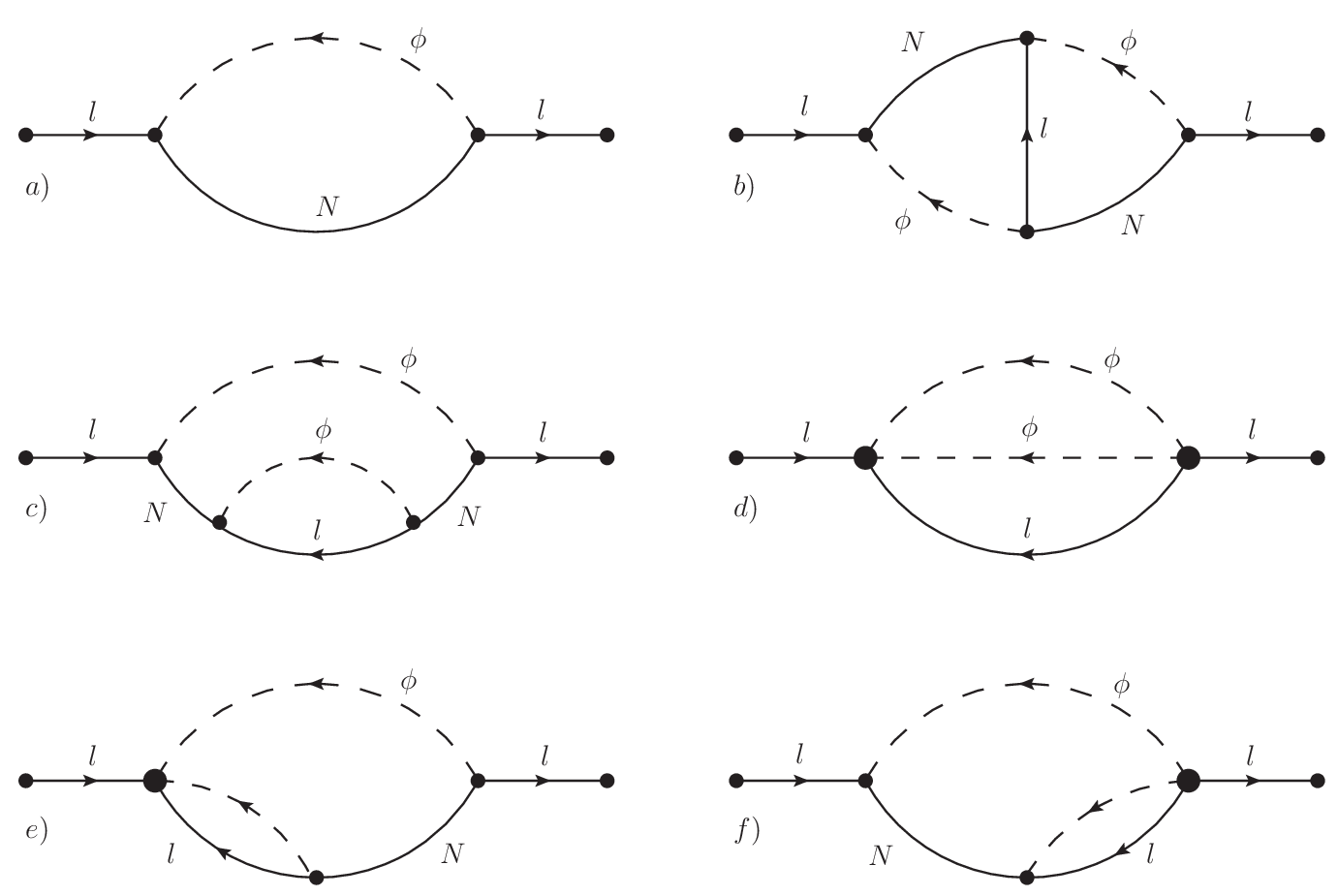}
\end{center}
\caption{One- and two-loop contributions to the lepton self-energy correspoding
to washout terms, a) - d), and CP violating terms which generate a lepton
asymmetry, e) and f).}
\label{washoutandproduction}
\end{figure}

The Standard Model interactions keep the bath in thermal equilibrium.
The corresponding time scale $\tau_{SM} \sim 1/(g^2T)$ at temperature 
$T \sim M$ is much shorter than the equilibration time 
$\tau_{N} \sim 1/(\lambda^2 M)$ of the heavy neutrino, which governs
the generation of the lepton asymmetry: $\tau_{SM}\ll\tau_{N}$. Lepton number
changing processes in the thermal bath are shown in 
Fig.~\ref{washoutandproduction}. As in the case of Boltzmann equations
discussed in Section~2, we focus on the CP violating interaction generating
the lepton asymmetry that correspond to Fig.~\ref{washoutandproduction} e) and
f).

To evaluate these graphs we need the correlation functions of lepton and Higgs
fields in the thermal bath. A system in thermal equilibrium is described by 
the density matrix 
\begin{align}\label{eqrho}
\varrho_{eq} = 
\frac{\exp\left(\beta\left(-\mathcal{H}+\mu_i\mathcal{Q}_i\right)\right)}
{{\rm Tr}\exp\left(\beta\left(-\mathcal{H}+\mu_i\mathcal{Q}_i\right)\right)}\ ,
\end{align}
where $\mathcal{H}$ is the Hamiltonian of the system, $\beta$ is the inverse 
temperature, $\mathcal{Q}_{i}$ are conserved charges and $\mu_{i}$ are the 
corresponding chemical potentials. As expected for an initial state after 
inflation, we set all chemical potentials equal to zero.

Equilibrium correlation functions of
a spatially homogeneous system only depend on space-time differences,
and it is convenient to consider the Fourier transforms,
\begin{align}
\Delta^{\pm}_{\bf q}(\omega) &= \int d^4x e^{i(\omega x^0 - {\bf q x})}
\Delta^{\pm}(x) \ ,\\
S^{\pm}_{\bf k}(\omega) &= \int d^4x e^{i(\omega x^0 - {\bf k x})}
S^{\pm}(x) \ .
\end{align}
The equilibrium density matrix (\ref{eqrho}) then corresponds to a shift
in imaginary time. This leads to the well-known 
Kubo-Martin-Schwinger (KMS) relations (cf.~\cite{LeB})
\begin{align}\label{preKMS}
\Delta^{<}_{\textbf{q}}(\omega) = 
e^{-\beta\omega}\Delta^{>}_{\textbf{q}}(\omega) \ , \quad
S^{<}_{\textbf{k}}(\omega) =
-e^{-\beta\omega}S^{>}_{\textbf{k}}(\omega) \ ,
\end{align}
which imply
\begin{align}
\Delta^{+}_{\textbf{q}}(\omega) &= 
-i\left(\frac{1}{2}+f_{\phi}(\omega)\right)
\Delta^{-}_{\textbf{q}}(\omega)
= -\frac{i}{2}\coth\left(\frac{\beta\omega}{2}\right)
\Delta^{-}_{\textbf{q}}(\omega) \ , \label{DKMS}\\
S^{+}_{\textbf{k}}(\omega) &= 
-i\left(\frac{1}{2}-f_{l}(\omega)\right)S^{-}_{\textbf{k}}(\omega)
= -\frac{i}{2}\tanh\left(\frac{\beta\omega}{2}\right)
S^{-}_{\textbf{k}}(\omega)\label{SKMS} \ ,
\end{align}
where 
\begin{align}
f_{\phi}(\omega) = \frac{1}{e^{\beta\omega}-1} \ , \quad 
f_{l}(\omega) = \frac{1}{e^{\beta\omega}+1} \ ,
\end{align}
are Bose-Einstein and Fermi-Dirac distribution functions, respectively. Note 
that the energy $\omega$ is not on-shell.

\begin{figure}[t]
\centering
\includegraphics[width=12cm]{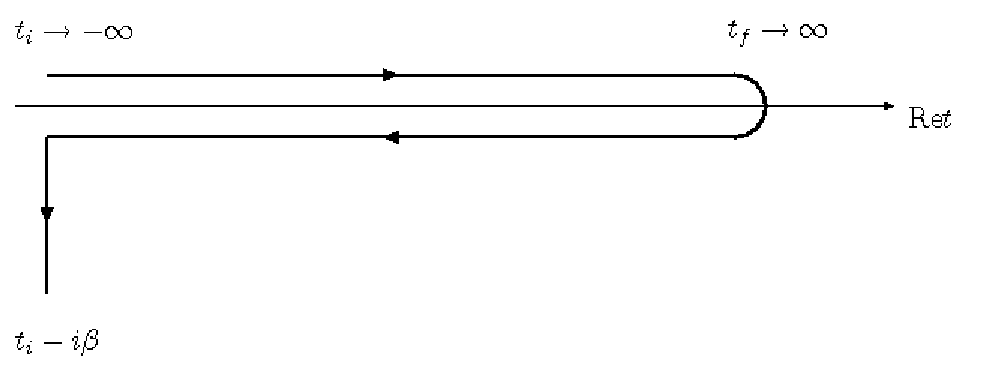}
\caption{Path $\C_{\beta}$ in the complex time plane for equilibrium 
correlation functions.} 
\label{contour2}
\end{figure}

Equilibrium Green's functions can be calculated in the real-time formalism
using the contour $\C_{\beta}$ in the complex time plane, which is shown 
in Fig.~\ref{contour2}. For the free equilibrium propagators of massless 
lepton and Higgs fields one obtains ($q = |{\bf q}|$, $k = |{\bf k}|$,
cf.~\cite{LeB}), 
\begin{align}
\Delta^-_{\textbf{q}}(y) &= \frac{1}{q}\sin(q y)\ ,\label{DeltaMinusFree}\\
\Delta^+_{\textbf{q}}(y) &= \frac{1}{2q}\coth\left(\frac{\beta q}{2}\right)
\cos(q y)\ ,\\
S^-_{L\textbf{k}}(y) &= P_L\left(i\gamma_0\cos(k y) -
\frac{\textbf{k}\pmb{\gamma}}{k}\sin(ky)\right)\label{SminusBath}\ ,\\
S^{+}_{L\textbf{k}}(y) &= -\frac{1}{2}P_L\tanh\left(\frac{\beta k}{2}\right)
\left(i\gamma_0\sin(ky)+\frac{\textbf{k}\pmb{\gamma}}{k}
\cos(ky)\right) \ .
\end{align}
All other propagators can be obtained as linear combinations using the 
relations described in the previous paragraph. A complete list is given in
Appendix~A.

In the following sections we shall see that the calculation of the lepton 
asymmetry represents an initial value problem which can be treated based
on the real time formalism together with the Keldysh contour 
Fig.~\ref{contour1}. Thermal and nonthermal properties of the system are
then encoded in the initial values of the various Green's functions.

\section{Nonequilibrium correlation functions}
		\label{Nsolution}
The assumption of weak coupling to a large thermal bath with negligible 
backreaction in the framework of Kadanoff-Baym equations implies 
that self-energies for the heavy neutrinos $N$ are computed from equilibrium 
propagators of bath fields only. This also corresponds to a leading order 
perturbative expansion in the coupling constant. 

Perturbative expansions of Boltzmann equations in multiscale problems are 
known to suffer from uncertainties, so-called secular terms, at late times. 
The Kadanoff-Baym equations (\ref{kb1}) and (\ref{kb2}) in full generality 
are free of secular terms and consistently include all memory effects.
Nevertheless, the neglect of backreaction in the computation of $\Sigma$
corresponds to a truncation in the perturbative expansion in the Yukawa 
couplings $\lambda$, which might introduce similar uncertainties related to 
the multiscale nature of the problem. However, in the system of consideration 
contributions of higher order in $\lambda$ are not only suppressed by the 
smallness of the coupling, but also by the number of degrees of freedom 
in the bath that justify the neglect of backreaction. Hence, we expect
potential problems due to secular terms not to be relevant. 

The assumption that the background medium equilibrates instantaneously 
on the time scale of the asymmetry generation leaves open the details of the 
equilibration process. In reality, there are effects related to the finite 
equilibration time and the finite size of the quasi-particles. As we shall 
see in Section~5, these quantities play a crucial role in the Kadanoff-Baym
result for the lepton asymmetry.

\begin{figure}[t]
    \psfrag{p}{$\phi$}\psfrag{wp}{$(\omega_{\bf p}, {\bf p})$}
        \psfrag{l}{$l$}
            \psfrag{N}{$N$}
   \centering
   \includegraphics[width=3in]{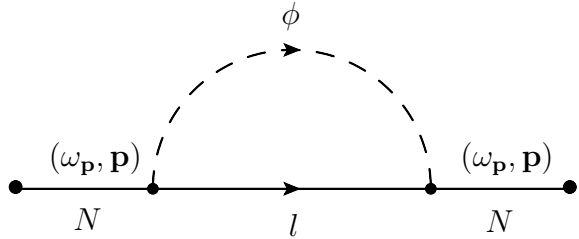}
\caption{One-loop contribution to the self-energies $C\Sigma^{\pm}_{\bf p}$ 
of the Majorana neutrino $N$.}
\label{fig:1loop}
\end{figure}

The self-energy for the heavy neutrino $N$ to leading order in $\lambda$ is 
given by the diagram in Fig.~\ref{fig:1loop}.
It contains time-translation invariant propagators of bath fields only, and 
hence it is also time-translation invariant.
As shown in \cite{Anisimov:2008dz}, this implies that also the spectral 
function is time-translation invariant, 
$G^{-}_{\bf p}(t_1,t_2) \equiv G^{-}_{\bf p}(y)$, $y=t_1-t_2$.
In this case we can find the general solutions to the Kadanoff-Baym equations 
without further approximations.

\subsection{Equation for the spectral function}

\begin{figure}[t]
\centering
\includegraphics[width=10 cm]{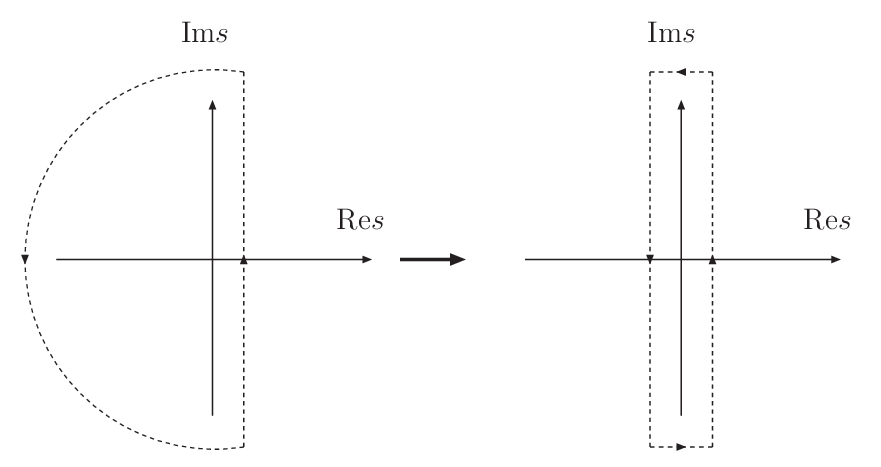}
\caption{Bromwich contour}
\label{Bromwich}
\end{figure}

Let us now consider the equation for the spectral function of the Majorana 
neutrino. After an obvious change of variables, the Kadanoff-Baym equation 
(\ref{kb1}) becomes, 
\begin{align}\label{SpectralEquation}
C(i\gamma^{0}\partial_{y}-\textbf{p}\pmb{\gamma}-M)G_{\textbf{p}}^{-}(y)
-\int_{0}^{y}dy' C\Sigma_{\textbf{p}}^{-}(y-y')G^{-}_{\textbf{p}}(y')=0 \ .
\end{align}
Defining the Laplace transform
\begin{align}
\tilde{G}_{\textbf{p}}^{-}(s)=
\int_{0}^{\infty}dy e^{-sy}G^{-}_{\textbf{p}}(y)\ , \quad
\tilde{\Sigma}_{\textbf{p}}^{-}(s)=
\int_{0}^{\infty}dy e^{-sy}\Sigma^{-}_{\textbf{p}}(y)\ , 
\end{align}
one obtains from Eq.~(\ref{SpectralEquation})
\begin{align}
\left(i\gamma^{0}s -\textbf{p}\pmb{\gamma} - M
- \tilde{\Sigma}^{-}_{\textbf{p}}(s)\right)
\tilde{G}^{-}_{\textbf{p}}(s)
= i\gamma^{0}G^{-}_{\textbf{p}}(0) \ .
\end{align}
Using the boundary condition (\ref{Gboundary}),
\begin{equation}
G^{-}_{\textbf{p}}(0)=i\gamma^{0}C^{-1}\ ,
\end{equation}
this leads to 
\begin{align}\label{GsGleichung}
\tilde{G}^{-}(s)=-\left(i\gamma^{0}s-\textbf{p}\pmb{\gamma}-M 
- \tilde{\Sigma}^{-}(s)\right)^{-1}C^{-1} \ .
\end{align}
The inverse Laplace transform is given by
\begin{align}\label{backtrans}
G^{-}_{\textbf{p}}(y)=\int_{\mathcal{C}_{B}}\frac{ds}{2\pi i} 
e^{sy}\tilde{G}^{-}_{\textbf{p}}(s)\ ,
\end{align}
where $\C_B$ is the Bromwich contour (see Fig.~\ref{Bromwich}): The part 
parallel to the 
imaginary axis is chosen such that all singularities of the integrand are to 
its left; the second part is the semicircle at infinity which closes the 
contour at ${\rm Re}(s)<0$. 

From the definition of the Laplace transform one can see that the self-energy
$\tilde\Sigma^{-}_{\textbf{p}}(s)$ is analytic on the real $s$ axis, 
but has a discontinuity across the imaginary axis. 
This gives rise to the spectral representation
\begin{align}
\tilde{\Sigma}^{-}_{\textbf{p}}(s) = i\int_{-\infty}^{\infty} 
\frac{dp_{0}}{2\pi}\frac{\Sigma^{-}_{\textbf{p}}(p_{0})}{is-p_{0}}\ .
\label{SpecRepMin}
\end{align}
Note that the retarded and advanced self-energies are given by
\begin{align}
\tilde{\Sigma}^{-}_{\textbf{p}}(-i\omega+\epsilon)
&= \Sigma^{R}_{\textbf{p}}(\omega)\ ,\label{LapR}\\
\tilde{\Sigma}^{-}_{\textbf{p}}(-i\omega-\epsilon)
&= \Sigma^{A}_{\textbf{p}}(\omega)\ .\label{LapA}
\end{align}
These self-energies are determined by the discontinuity of 
$\tilde{\Sigma}^{-}_{\textbf{p}}(s)$,
\begin{align}
\mathrm{disc}\tilde{\Sigma}^{-}_{\textbf{p}}(-i\omega)=
\tilde{\Sigma}^{-}_{\textbf{p}}(-i\omega+\epsilon)
-\tilde{\Sigma}^{-}_{\textbf{p}}(-i\omega-\epsilon)
=\Sigma^{-}_{\textbf{p}}(\omega)\ ,
\end{align}
with the real part given by the principal value, i.e.,
\begin{align}
\tilde{\Sigma}^{-}_{\textbf{p}}(-i\omega\pm\epsilon)=
i\mathcal{P}\int_{-\infty}^{\infty} \frac{dp_{0}}{2\pi}
\frac{\Sigma^{-}(p_{0})}{\omega-p_{0}}
\pm\frac{1}{2}\Sigma^{-}_{\textbf{p}}(\omega) \ .
\end{align}
This representation of the self-energy is familiar from the theory at zero
temperature.

We are now ready to calculate the spectral function in terms of the
self-energy $\Sigma^{-}_{\textbf{p}}(\omega)$. Its Laplace transform 
has singularities only on the imaginary axis. Hence the Bromwich contour can 
be deformed as 
$\C_B\to \int^{i\infty+\e}_{-i\infty+\e}+\int^{-i\infty-\e}_{i\infty-\e}$
(see Fig.~\ref{Bromwich}),
which yields for the spectral function 
\begin{align}
G^{-}_{\textbf{p}}(y) &=\int_{\mathcal{C}_{B}}\frac{ds}{2\pi i} 
e^{sy}\tilde{G}^{-}_{\textbf{p}}(s) \nonumber\\
&=\int_{-\infty}^{\infty}\frac{d\omega}{2\pi}
e^{(i\omega+\epsilon)y}\tilde{G}^{-}_{\textbf{p}}(i\omega+\epsilon) 
+ \int_{\infty}^{-\infty}\frac{d\omega}{2\pi}
e^{(i\omega-\epsilon)y}\tilde{G}^{-}_{\textbf{p}}(i\omega-\epsilon)\nonumber\\
&=\int_{-\infty}^{\infty}\frac{d\omega}{2\pi}
e^{-i\omega y}\left(\tilde{G}^{-}_{\textbf{p}}(-i\omega+\epsilon)
-\tilde{G}^{-}_{\textbf{p}}(-i\omega-\epsilon)\right)\ .
\end{align}
The Fourier transform of the spectral function,
\begin{align}
\rho_{\textbf{p}}(\omega) = 
-i \int_{-\infty}^{\infty}dy e^{i\omega y}G^{-}_{\textbf{p}}(y) \ ,
\end{align}
is then given by
\begin{align}\label{spectral1}
\rho_{\textbf{p}}(\omega) = 
\left(\frac{-i}{\Slash{p}-M-\frac{1}{2}\Sigma^{-}_{\textbf{p}}(\omega)}
-\frac{-i}
{\Slash{p}-M+\frac{1}{2}\Sigma^{-}_{\textbf{p}}(\omega)}\right)C^{-1}\ .
\end{align}
Here we have assumed that the divergent contribution of the real part has 
already been absorbed into mass and wave function renormalization, so that
$\rho_{\textbf{p}}(\omega)$ represents the renormalized spectral density.
The finite part of the self-energy is negligable because of the small Yukawa 
coupling.

A straightforward calculation yields for the self-energy 
(cf.~\cite{Weldon:1983jn}),
\begin{align}
\Sigma^{-}_{\textbf{p}}(\omega)=
2i\left(\lambda^{\dagger}\lambda\right)_{11}\int_{\bf k,q} \Slash{k}\ 
\sigma(p;k,q) \ ,
\end{align}
where we have defined
\begin{align}
\sigma(p;k,q) = &\; f_{l\phi}(k,q)(2\pi)^4\left(\delta^4(p-k-q)  
+ \delta^4(p+k+q)\right) \nonumber\\
&+ \bar{f}_{l\phi}(k,q)(2\pi)^4\left(\delta^4(p+k-q) 
+ \delta^4(p-k+q)\right)\ ,
\end{align}
with the statistical factors 
\begin{align}
f_{l\phi}(k,q) = 1-f_l(k)+f_{\phi}(q) \ , \quad 
\bar{f}_{l\phi}(k,q) = f_{\phi}(q) + f_l(k) \ .
\end{align}
Note that $k$ and $q$ are on-shell, i.e., $k=(k,{\bf k})$ and  $q=(q,{\bf q})$,
whereas $p = (\omega, {\bf p})$ is off-shell. The properties of the Dirac 
matrices and rotational invariance imply
\begin{align}\label{Sigmarotinv}
\Sigma^{-}_{\textbf{p}}(\omega)=
ia_{\textbf{p}}(\omega)\gamma^{0} 
+ ib_{\textbf{p}}(\omega)\textbf{p}\pmb{\gamma}\ ,
\end{align}
where
\begin{align}
a_{\textbf{p}}(\omega) &=
2\left(\lambda^{\dagger}\lambda\right)_{11}\int_{\bf q,k} {k}\ \sigma(p;k,q)
\ , \\
b_{\textbf{p}}(\omega) &=
-2\left(\lambda^{\dagger}\lambda\right)_{11}\frac{1}{\textbf{p}^2}
\int_{\bf q,k} \textbf{p}\textbf{k}\ \sigma(p;k,q) \ .
\end{align}
These functions satisfy the relations
\begin{align}\label{ab}
a_{\textbf{p}}(-\omega_{\textbf{p}}) = a_{\textbf{p}}(\omega_{\textbf{p}})\ ,
\quad 
b_{\textbf{p}}(-\omega_{\textbf{p}}) = - b_{\textbf{p}}(\omega_{\textbf{p}})\ .
\end{align}

Using Eq.~(\ref{Sigmarotinv}) and linearising the denominators in 
Eq.~(\ref{spectral1}) in the small quantities $a_{\textbf{p}}(\omega)$ and 
$b_{\textbf{p}}(\omega)$, one obtains for the spectral density
\begin{align}\label{spectral2}
\rho_{\textbf{p}}(\omega) = 
\frac{2\omega\Gamma_{\textbf{p}}(\omega)}
{\left(\omega^2-\omega_{\textbf{p}}^2\right) + 
\left(\omega\Gamma_{\textbf{p}}(\omega)\right)^2}
\left(\Slash{p}+M\right)C^{-1} \ ,
\end{align}
where
\begin{align}
\omega \Gamma_{\textbf{p}}(\omega) &= \omega a_{\textbf{p}}(\omega)
+ \textbf{p}^2 b_{\textbf{p}}(\omega) \nonumber\\
&= 2 \left(\lambda^{\dagger}\lambda\right)_{11}
\int_{\bf q,k}\ p\cdot k\ \sigma(p;k,q)\ . 
\end{align}
On-shell, only the first of the $\delta$-functions in $\sigma(p;k,q)$ 
contributes,
and one obtains the width appearing in the Boltzmann equations,
\begin{align}\label{equilgamma}
\Gamma_{\textbf{p}}(\omega_{\textbf{p}}) =
\left(\lambda^{\dagger}\lambda\right)_{11}\frac{2}{\omega_{\textbf{p}}} 
\int_{\bf q,k}\ p\cdot k\ f_{l\phi}(k,q)(2\pi)^4 \delta^4(p-k-q)
\equiv \Gamma_{\textbf{p}}\ ,
\end{align}
which satisfies the relations
\begin{align}
\Gamma_{\textbf{p}}(-\omega_{\textbf{p}}) =
\Gamma_{-\textbf{p}}(\omega_{\textbf{p}}) =
\Gamma_{\textbf{p}}(\omega_{\textbf{p}}) \ .
\end{align}
In the zero-width limit the spectral function (\ref{spectral2}) reduces to
the familiar expression in vacuum,
\begin{equation}
\rho_{\textbf{p}}(\omega) = 
2\pi{\rm sign}(\omega)\delta(p^{2}-M^{2})(\Slash{p}+M)C^{-1} \ .
\end{equation}

The spectral propagator is now obtained by evaluating the Fourier transform
of the spectral function (\ref{spectral2}),
\begin{align}
G_{\bf p}^-(y) = i\int^{\infty}_{-\infty}\frac{d\omega}{2\pi} 
e^{-i\omega y}\rho_{{\bf p}}(\omega)\ ,
\end{align}
which yields the final result
\begin{align}\label{gmfinal}
G_{\bf p}^-(y) = \left(i\gamma_0\cos(\omega_{\bf p}y)
+\frac{M-\textbf{p}\pmb{\gamma}}{\omega_{\bf p}}
\sin(\omega_{\bf p}y)\right)e^{-\Gamma_{\bf p}|y|/2}C^{-1}\ .
\end{align}
Compared to the free spectral function only an exponential damping factor
appears. This is a feature of the narrow-width approximation, analogous to
the scalar field case discussed in \cite{Anisimov:2008dz}.

\subsection{Equation for the statistical propagator}

We now proceed to the solution of the second Kadanoff-Baym equation 
(\ref{kb2}) which, choosing $t_i=0$, reads
\begin{align}\label{kadba}
C(i\gamma^0\partial_{t_1}-\textbf{p}\pmb{\gamma}-M)G_{\bf
p}^+(t_1,t_2)-\int^{t_1}_{0}dt' C\Sigma^-_{\bf p}(t_1-t')G_{\bf
p}^{+}(t',t_2)=\zeta_{\bf p}(t_1-t_2)\ ,
\end{align}
with the source term
\begin{align}
\zeta_{\bf p}(t_1-t_2)=-\int^{t_2}_0dt'\Sigma_{\bf p}^{+}(t_1-t')
G_{\bf p}^-(t'-t_2)\ .
\end{align}
The general solution of (\ref{kadba}) takes the form
\begin{align}\label{homplusmem}
G_{\bf p}^+(t_1,t_2)=\hat{G}_{\bf p}^+(t_1,t_2)
+G_{{\bf p},{\rm mem}}^+(t_1,t_2)\ ,
\end{align}
where $\hat{G}_{\bf p}^+(t_1,t_2)$ is the general solution of the homogeneous 
equation
\begin{align}\label{kadba0}
C(i\gamma^0\partial_{t_1}-\textbf{p}\pmb{\gamma}-M)\hat{G}_{{\bf
p}}^+(t_1,t_2)-\int^{t_1}_{0}dt' C\Sigma^-_{\bf p}(t_1-t')\hat{G}_{\bf
p}^{+}(t',t_2)=0\ ,
\end{align}
and the `memory integral', which contains non-Markovian effects, is given by
\begin{equation}\label{uiop}
G^+_{{\bf p},{\rm mem}}(t_1,t_2)=\int^{t_1}_0dt'\int^{t_2}_0dt''
G_{\bf p}^-(t_1-t')\Sigma_{\bf p}^{+}(t'-t'')
G_{\bf p}^-(t''-t_2) \ . \label{xisol}
\end{equation}
One easily verifies that the memory integral is a special solution of the
inhomogeneous equation. 

In order to evaluate the memory integral we perform a Fourier transform 
of the self-energy ($y=t_1-t_2$),
\begin{align}\label{gpluslorentz}
G_{{\bf p},{\rm mem}}^{+}&(t_1,t_2) =  \nonumber\\
&\int\frac{d\omega}{2\pi}\left(\int_{0}^{t_{1}}d\tm_{1}
G_{\bf p}^{-}(\tm_{1})e^{i\omega\tm_{1}}\right)\Sigma_{\bf p}^{+}(\omega)
\left(\int_{0}^{t_{2}}d\tm_{2}G_{\bf p}^{-}(-\tm_{2})e^{-i\omega\tm_{2}}\right)
e^{-i\omega y}\ .
\end{align}
Since the self-energy is computed with fields in thermal equilibrium, it 
satisfies the KMS condition (cf.~(\ref{SKMS}))
\begin{equation}
\Sigma_{\bf
p}^{+}(\omega)=-\frac{i}{2}\tanh\left(\frac{\beta\omega}{2}\right)
\Sigma_{\bf p}^{-}(\omega)\ .
\end{equation}

Using the expressions (\ref{Sigmarotinv}) and (\ref{gmfinal}) for 
self-energy and spectral function, respectively, which were derived in 
the previous 
section, it is now straightforward to calculate the memory integral explicitly.
Neglecting terms $\mathcal{O}(\Gamma_{\textbf{p}})$ in the numerator,
one finds
\begin{align}
\int_{0}^{t}dy e^{i\omega y}G_{\bf p}^{-}(y)\ &=
\frac{1}{\omega_{\textbf{p}}^{2}-(\omega+i\Gamma_{\bf p}/2)^{2}}\ \times \\
&\Big(i\gamma^{0}\big[\big(\omega_{\textbf{p}}\sin(\omega_{\textbf{p}}t)
+i\omega \cos(\omega_{\textbf{p}}t)\big)e^{i(\omega+i\Gamma_{\bf p}/2)t}
-i\omega\big]\nonumber\\
&+\frac{M-{\bf p}\pmb{\gamma}}{\omega_{\textbf{p}}}
\big[i\omega\big(\sin(\omega_{\textbf{p}}t)
-\omega_{\textbf{p}}\cos(\omega_{\textbf{p}}t)\big)
e^{i(\omega+i\Gamma_{\bf p}/2)t}
+\omega_{\textbf{p}}\big]\Big)C^{-1}\ , \nonumber\\
\int_{0}^{t}dy e^{-i\omega y}G_{\bf p}^{-}(-y)\ &=
\frac{1}{\omega_{\textbf{p}}^{2}-(\omega-i\Gamma_{\bf p}/2)^{2}}\ \times \\
&\Big(i\gamma^{0}\big[\big(\omega_{\textbf{p}}\sin(\omega_{\textbf{p}}t)
-i\omega \cos(\omega_{\textbf{p}}t)\big)e^{-i(\omega-i\Gamma_{\bf p}/2)t}
-i\omega\big]\nonumber\\
&+\frac{M-{\bf p}\pmb{\gamma}}{\omega_{\textbf{p}}}
\big[i\omega\big(\sin(\omega_{\textbf{p}}t)
-\omega_{\textbf{p}}\cos(\omega_{\textbf{p}}t)\big)
e^{i(\omega+i\Gamma_{\bf p}/2)t}
+\omega_{\textbf{p}}\big]\Big)C^{-1}\ .\nonumber
\end{align}
After inserting these expressions in Eq.~(\ref{gpluslorentz}) one can perform
the $\omega$-integration using Cauchy's theorem. The integrand has two 
poles\footnote{There are
further poles at $\omega_n = \pm i\pi(1+2n)/\beta$, $n$ integer. However,
their contribution to $G_{{\bf p},{\rm mem}}^{+}$ is 
${\cal O}(\Gamma_{\bf p}/M)$ and therefore negligible.}
in the upper-half plane at 
$\omega = i\Gamma_{\bf p}/2 \pm \omega_{\textbf{p}}$, and two poles in the
lower-half plane
$\omega = -i\Gamma_{\bf p}/2 \pm \omega_{\textbf{p}}$.
The choice of the
contour depends on the sign of the time variables in the exponent.
The result is a sum of the contributions from all four poles. The 
expressions appearing in the numerator can be simplified by means of 
Eqs.~(\ref{ab}) and (\ref{equilgamma}) for self-energy and equilibration 
width, respectively,
\begin{align}
\Big(\gamma^{0} + \frac{M-{\bf p}\pmb{\gamma}}{\omega_{\textbf{p}}}\Big)
\Sigma^{-}_{\textbf{p}}(\omega_{\textbf{p}})
\Big(\gamma^{0} + \frac{M-{\bf p}\pmb{\gamma}}{\omega_{\textbf{p}}}\Big)
&= 2i\Gamma_{\bf p}
\Big(\gamma^{0} + \frac{M-{\bf p}\pmb{\gamma}}{\omega_{\textbf{p}}}\Big)\ ,\\
\Big(\gamma^{0} - \frac{M-{\bf p}\pmb{\gamma}}{\omega_{\textbf{p}}}\Big)
\Sigma^{-}_{\textbf{p}}(-\omega_{\textbf{p}})
\Big(\gamma^{0} - \frac{M-{\bf p}\pmb{\gamma}}{\omega_{\textbf{p}}}\Big)
&= 2i\Gamma_{\bf p}
\Big(\gamma^{0} - \frac{M-{\bf p}\pmb{\gamma}}{\omega_{\textbf{p}}}\Big)\ .
\end{align}
Using these expressions one finally obtains for the memory integral,
changing variables from $(t_1,t_2)$ to $(t,y)$,
\begin{align}
&G^{+}_{\textbf{p},{\rm mem}}(t,y) = \label{Gmem}\\
&\; -\frac{1}{2}\tanh\left(\frac{\beta\omega_{\textbf{p}}}{2}\right)
\left(i\gamma_0\sin(\omega_{
\textbf{p}}y)-\frac{M-\textbf{p}\pmb{\gamma}}{\omega_{\textbf{p}}}
\cos(\omega_{\textbf{p}}y)\right)\left(e^{-\Gamma_{{\bf q}}|y|/2}
-e^{-\Gamma_{{\bf q}} t}\right)C^{-1}\ .\nonumber
\end{align}

Asymptotically, for $t\rightarrow \infty$, the memory integral becomes
\begin{align}\label{Geq}
G^{+\mathrm{eq}}_{\textbf{p}}(t,y) = 
-\frac{1}{2}\tanh\left(\frac{\beta\omega_{\textbf{p}}}{2}\right)
\left(i\gamma_0\sin(\omega_{\textbf{p}}y) - 
\frac{M-\textbf{p}\pmb{\gamma}}{\omega_{\textbf{p}}}
\cos(\omega_{\textbf{p}}y)\right) e^{-\Gamma_{{\bf q}}|y|/2} C^{-1}\ .
\end{align}
One easily verifies that $G^{+ \mathrm{eq}}_{\textbf{p}}(t,y)$ indeed
represents the equilibrium statistical propagator. For the Fourier transform 
one obtains
\begin{align}
G^{+ \mathrm{eq}}_{\textbf{p}}(\omega) &= 
\int_{-\infty}^{\infty}dy e^{i\omega y}G^{+ \mathrm{eq}}_{\textbf{p}}(y)
\nonumber\\
& = \frac{1}{2}\tanh\left(\frac{\beta\omega}{2}\right)
\frac{2\omega\Gamma_{\textbf{p}}(\omega)}
{\left(\omega^2-\omega_{\textbf{p}}^2\right) + 
\left(\omega\Gamma_{\textbf{p}}(\omega)\right)^2}
\left(\Slash{p}+M\right)C^{-1} \nonumber \\
& = \frac{1}{2}\tanh\left(\frac{\beta\omega}{2}\right)
\rho_{\textbf{p}}(\omega) \\
& = -\frac{i}{2}\tanh\left(\frac{\beta\omega}{2}\right)
G^{-}_{\textbf{p}}(\omega) \ ,
\end{align}
i.e., the KMS condition (cf.~(\ref{SKMS})) is indeed satisfied.

In order to obtain the general solution of the inhomogeneous Kadanoff-Baym
equation we have to add to the memory integral the general solution of the
homogeneous equation (\ref{kadba0}). This equation is identical to the
Kadanoff-Baym equation for the spectral function (\ref{SpectralEquation}) 
with $t_{2}$ playing the role of an additional parameter. Hence, the 
functional dependence of $\hat{G}^+_{\bf p}(t_1,t_2)$ 
on the first argument $t_1$  can be obtained in the same way as for the 
spectral function. Applying the Laplace transform to (\ref{kadba0}) one finds
\begin{equation}
{\tilde G}^+_{{\bf p}}(s,t_2)={1\over i\gamma_0 s-{\bf
p}\gamma-M-{\tilde\Sigma^-}(s)}i\gamma_0\hat{G}^+_{\bf p}(0,t_2)\ .
\end{equation}
The inverse Laplace transform then gives
\begin{equation}
 \hat{G}^+_{{\bf p}}(t_1,t_2) = 
-G^-_{\bf p}(t_1)Ci\gamma_0\hat{G}^+_{\bf p}(0,t_2)\ .
\end{equation}
The function $\hat{G}^+_{\bf p}(0,t_2)$ can now be determined by the symmetries
(\ref{GminusSymmetry}) and (\ref{GplusSymmetry}) of 
$\hat{G}^{\pm}_{\bf p}(t_1,t_2)$ , which imply
\begin{align}
\hat{G}^+_{{\bf p}}(t_1,t_2)^T = -\hat{G}^+_{{\bf p}}(t_2,t_1)\ .
\end{align}
This yields the result
\begin{align}\label{homsol}
\hat{G}^+_{{\bf p}}(t_1,t_2) = 
-G^-_{{\bf p}}(t_1)C\gamma_0G^+_{{\bf p}}(0,0)\gamma_0C^{-1}
G^-_{{\bf p}}(-t_2)\ ,
\end{align}
where $G^+_{{\bf p}}(0,0)$ is an antisymmetric matrix.

Let us first consider the case of thermal initial condition,
\begin{align}
G^{+{\rm eq}}_{{\bf p}}(0,0) = 
\frac{M-\textbf{p}\pmb{\gamma}}{2\omega_{{\bf p}}}
\tanh\left(\frac{\beta\omega}{2}\right)C^{-1}.
\end{align}
From Eq.~(\ref{homsol}) one then obtains
\begin{align}
\hat{G}^{+{\rm eq}}_{{\bf p}}(t_1,t_2) = -\frac{1}{2} 
\left(i\gamma_0\sin(\omega_{\textbf{p}}y)
-\frac{M-\textbf{p}\pmb{\gamma}}{\omega_{\textbf{p}}}
\cos(\omega_{\textbf{p}}y)\right)
\tanh\left(\frac{\beta\omega_{\textbf{p}}}{2}\right)
e^{-\Gamma_{{\bf q}}(t_1+t_2)/2}C^{-1}\ . 
\end{align}
Adding this expression to the memory integral 
$\hat{G}^{+}_{{\bf p},\mathrm{mem}}$ one obtains the
equilibrium statistical propagator $\hat{G}^{+{\rm eq}}_{{\bf p}}$ 
which is independent of $t=(t_1+t_2)/2$. Hence, as expected, the equilibrium 
statistical propagator is a solution of the full Kadanoff-Baym equation.

We are particularly interested in the case of vacuum initial condition,
which corresponds to zero initial abundance for heavy neutrinos in the
Boltzmann case. The vacuum propagators are obtained from the equilibrium
ones in the limit $\beta \rightarrow \infty$. Hence we choose 
\begin{align}
G^{+{\rm vac}}_{{\bf p}}(0,0) = 
\frac{M-\textbf{p}\pmb{\gamma}}{2\omega_{{\bf p}}}C^{-1} \ .
\end{align}
From Eqs.~(\ref{homplusmem}), (\ref{Gmem}) and (\ref{homsol}) one then
obtains the full solution for the statistical propagator, which interpolates
between vacuum at $t=0$ and equilibrium for $t\rightarrow\infty$, 
\begin{align}
G^{+}_{ \textbf{p}}(t,y) &= - \left(i\gamma_0\sin(\omega_{{\bf p}} y)
-\frac{M-\textbf{p}\pmb{\gamma}}{\omega_{{\bf p}}}
\cos(\omega_{{\bf p}}y)\right)\label{g+} \nonumber \\ 
&\quad\times\left[\frac{1}{2}\tanh\left(\frac{\beta\omega_{{\bf p}}}{2}\right)
e^{-\Gamma_{\bf p}|y|/2} + f_N^{eq}(\omega_{{\bf p}}) e^{-\Gamma_{\textbf{p}}
t}\right]C^{-1} . 
\end{align}
This result will be the basis for the calculation of the lepton asymmetry in
the next section. All heavy neutrino propagators can be obtained as linear
combinations of the spectral function $G^{-}_{ \textbf{p}}(y)$ and the
statistical propagator $G^{+}_{ \textbf{p}}(t,y)$. A full list is given
in Appendix~A.  

Finally, let us emphasize that the solution of the Kadanoff-Baym equation for 
the statistical propagator is not related to the equilibrium propagator
by a simple change of the distribution function from $f_N^{\rm eq}(\omega)$
to some nonequilibrium function  $f_N^{\rm eq}(t,\omega)$. This is in contrast
to the assumption made in the derivation of Quantum Boltzman equations 
\cite{DeSimone:2007rw,Garny:2009rv,Cirigliano:2009yt,Beneke:2010wd}. 
For a system close to equilibrium this assumption leads to a valid 
approximation of the Kadanoff-Baym equations \cite{Buchmuller:2000nd}, but
in general it is not justified.

\section{Lepton asymmetries}
		\begin{figure}[t]
    \psfrag{p}{$\phi$}
    \psfrag{p}{$\phi$}
    \psfrag{k}{$\bf{k}$}
    \psfrag{q}{$\bf{q}$}
    \psfrag{kp}{$\bf{k}'$}
    \psfrag{qp}{$\bf{q}'$}
    \psfrag{omega}{$(\omega_{\bf{p}},\bf{p})$}
    \psfrag{t1}{$t$}
    \psfrag{t2}{$t'$}
    \psfrag{t3}{$t_1$}
    \psfrag{t4}{$t_3$}
    \psfrag{t5}{$t_2$}
    \psfrag{l}{$l$}
    \psfrag{N}{$N$}
    \psfrag{+}{$+$}
   \centering
   \includegraphics[width=4in]{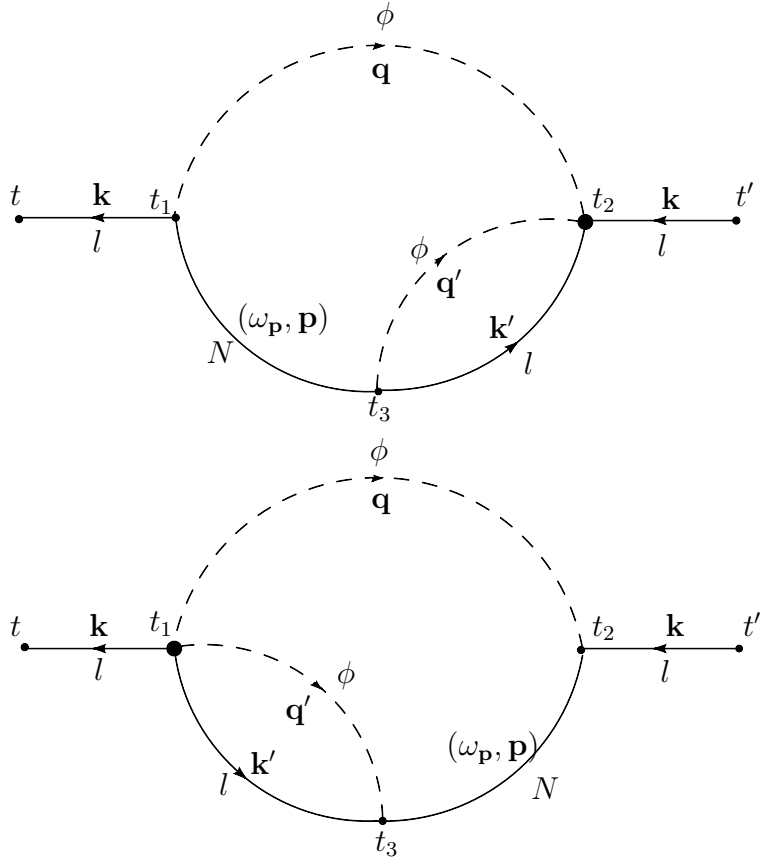} 
   \caption{Two-loop contributions to the lepton self-energies 
$\Pi^{\pm}_{{\bf k}}$, which lead to a nonzero lepton number densities.}
\label{2loopb}
\end{figure}

We are now ready to calculate the lepton asymmetry which is generated during
the approach of the heavy Majorana neutrino $N$ to thermal equilibrium. Our
starting point is the flavour non-diagonal lepton current, which is obtained 
from the statistical propagator,
\begin{align}
j^{\mu}_{ij}(x)=-{\rm tr}[\gamma^{\mu}
S^+_{Lij}(x,x')]_{x'\rightarrow x}\ .
\end{align}
Since we consider a spatially homogeneous system, $S^+_{ij}(x,x')$ only depends
on the difference $\vec{x}-\vec{x}'$, and it is convenient to perform a Fourier
transform. The zeroth component of the current, the `lepton number matrix',
is given by 
\begin{align}
L_{{\bf k}ij}(t,t') = -{\rm tr}[\gamma_0S_{L{\bf k}ij}^+(t,t')]\ . 
\end{align}
One easily verifies that for free fields in equilibrium
\begin{align}
L_{{\bf k}ii}(t,t) = f_{li}(k) - f_{\bar{l}i}(k)\ ,
\end{align}
where $f_{li}$ and $f_{\bar{l}i}$ are the distribution functions of leptons
and anti-leptons, respectively.

The lepton number matrix $L_{{\bf k}ij}(t,t')$ can be directly computed from 
the self-energy corrections to the statistical propagator shown in  
Fig.~\ref{washoutandproduction}: the external lepton
couples to Majorana neutrino and Higgs boson, and also to Higgs boson and
Higgs-lepton pair. Complex Yukawa couplings and quantum interference then lead
to a non-vanishing lepton asymmetry.   

For a homogeneous system, the Kadanoff-Baym equation for the statistical 
propagator (cf.~(\ref{kb2})) yields for each Fourier mode the equations
\begin{align}
(i\gamma^0\partial_{t}-{\bf k}\pmb{\gamma})S_{L{\bf k}}^+(t,t') =
&\int^{t}_{0}dt_1\Uppi_{\bf k}^{-}(t,t_1) S_{L{\bf k}}^+(t_1,t')\nonumber\\
&-\int^{t'}_{0}dt_1 \Uppi^+_{\bf k}(t,t_1)S_{L{\bf k}}^{-}(t_1,t') \ ,
\label{kdbleptonse2} \\
S_{L{\bf k}}^+(t,t')(-i\gamma^0\!\stackrel{\leftarrow}{\partial_{t'}}
-{\bf k}\pmb{\gamma}) =
&-\int^{t'}_{0}dt_1 S_{L{\bf k}}^+(t,t_1)\Uppi_{\bf k}^{-}(t_1,t') \nonumber\\
&+\int^{t}_{0}dt_1 S_{L{\bf k}}^{-}(t,t_1) \Uppi^+_{\bf k}(t_1,t')\ .
\label{kdbleptonse3} 
\end{align}
One then obtains for the time derivative of the lepton number matrix, dropping
flavour indices (cf.~\cite{Garny:2009qn}),\footnote{We thank C. Weniger for 
helpful discussions.}
\begin{align}\label{derivativeL}
\partial_{t}L_{{\bf k}}(t,t) 
&= i{\rm tr}\left[(i\gamma_0\partial_{t} + i\gamma_0\partial_{t'})
S^+_{L{\bf k}}(t,t')\right]_{t=t'}  \nonumber\\
&= i{\rm tr}\left[(i\gamma_0\partial_{t}-{\bf k}\pmb{\gamma})
S^+_{L{\bf k}}(t,t')
+ S_{L{\bf k}}^+(t,t')(i\gamma^0\!\stackrel{\leftarrow}{\partial_{t'}}
+{\bf k}\pmb{\gamma}) \right]_{t=t'} \nonumber \\
&= i{\rm tr}\bigg[\int^{t}_0dt_1\Uppi^-_{\bf k}(t,t_1)S^+_{L{\bf k}}(t_1,t')
-\int^{t'}_0dt_1\Uppi^+_{\bf k}(t,t_1)S^-_{L{\bf k}}(t_1,t') \nonumber\\
&\quad\quad + \int^{t'}_0dt_1 S^+_{L{\bf k}}(t,t_1)\Uppi^-_{\bf k}(t_1,t')
-\int^{t}_0dt_1 S^-_{L{\bf k}}(t,t_1)\Uppi^+_{\bf k}(t_1,t')\bigg]_{t=t'}\ .
\end{align}
Using properties of the trace and the identity between integration domains
\begin{align}
\int_{0}^{t}dt_1\int_{0}^{t_1}dt_2 \cdots + 
\int_{0}^{t}dt_2\int_{0}^{t_2}dt_1 \cdots 
=\int_{0}^{t}dt_1\int_{0}^{t}dt_2 \cdots\ ,
\end{align}
one finds
\begin{align}\label{lasymfinal1}
L_{\bf k}(t,t)=i\int^t_0dt_1\int^{t}_0dt_2\  
\mathrm{tr}\left[\Uppi_{\bf k}^-(t_1,t_2)S^+_{L{\bf k}}(t_2,t_1)
-\Uppi_{\bf k}^+(t_1,t_2)S^-_{L{\bf k}}(t_2,t_1)\right]\ .
\end{align}
Note that $\Uppi_{\bf k}^{\pm}$ and $S^{\pm}_{\bf k}$ are self-energies and
propagators of the full theory including gauge interactions of lepton and
Higgs fields. 

Using the relations for propagators and self-energies 
\begin{align}
S^+_{L{\bf k}} &= \frac{1}{2}\left(S^>_{L{\bf k}}+S^<_{L{\bf k}}\right)\ ,\quad
S^-_{L{\bf k}} = i\left(S^>_{L{\bf k}} - S^<_{L{\bf k}}\right)\ ,\\
\Uppi_{\bf k}^+ &=\frac{1}{2}\left(\Uppi_{\bf k}^>+\Uppi_{\bf k}^<\right)\ ,
\quad \Uppi_{\bf k}^- = i\left(\Uppi_{\bf k}^> - \Uppi_{\bf k}^<\right)\ ,
\end{align}
one obtains from Eq.~(\ref{lasymfinal1}) an equivalent useful 
expression for the lepton number matrix,
\begin{align}\label{lasymfinal2}
L_{\bf k}(t,t)=-\int^t_0dt_1\int^{t}_0dt_2 
{\rm tr}\left[\Uppi_{\bf k}^>(t_1,t_2)S^<_{L{\bf k}}(t_2,t_1)
-\Uppi_{\bf k}^<(t_1,t_2)S^>_{L{\bf k}}(t_2,t_1)\right]\ .
\end{align}

We want to calculate the lepton asymmetry to leading order in the small
Yukawa coupling $\lambda$, which can be achieved in a perturbative expansion.
For the heavy neutrino propagator appearing in the loop, the departure from 
the equilibrium propagator is important,\footnote{We show in Appendix~D that
the equilibrium part of the propagator does indeed not contribute to the
asymmetry.} which has been evaluated in the previous section,
\begin{align}
G_{\bf p}(t_1,t_2) = G^{\rm eq}_{\bf p}(t_1-t_2) 
+ \bar{G}_{\bf p}(t_1,t_2) \ .
\end{align}
Lepton propagators and self-energies have large equilibrium contributions
dominated by gauge interaction, with small corrections 
$\mathcal{O}(\lambda^2)$, 
\begin{align}
S_{L{\bf k}}(t_1,t_2) &= S^{\rm eq}_{L{\bf k}}(t_1-t_2) 
+ \delta S_{L{\bf k}}(t_1,t_2) \ , \\ 
\Uppi_{\bf k}(t_1,t_2) &= \Uppi^{\rm eq}_{\bf k}(t_1-t_2) 
+ \delta\Uppi_{\bf k}(t_1,t_2) \ ,
\end{align}
which include CP-violating source terms and washout terms. Clearly,
inserting $\Uppi^{\rm eq}_{\bf k}$ and $S^{\rm eq}_{\bf k}$ in 
Eq.~(\ref{lasymfinal1}) must yield $L_{\bf k}^{\rm eq}(t,t) = 0$, since no 
asymmetry
is generated in thermal equilibrium.\footnote{Note that thermal equilibrium
does not correspond to a Gaussian state \cite{Garny:2009ni}. Therefore one
has to include contributions from $n$-point functions which are not determined
by equilibrium 2-point functions. However, such terms do not contribute to
leading order in the Yukawa coupling $\lambda$.}  
As discussed in Section~2, we also neglect washout terms for simplicity. 
One then obtains for the lepton number matrix $L_{\bf k}(t,t)$ to leading 
order in $\lambda$,
\begin{align}\label{lasymfinal3}
L_{\bf k}(t,t) = i\int^t_0dt_1\int^{t}_0dt_2 
{\rm tr}\big[
&\delta\Uppi_{\bf k}^-(t_1,t_2)S^{\mathrm{eq}+}_{L{\bf k}}(t_2-t_1)
\nonumber\\
-&\delta\Uppi_{\bf k}^+(t_1,t_2)S^{\mathrm{eq}-}_{L{\bf k}}(t_2-t_1)\big]\ .
\end{align}
Here $\delta\Uppi_{\bf k}$ is given by the two-loop graphs shown in 
Fig.~\ref{2loopb}, which have to be evaluated with equilibrium propagators 
for lepton and Higgs fields and the nonequilibrium Majorana neutrino 
propagator. 

The equilibrium propagators with standard model gauge interactions remain to 
be evaluated. In the quasi-particle approximation one simply replaces 
energies $k$ by complex quasi-particle energies
$\Omega_{\bf k} = ({\bf k}^2 + m_{\rm th}^2)^{1/2} + i\gamma(k)$.
In the following we shall consider two approximations: free equilibrium 
propagators with zero chemical potential as
given in Eqs.~(\ref{D-}), (\ref{D+}) and (\ref{S-}), (\ref{S+}),
\begin{align}
\Delta^{\mathrm{eq}\pm}_{\bf k}(y) = \Delta^{\pm}_{\bf k}(y) \ , \quad
S^{\mathrm{eq}\pm}_{L{\bf k}}(y) = S^{\pm}_{L{\bf k}}(y) \ , 
\end{align}
and, as a rough approximation to full thermal propagators, free equilibrium 
propagators modified by thermal damping rates,
\begin{align}\label{damping}
\Delta^{\mathrm{eq}\pm}_{\bf k}(y) = 
\Delta^{\pm}_{\bf k}(y)e^{-\gamma_{\Phi}|y|} \ , \quad
S^{\mathrm{eq}\pm}_{L{\bf k}}(y) = S^{\pm}_{L{\bf k}}(y)e^{-\gamma_l|y|} \ .
\end{align}
Remarkably, thermal widths turn out to be qualitatively more important than
thermal masses, as we shall explain in Section~6.

The two contributions to the self-energy $\delta\Uppi_{{\bf k}ij}$  
(cf.~Fig.~\ref{2loopb}),
\begin{align}\label{pipplus}
\delta\Uppi_{{\bf k}ij}(t_1,t_2) = \Uppi^{(1)}_{{\bf k}ij}(t_1,t_2)
+\Uppi^{(2)}_{{\bf k}ij}(t_1,t_2)\ ,
\end{align}
factorize into a product of Yukawa couplings, which contains
the flavour dependence, and a trace of thermal propagators,
\begin{align}
\Uppi^{(1)}_{{\bf k}ij}(t_1,t_2) &=
-3i \lambda^*_{i1}\left(\eta\lambda^*\right)_{j1}
\Uppi^{(1)}_{\bf k}(t_1,t_2) \ ,\\
\Uppi^{(2)}_{{\bf k}ij}(t_1,t_2) &=
3i \left(\eta^*\lambda\right)_{i1}\lambda_{j1}
\Uppi^{(2)}_{\bf k}(t_1,t_2) \ .
\end{align}
In the case of free equilibrium propagators for lepton and Higgs fields, 
we obtain for the self-energies $\Uppi_{\textbf{k}}^{(1,2)>}$ and 
$\Uppi_{ \textbf{k}}^{(1,2)<}$:
\begin{align}
\Uppi_{ \textbf{k}}^{(1)>}(t_1,t_2)\ =\ &\int_0^{\infty}dt_3
\int\frac{d^3 {\bf q}}{(2\pi)^3}\frac{d^3 {\bf q}'}{(2\pi)^3} \nonumber\\
&\times[\tilde{G}^>_{\textbf{p}}(t_1,t_3)S^{11}_{{\bf k}'}(t_2-t_3)
\Delta^{11}_{{\bf q}'}(t_2-t_3)\Delta^{<}_{\bf q}(t_2-t_1)\nonumber\\
&-\tilde{G}^{22}_{\textbf{p}}(t_1,t_3)S^{<}_{{\bf k}'}(t_2-t_3)
\Delta^{<}_{ {\bf q}'}(t_2-t_3)\Delta^{<}_{\bf q}(t_2-t_1)]P_L\ ,\\
\Uppi_{\textbf{k}}^{(1)<}(t_1,t_2)\ =\ &\int_0^{\infty}dt_3
\int\frac{d^3 {\bf q}}{(2\pi)^3}\frac{d^3 {\bf q}'}{(2\pi)^3} \nonumber\\
&\times[\tilde{G}^{11}_{\textbf{p}}(t_1,t_3)S^{>}_{{\bf k}'}(t_2-t_3)
\Delta^{>}_{{{\bf q}'}}(t_2-t_3)\Delta^{>}_{\bf q}(t_2-t_1)\nonumber\\
&-\tilde{G}^{<}_{\textbf{p}}(t_1,t_3)S^{22}_{{\bf k}'}(t_2-t_3)
\Delta^{22}_{ {\bf q}'}(t_2-t_3)\Delta^{>}_{ q}(t_2-t_1)]P_L\ ,\\
\Uppi_{\textbf{k}}^{(2)>}(t_1,t_2)\ =\ &\int_0^{\infty}dt_3
\int\frac{d^3 {\bf q}}{(2\pi)^3}\frac{d^3 {\bf q}'}{(2\pi)^3} \nonumber\\
&\times[\tilde{G}^{<}_{\textbf{p}}(t_2,t_3)S^{22}_{{\bf k}'}(t_3-t_1)
\Delta^{22}_{{\bf q}'}(t_3-t_1)\Delta^{<}_{\bf q}(t_2-t_1)\nonumber\\
&-\tilde{G}^{11}_{\textbf{p}}(t_2,t_3)S^{<}_{{\bf k}'}(t_3-t_1)
\Delta^{<}_{ {\bf q}'}(t_3-t_1)\Delta^{<}_{ \textbf{q}}(t_2-t_1)]P_L\ ,\\
\Uppi_{\textbf{k}}^{(2)<}(t_1,t_2)\ =\ &\int_0^{\infty} dt_3
\int\frac{d^3 {\bf q}}{(2\pi)^3}\frac{d^3 {\bf q}'}{(2\pi)^3} \nonumber\\
&\times[\tilde{G}^{22}_{\textbf{p}}(t_2,t_3)S^{>}_{{\bf k}'}(t_3-t_1)
\Delta^{>}_{{\bf q}'}(t_3-t_1)\Delta^{>}_{\textbf{q}}(t_2-t_1)\nonumber\\
&-\tilde{G}^{>}_{\textbf{p}}(t_2,t_3)S^{11}_{{\bf k}'}(t_3-t_1)
\Delta^{11}_{{\bf q}'}(t_3-t_1)\Delta^{>}_{\textbf{q}}(t_2-t_1)]P_L\;.
\end{align}
Due to the chiral projections at the vertices, only the scalar parts of the
nonequilibrium Majorana propagators contribute, which are the same for
$\tilde{G}_{{\bf p}}^>$, $\tilde{G}_{{\bf p}}^<$, $\tilde{G}_{{\bf p}}^{11}$
and $\tilde{G}_{{\bf p}}^{22}$ (cf.~Eqs. (\ref{Gne1}) - (\ref{Gne6})), 
\begin{align}
P_L\bar{G}_{{\bf p}}(t,t') C P_L = 
\tilde{G}_{{\bf p}}(t,t') P_L \ , \quad
\tilde{G}_{{\bf p}}(t,t') = 
\frac{M}{\omega_{\textbf{p}}}\cos(\omega_{\textbf{p}}(t-t'))
f_N^{eq}(\omega_{\bf p})e^{-\Gamma_{{\bf p}} (t+t')/2}\ .
\end{align}

The number of terms which contribute to the asymmetry $L_{\bf k}(t,t)$ can be 
significantly reduced by means of the following symmetry properties of the 
massless propagators:
\begin{align}
S^{>}_{\bf k}(y)^* &= CS^{<}_{\bf k}(y)C^{-1}\ , &
S^{11}_{\bf k}(y)^* &= CS^{22}_{\bf k}(y)C^{-1}\ , \\
\Delta^{>}_{\bf q}(y)^* &= \Delta^{<}_{\bf q}(y)\ , &
\Delta^{11}_{\bf q}(y)^* &= \Delta^{22}_{\bf q}(-y)\ , \\
S^{<}_{\bf k}(y) &= \gamma_5 S^{>}_{\bf -k}(-y)\gamma_5\ , &
S^{11}_{\bf k}(y) &= \gamma_5 S^{11}_{\bf -k}(-y)\gamma_5\ , \\
\Delta^{<}_{\bf q}(y) &= \Delta^{>}_{\bf q}(-y)\ , &
\Delta^{11}_{\bf q}(y) &= \Delta^{11}_{\bf q}(-y)\ .  
\end{align}
Employing these transformation properties one can derive the following useful
relations among different contributions to the integrand of 
Eq.~(\ref{lasymfinal2}):
\begin{align}
\mathrm{tr}\left[
\Uppi_{\textbf{k}}^{(1,2)>}(t_1,t_2)S^{<}_{{\bf k}}(t_2-t_1)\right] &=
-\mathrm{tr}\left[
\Uppi_{\textbf{k}}^{(1,2)<}(t_1,t_2)S^{>}_{{\bf k}}(t_2-t_1)\right]^* \ ,\\
\mathrm{tr}\left[
\Uppi_{\textbf{k}}^{(1)>}(t_1,t_2)S^{<}_{{\bf k}}(t_2-t_1)\right] &=
-\mathrm{tr}\left[
\Uppi_{\textbf{k}}^{(2)<}(t_2,t_1)S^{>}_{{\bf k}}(t_1-t_2)\right]\ .
\end{align}

Using these relations one obtains from Eq.~(\ref{lasymfinal2}) the compact 
expression for the lepton asymmetry
\begin{align}
L_{{\bf k}ii}(t,t) = \ 
&12\ \mathrm{Im}\{\lambda^*_{i1}\left(\eta\lambda^*\right)_{i1}\}
\nonumber\\
&\times \int_0^{t}dt_1\int_0^{t}dt_2\ \mathrm{Re}\left(
\mathrm{tr}\left[
\Uppi_{\textbf{k}}^{(2)>}(t_1,t_2)S^{<}_{{\bf k}}(t_2-t_1)\right]\right) \ .
\end{align}
Since $\mathrm{Im}\{\lambda^*_{i1}\left(\eta\lambda^*\right)_{j1}\}
= 16\pi\epsilon_{ij}/(3M)$ (cf.~Eq.~(\ref{cpeps})),
the leading dependence of the flavour-diagonal lepton asymmetry 
$L_{{\bf k}ii}(t,t)$ on the Yukawa couplings is identical to the dependence 
of the difference $f_{\mathrm{L}i}(t,k)$ of lepton and anti-lepton 
distribution functions appearing in the Boltzmann equations.

To proceed further in the evaluation of $L_{{\bf k}ii}(t,t)$, the following
relation can be used to simplify the integrand,
\begin{align}
&S^{22}_{{\bf k}}(y)\Delta^{22}_{\bf q}(y)
- S^{<}_{{\bf k}}(y)\Delta^{<}_{\bf q}(y) = \nonumber\\
&\ \frac{\Theta(-y)}{2 q}\bigg[\gamma^0\left(
\coth\left(\frac{\beta q}{2}\right)\cos(ky)\cos(qy)
-\tanh\left(\frac{\beta k}{2}\right)\sin(ky)\sin(qy)\right)\nonumber\\
&\quad -i \frac{M-{\bf k}\gamma}{k}\left(
\tanh\left(\frac{\beta k}{2}\right)\cos(ky)\sin(qy)
+\coth\left(\frac{\beta q}{2}\right)\sin(ky)\cos(qy)\right)\bigg]\ .
\end{align}
One then obtains for the real part of the sum of products of thermal lepton 
and Higgs propagators
($y_{ij} = t_i-t_j$),
\begin{align}
\mathrm{Re}\big(
\mathrm{tr}\big[\big(S^{22}_{{\bf k'}}(y_{31})&\Delta^{22}_{\bf q'}(y_{31})
- S^{<}_{{\bf k'}}(y_{31})\Delta^{<}_{\bf q'}(y_{31})\big) 
S^{<}_{{\bf k}}(y_{21})\big]\Delta^{<}_{\bf q}(y_{21})\big) = \nonumber\\
-\frac{\Theta(y_{13})}{16qq'}
\bigg[&\left(\coth\left(\frac{\beta q}{2}\right)
\Big(\cos((k+q)y_{21})+\cos((k-q)y_{21})\Big)\right.\nonumber\\
&\left.+\tanh\left(\frac{\beta k}{2}\right)
\Big(\cos((k+q)y_{21})-\cos((k-q)y_{21})\Big)\right)\nonumber\\
\times&\left(\coth\left(\frac{\beta q'}{2}\right)
\Big(\cos((k'+q')y_{31})+\cos((k'-q')y_{31})\Big)\right.\nonumber\\
&+\left.\tanh\left(\frac{\beta k'}{2}\right)
\Big(\cos((k'+q')y_{31})-\cos((k'-q')y_{31})\Big)\right)\nonumber\\
+\frac{{\bf k}\cdot{\bf k'}}{kk'}&\left(\coth\left(\frac{\beta q}{2}\right)
\Big(\sin((k+q)y_{21})+\sin((k-q)y_{21})\Big)\right.\nonumber\\
&\left.+\tanh\left(\frac{\beta k}{2}\right)
\Big(\sin((k+q)y_{21})-\sin((k-q)y_{21})\Big)\right) \nonumber\\
\times&\left(\coth\left(\frac{\beta q'}{2}\right)
\Big(\sin((k'+q')y_{31})+\sin((k'-q')y_{31})\Big)\right.\nonumber\\
&+\left.\tanh\left(\frac{\beta k'}{2}\right)
\Big(\sin((k'+q')y_{31})-\sin((k'-q')y_{31})\Big)\right)\bigg] \ .
\end{align}
Defining the linear combinations of lepton and Higgs distribution functions
(cf.~(\ref{leptonhiggs})),
\begin{align}
f_{l\phi}(k,q)=1-f_l(k)+f_{\phi}(q)\ , \quad
\bar{f}_{l\phi}(k,q)=f_{l}(k)+f_{\phi}(q)\ ,
\end{align}
and using the relations
\begin{align}
\coth\left(\frac{\beta q}{2}\right)+\tanh\left(\frac{\beta k}{2}\right)
&= 2f_{l\phi}(k,q)\ ,\\
\coth\left(\frac{\beta q}{2}\right)-\tanh\left(\frac{\beta k}{2}\right)
&= 2\bar{f}_{l\phi}(k,q)\ ,
\end{align}
one finds
\begin{align}
\label{LKB}
L_{{\bf k}ii}(t,t)= -\epsilon_{ii}\ &32\pi
\int_0^{t}dt_1\int_0^{t}dt_2\int_0^{t_2}dt_3\int_{{\bf q},{\bf q'}}
\frac{1}{\omega_{\bf p}}f^{eq}_N(\omega_{\bf p})e^{-\frac{\Gamma}{2}(t_1+t_3)}
\cos(\omega_{\bf p} y_{31})\nonumber\\
\times\bigg[&\Big(f_{l\phi}(k,q)\cos((k+q)y_{21})
+\bar{f}_{l\phi}(k,q)\cos((k-q)y_{21})\Big)\nonumber\\
\times &\Big(f_{l\phi}(k',q')\cos((k'+q')y_{23})
+\bar{f}_{l\phi}(k',q')\cos((k'-q')y_{23})\Big)\nonumber\\
+\frac{{\bf k}\cdot{\bf k'}}{kk'}
\bigg(&\Big(f_{l\phi}(k,q)\sin((k+q)y_{21})
+\bar{f}_{l\phi}(k,q)\sin((k-q)y_{21})\Big) \nonumber\\
\times &\Big(f_{l\phi}(k',q')\sin((k'+q')y_{23})
+\bar{f}_{l\phi}(k',q')\sin((k'-q')y_{23})\Big)\bigg)\bigg] \ ,
\end{align}
where we have again used the notation
\begin{align}
\int_{\bf q}\cdots=\int \frac{d^3{\bf q}}{(2\pi)^32q}\cdots\ . \nonumber
\end{align}

The functions $f_{l\phi}$ and $\bar{f}_{l\phi}$ are well known from Weldon's
analysis of discontinuities in finite-temperature field theory 
\cite{Weldon:1983jn}.
The sum of statistical factors
\begin{align}
f_{l\phi}(k,q) = (1-f_l(k))(1+f_{\phi}(q)) +  f_l(k)f_{\phi}(q) 
\end{align}
corresponds to decays and inverse decays of the massive Majorana neutrinos
whereas 
\begin{align}
\bar{f}_{l\phi}(k,q) = f_{\phi}(q)(1-f_l(k)) +  f_l(k)(1 + f_{\phi}(q))
\end{align}
accounts for their disappearance or appearance where a single quant, lepton or
Higgs, is absorbed from or emitted into the thermal bath. The function 
$f_{l\phi}$ contains the vacuum contribution, i.e., $f_{l\phi} \rightarrow 1$
as $\beta \rightarrow \infty$, whereas $\bar{f}_{l\phi} \rightarrow 0$.
 
We now have to perform the three time integrations in Eq.~(\ref{LKB}). It
is convenient to express the products of cosine's and sine's as sum of
products of exponentials. Each term then becomes a sum of four exponentials, 
where the energies $\omega$, $k\pm q$ and $k'\pm q'$ appear in different
linear combinations, and the four complex conjugate exponentials.
As an example, consider the integral
\begin{align}
\mathcal{I}(t) = \int_0^{t}dt_1\int_0^{t}dt_2\int_0^{t_2}dt_3
e^{-i\Omega_1t_1+i\Omega_2t_2+i\Omega_3t_3}e^{-\frac{\Gamma}{2}(t_1+t_3)}\ ,
\label{I}
\end{align}
with $\Omega_1=\omega_{\bf p}-k-q$,  $\Omega_3=\omega_{\bf p}-q'-k'$, and 
$\Omega_2 = \Omega_1 - \Omega_3 = k'+q'-k-q$. A straightforward calculation
yields
\begin{align}\label{ReI}
\mathcal{I}(t) + \mathcal{I}^*(t) =
\frac{-\Gamma\left(e^{-\Gamma t}+\cos(\Omega_2 t) - e^{-\frac{\Gamma t}{2}}
\left(\cos(\Omega_1 t) + \cos(\Omega_3 t)\right)\right) + \mathcal{O}(t)}
{(\Omega_1^2+\frac{\Gamma^2}{4})(\Omega_3^2 +\frac{\Gamma^2}{4})} \ ,
\end{align}
where 
\begin{align}\label{O}
\mathcal{O}(t) =
\frac{2\Omega_1\Omega_3 +\frac{\Gamma^2}{2}}{\Omega_2}
\left(\sin(\Omega_2 t) - e^{-\frac{\Gamma t}{2}}\left(\sin(\Omega_1 t)
-\sin(\Omega_3 t)\right)\right)
\end{align}
is of higher order in $\Gamma$ at $\Omega_{1,3} = 0$. Hence, this term
does not contribute to the lepton asymmetry at leading order in $\Gamma$,
i.e., in the Yukawa couplings.

The two contributions in Eq.~(\ref{LKB}), without and with the prefactor
${\bf k}\cdot{\bf k'}/(kk')$, add up to a single term proportional to
$k\cdot k'/(kk')$ where $k\cdot k'$ denotes the product of 4-vectors.
This is a consequence of Lorentz invariance of the vacuum contribution.
The full result is now easily obtained from Eqs.~(\ref{LKB}) and
(\ref{ReI}) by adding the contributions with
reversed sign of $q$ and/or $q'$, accompanied by the corresponding 
substitution $f_{l\Phi} \rightarrow \bar{f}_{l\Phi}$. Omitting the
subleading terms $\mathcal{O}$ (cf.~(\ref{O})), one finally obtains
\begin{align} \label{lkb1}
L_{{\bf k}ij}(t,t)\ =\ \sum_{a=1}^4  L^a_{{\bf k}ij}(t,t) \ , 
\end{align}
where
\begin{align}\label{lkb2}
L^a_{{\bf k}ii}(t,t) =  -\epsilon_{ii}\ 8\pi 
\int_{\bf q,q'} \frac{k\cdot k'}{kk'\omega_{\bf p}}\ f_N^{eq}(\omega_{\bf p})\ 
\frac{1}{2}\Gamma
\sum_{\alpha,\beta=\pm}\hat{L}^a_{{\bf k,q,q'}}(t;\alpha,\beta) 
\end{align}
and
\begin{align}
\hat{L}^1_{{\bf k,q,q'}}(t;\alpha,\beta) &=  
\frac{f_{l\phi}(k,q) f_{l\phi}(k',q')}
{((\omega_{\bf p}-\alpha(k+q))^2+\frac{\Gamma^2}{4})
((\omega_{\bf p}-\beta(k'+q'))^2+\frac{\Gamma^2}{4})} \nonumber\\
&\times\Big(e^{-\Gamma t}+\cos[(\alpha(k+q)-\beta(k'+q'))t] \nonumber\\
&\; -e^{-\frac{\Gamma t}{2}}\big(\cos[(\omega_{\bf p}-\alpha(k+q))t]
+\cos[(\omega_{\bf p}-\beta(k'+q'))t]\big)\Big) \ , \label{lkb3}\\
\hat{L}^2_{{\bf k,q,q'}}(t;\alpha,\beta) &=  
\frac{\bar{f}_{l\phi}(k,q) f_{l\phi}(k',q')}
{((\omega_{\bf p}-\alpha(k-q))^2+\frac{\Gamma^2}{4})
((\omega_{\bf p}-\beta(k'+q'))^2+\frac{\Gamma^2}{4})} \nonumber\\
&\times\Big(e^{-\Gamma t}+\cos[(\alpha(k-q)-\beta(k'+q'))t] \nonumber\\
&\; -e^{-\frac{\Gamma t}{2}}\big(\cos[(\omega_{\bf p}-\alpha(k-q))t]
+\cos[(\omega_{\bf p}-\beta(k'+q'))t]\big)\Big) \ , \label{lkb4}\\
\hat{L}^3_{{\bf k,q,q'}}(t;\alpha,\beta) &=  
\frac{f_{l\phi}(k,q) \bar{f}_{l\phi}(k',q')}
{((\omega_{\bf p}-\alpha(k+q))^2+\frac{\Gamma^2}{4})
((\omega_{\bf p}-\beta(k'-q'))^2+\frac{\Gamma^2}{4})} \nonumber\\
&\times\Big(e^{-\Gamma t}+\cos[(\alpha(k+q)-\beta(k'-q'))t] \nonumber\\
&\; -e^{-\frac{\Gamma t}{2}}\big(\cos[(\omega_{\bf p}-\alpha(k+q))t]
+\cos[(\omega_{\bf p}-\beta(k'-q'))t]\big)\Big) \ , \label{lkb5}\\
\hat{L}^4_{{\bf k,q,q'}}(t;\alpha,\beta) &=  
\frac{f_{l\phi}(k,q) f_{l\phi}(k',q')}
{((\omega_{\bf p}-\alpha(k-q))^2+\frac{\Gamma^2}{4})
((\omega_{\bf p}-\beta(k'-q'))^2+\frac{\Gamma^2}{4})} \nonumber\\
&\times\Big(e^{-\Gamma t}+\cos[(\alpha(k-q)-\beta(k'-q'))t] \nonumber\\
&\; -e^{-\frac{\Gamma t}{2}}\big(\cos[(\omega_{\bf p}-\alpha(k-q))t]
+\cos[(\omega_{\bf p}-\beta(k'-q'))t]\big)\Big) \ . \label{lkb6}
\end{align}
This expression contains off-shell and memory effects which are not contained
in Boltzmann equations. A detailed comparison will be given in the following 
section.

So far we have neglected the thermal damping widths of lepton and Higgs
fields due to gauge interactions, which are known to be much larger than 
the width of the heavy Majorana neutrino, 
$\gamma_{l} \sim \gamma_{\Phi} \sim g^2 T \gg \lambda^2 M \sim 
\Gamma$, for $M\lsim T$. To estimate their effect we replace the free 
equilibrium propagators by
\begin{align}
\Delta^{\mathrm{eq}\pm}_{\bf k}(y) = 
\Delta^{\pm}_{\bf k}(y)e^{-\gamma_{\Phi}|y|} \ , \quad
S^{\mathrm{eq}\pm}_{\bf k}(y) = S^{\pm}_{\bf k}(y)e^{-\gamma_l|y|} \ . 
\end{align}
This has a drastic effect on the calculation described above. For the
dominant term in Eq.~(\ref{lkb2}), $\hat{L}^1_{{\bf k,q,q'}}$ with 
$\alpha = \beta = 1$, where the energy dominators can be 
$\mathcal{O}(\Gamma^2)$, one now finds ($\gamma = \gamma_l + \gamma_{\Phi}$),
\begin{align}\label{lkbtherm}
\bar{L}_{{\bf k}ii}(t,t) = -\epsilon_{ii}\ 16\pi &\int_{{\bf q,q'}}
\frac{k\cdot k'}{kk'\omega_{\bf p}}\nonumber\\
\times&\frac{\gamma\gamma'}
{((\omega_{\bf p}-k-q)^2+ \gamma^{2})
((\omega_{\bf p}-k'-q')^2+ \gamma^{'2})}\nonumber\\
\times &f_{l\phi}(k,q) f_{l\phi}(k',q') f_N^{eq}(\omega_{\bf p})\nonumber\\
\times&\frac{1}{\Gamma}\left(1-e^{-\Gamma t}\right)\ ,
\end{align}
where
$\gamma=\gamma(k,q)$ and $\gamma'=\gamma'(k',q')$.
Note that now all memory effects have disappeared.

\section{Boltzmann vs Kadanoff-Baym}
Let us now consider in detail the relation between the two results obtained
for the lepton asymmetry: Eq.~(\ref{soll1}) from the Boltzmann equations and 
Eqs.~(\ref{lkb1}) - (\ref{lkb6}) and (\ref{lkbtherm}) from the 
Kadanoff-Baym equations. 

Clearly, the overall CP asymmetry is identical in both cases and also the 
momentum integrations
are very similar. Compared to the Boltzmann result the Kadanoff-Baym result
has an additional statistical lepton-Higgs factor and expected off-shell energy
denominators. Furthermore, there are 16 different terms corresponding to
the various combinations of decay and inverse decay, appearance and
dissappearance. The most striking difference is the time dependence
of the integrand: the Boltzmann result has a simple exponential behaviour
whereas the Kadanoff-Baym result has terms rapidly oscillating with time
with frequencies $\mathcal{O}(M) \gg \Gamma$, a manifestation of memory 
effects. 

The time-dependence is contained in the integral $\mathcal{I}(t)$ given in
Eq.~(\ref{I}). Defining
\begin{align}
\bar{\Omega}_1 = \Omega_1 + \frac{i}{2}\Gamma \ , \quad
\bar{\Omega}_3 = \Omega_3 + \frac{i}{2}\Gamma \ , 
\end{align}
and using the identities $t_3 = t_1 + (t_2 - t_1) + (t_3 - t_2)$ and
$\Omega_2 = \Omega_1 - \Omega_3$, one has
(cf.~(\ref{I})),
\begin{align}
\mathcal{I}(t) = \int_0^{t}dt_1 e^{-\Gamma t_1}
\int_{-t_1}^{t-t_1} dt_{21}\int^0_{-t_{2}} dt_{32}\
e^{i\bar{\Omega}_1t_{21} + i\bar{\Omega}_3 t_{32}}\ ,
\label{Ibetter}
\end{align}
where $t_{ij} = t_i-t_j$. After performing the time-integrations, one
obtains the result
\begin{align}
\mathcal{I}(t) = \frac{1}{i\bar{\Omega}_3}\left[\frac{1}{|\bar{\Omega}_1|^2}
\left(e^{i\bar{\Omega}_1t} - 1\right)\left(e^{-i\bar{\Omega}_1^* t} - 1\right)
- \frac{1}{\Omega_2 \bar{\Omega}_1^*} 
\left(e^{i\Omega_2 t} - 1\right)\left(e^{-i\bar{\Omega}_1^* t} - 1\right) 
\right]\ ,
\label{Icompact}
\end{align}
which satisfies
\begin{align}
\mathcal{I}(0) = \mathcal{I}'(0) = \mathcal{I}''(0) = 0 \ , \quad 
\mathcal{I}'''(0) \neq 0 \ .
\end{align}
For large times, $t \gg 1/\Gamma$, there remains a term oscillating with time,
\begin{align}
\mathcal{I}(t) \sim \frac{1}{i\bar{\Omega}_3}\left[\frac{1}{|\bar{\Omega}_1|^2}
+ \frac{1}{\Omega_2 \bar{\Omega}_1^*} 
\left(e^{i\Omega_2 t} - 1\right) \right]\ .
\label{Iosc}
\end{align}
This is in contrast to the Boltzmann result whose time-dependence is  given by
\begin{align}
\mathcal{I}_B(t) = \frac{1-e^{-\Gamma t}}{\Gamma}\ ,
\end{align}
with 
\begin{align}
\mathcal{I}_B(0) = 0 \ , \quad \mathcal{I}'_B(0) \neq 0 \ , 
\end{align}
and $\mathcal{I}_B(t) \sim 1/\Gamma = \mathrm{const}$ for large times 
$t \gg 1/\Gamma$.

Where is the Boltzmann result hidden in the Kadanoff-Baym result, and in
which limit is it recovered? To answer this question it is instructive to
consider a modified integral $\bar{\mathcal{I}}(t)$, where thermal damping 
rates $\gamma \sim \gamma' \sim g^2 T$ are included, which affect the 
dependence on the time differences $|t_2-t_1|$ and $|t_3-t_2|$ 
(cf.~Fig.~\ref{2loopb}),
\begin{align}
\bar{\mathcal{I}}(t) = \int_0^{t}dt_1 e^{-\Gamma t_1}
\int_{-t_1}^{t-t_1} dt_{21}\int^0_{-t_{2}} dt_{32}\
e^{i\bar{\Omega}_1t_{21} - \gamma |t_{21}|} \ 
e^{i\bar{\Omega}_3 t_{32} - \gamma' |t_{32}|}\ .
\label{Ibettertherm}
\end{align}
Compared to Eq.~(\ref{I}) the main difference is that the damping term in
the $t_{21}$-integration changes sign at $t_{21}=0$. This is in contrast to
the damping due to the Majorana neutrino decay width $\Gamma$. 

Carrying out the time-integrations one now obtains the result
\begin{align}
\bar{\mathcal{I}}(t) = \frac{1}{i\bar{\Omega}_3 + \gamma'}
\bigg[&\frac{1}{(i\bar{\Omega}_1-\gamma)(-i\bar{\Omega}_1^* + \gamma)}
e^{(i\bar{\Omega}_1 - \gamma)t}
\left(e^{(-i\bar{\Omega}_1^* + \gamma)t} - 1\right) \nonumber \\
-&\frac{1}{(i\bar{\Omega}_1+\gamma)(-i\bar{\Omega}_1^* - \gamma)}
\left(e^{(-i\bar{\Omega}_1^* - \gamma)t} - 1\right) \nonumber \\
- &\frac{1}{(i\Omega_2 - \gamma - \gamma')(-i\bar{\Omega}_1^* + \gamma)} 
e^{(i\Omega_2 - \gamma - \gamma')t}
\left(e^{(-i\bar{\Omega}_1^* + \gamma)t} - 1\right) \nonumber \\
+ &\frac{1}{(i\Omega_2 + \gamma - \gamma')(-i\bar{\Omega}_1^* - \gamma)} 
\left(e^{(-i\bar{\Omega}_1^* - \gamma)t} - 1\right) \nonumber \\
+ &\frac{2\gamma}{\bar{\Omega}_1^2 + \gamma^2}\frac{1-e^{-\Gamma t}}{\Gamma}
-\frac{2\gamma}{(i\bar{\Omega}_3^* + \gamma')
((i\Omega_2-\gamma')^2 - \gamma^2)}
\left(e^{(-i\bar{\Omega}_3^* - \gamma')t} - 1\right) \bigg]\ .
\label{Icompacttherm}
\end{align}
The first four terms reduce to Eq.~(\ref{Icompact}) for $\gamma = \gamma' = 0$.
Particularly interesting is the last line in Eq.~(\ref{Icompacttherm}), which
is a contribution from the point $t_{21} = t_2 - t_1 = 0$, where the damping
term changes sign. This local contribution contains the only term which is 
enhanced by $1/\Gamma$ and has Boltzmann-like time-dependence,
\begin{align}\label{IB}
\bar{\mathcal{I}}(t)\ \supset\ \mathcal{I}_B(t) =
\frac{2\gamma}{(i\bar{\Omega}_3 + \gamma')
(\bar{\Omega}_1^2 + \gamma^2)}\frac{1-e^{-\Gamma t}}{\Gamma} \ .
\end{align}
Note that as consequence of thermal damping all oscillatory terms are 
exponentially suppressed for times $t > 1/\gamma$,
\begin{align}
\bar{\mathcal{I}}(t) \sim \frac{1}{i\bar{\Omega}_3 + \gamma'}
\bigg[
&\frac{2\gamma}{\bar{\Omega}_1^2 + \gamma^2}\frac{1-e^{-\Gamma t}}{\Gamma}
+\frac{2\gamma}{(i\bar{\Omega}_3^* + \gamma')
((i\Omega_2-\gamma')^2 - \gamma^2)} \nonumber\\
+ &\frac{1}{(i\bar{\Omega}_1 + \gamma)(-i\bar{\Omega}_1^* - \gamma)}
- \frac{1}{(i\Omega_2 + \gamma - \gamma')(-i\bar{\Omega}_1^* -\gamma)}\bigg]\ .
\end{align}
The Boltzmann-like term (\ref{IB}), which originates from the point 
$t_2 = t_1$, vanishes for $\gamma = 0$. 

What is the order of magnitude of the lepton asymmetry (\ref{lkb1}) 
relative to the Boltzmann result in the case $\gamma = \gamma' = 0$? 
The Kadanoff-Baym result depends on $\tau = \Gamma t$, like the Boltzmann
result, and in addition on the dimensionless parameter $\Gamma/M \ll 1$.
In appendix~C we shown that
\begin{align}\label{zero}
\frac{L_{\bf k}(t,t)}{f_L(t,k)} \rightarrow 0\ , \qquad \mathrm{for}\quad
\frac{\Gamma}{M} \rightarrow 0\ ,\; \tau=\Gamma t\;\; \mathrm{fixed}\ .
\end{align}
Hence, in this zero-width limit, due to rapid oscillations of the integrand,
the lepton asymmetry obtained from the Kadanoff-Baym equation is at least
$\mathcal{O}(\Gamma/M)$ relative to the Boltzmann lepton asymmetry.

We are thus led to the conclusion that the lepton asymmetry obtained from
the Kadanoff-Baym equations does not contain the Boltzmann result as
limiting case as long as free equilibrium propagators are used for lepton
and Higgs fields. This may not be too surprising. After all, the underlying
assumption in our calculation has been that (gauge) interactions, much faster 
than heavy neutrino decay, establish kinetic equilibrium for leptons and   
Higgs particles. These interactions will unavoidably lead to thermal
damping widths much larger than $\Gamma$. If these interactions are not
taken into account in the calculation of the lepton asymmetry, one misses
the main contribution and obtains a misleading result. This means that at 
present the best estimate for the full quantum mechanical lepton asymmetry 
is given by Eq.~(\ref{lkbtherm}), which leads to a temperature dependent
suppression compared to the Boltzmann result.

Note that the proposed incorporation of thermal damping rates leads to a 
Boltzmann-like result, Eq.~(\ref{lkbtherm}), which is valid for 
$t\gsim 1/\Gamma$. For $t < 1/\Gamma$, all terms have to be kept, and one
has $\partial_t \bar{L}_{\bf k}(t,t)|_{t=0} = 0$, which is a property of 
the exact result (\ref{lasymfinal1}), contrary to the Boltzmann approximation.

\section{Numerical analysis}
Let us now quantitatively compare the Boltzmann result (\ref{soll2}) for the 
lepton asymmetry 
\begin{align}
f_{Li}(t,k) = f_{li}(t,k) - f_{\bar{l}i}(t,k) \nonumber
\end{align}
with the Kadanoff-Baym result for the lepton asymmetry
\begin{align}
L_{{\bf k}ii}(t,t) = -{\rm tr}[\gamma_0 S_{L{\bf k}ii}^+(t,t)]\ . 
\end{align}
For free fields in thermal equilibrium both expressions are identical.
For the Kadanoff-Baym result we use Eq.~(\ref{lkbtherm}) which includes
the estimated effect of thermal widths for lepton and Higgs fields.

As shown in Appendix~C, the Boltzmann result (\ref{soll2}) can be reduced
to a two-dimensional momentum integral (cf.~(\ref{intboltz})),
\begin{align}
f_{Li}(t,k) = - \frac{\epsilon_{ii}}{4\pi}\ F_{\rm B}(k,\beta)\
\frac{1}{\Gamma}\left(1-e^{-\Gamma t}\right)\ ,
\end{align}
where we have defined 
\begin{align}
F_{\rm B}(k,\beta) =
\frac{1}{k}&\int_{p_{\rm min}(k)}^{\infty}dp
\int_{k'_{\rm min}(p)}^{k'_{\rm max}(p)} k'dk' 
\frac{1}{\omega_{\bf p}}\nonumber\\
\times&\left(1-\frac{2\omega_{\bf p} k-M^2}{2pk}
\frac{2\omega_{\bf p} k'-M^{2}}{2pk'}\right)
 f_{l\phi}(k,\omega_{\bf p}-k)f_N^{eq}(\omega_{\bf p})\ ;
\end{align}
here $\omega_{\bf p} = \sqrt{M^2 + {\bf p}^2}$,
the bracket represents the product of 4-vectors divided by the corresponding 
energies, $k\cdot k'/(kk')$, and the integration boundaries are  
\begin{align}
p_{\rm min}(k)=\frac{|M^2-4k^2|}{4k}\ , \quad 
k'_{\rm min}=\frac{\omega_{\bf p}-p}{2}\ , \quad 
k'_{\rm max}(p)=\frac{\omega_{\bf p}+p}{2}\ .
\end{align}
The dependence on temperature ($\beta = 1/T$) enters through the
equilibrium distribution functions of Higgs particles and leptons,
\begin{align}
f_{l\phi}(k,q) = 1 - f_l(k) + f_{\phi}(q) \ , \quad 
q = \omega_{\bf p} - k\ ,
\end{align}
\begin{align}
&f_l(k) = \frac{1}{e^{\beta k} + 1} \ , \quad
f_{\phi}(q) = \frac{1}{e^{\beta q} - 1} \ ,\quad
f_N^{eq}(\omega_{\bf p}) = \frac{1}{e^{\beta\omega_{\bf p}}+1} \ .
\end{align}

The Kadanoff-Baym result (\ref{lkbtherm}) for the lepton asymmetry, which
includes effects of thermal damping, takes the same form as the Boltzmann
result 
\begin{align}
\bar{L}_{{\bf k}ii}(t,t) 
= -\frac{\epsilon_{ii}}{4\pi}\ F_{\rm KB}(k,\beta)\
\frac{1}{\Gamma}\left(1-e^{-\Gamma t}\right)\ .
\end{align}
Since the integrand of the momentum integrations contains two delta-functions
less than the expression for the Boltzmann result, the function  
$F_{\rm KB}(k,\beta)$ can only be written as a four-dimensional integral
(cf.~(\ref{kbint2})), 
\begin{align}
F_{\rm KB}(k,\beta) = \frac{1}{\pi^2}\frac{1}{k}
&\int_{p_{\rm min}(k)}^{\infty}dp\int_{k'_{\rm min}(p)}^{k'_{\rm max}(p)}k'dk
\int_{q_-}^{q_+}dq\int_{q'_-}^{q'_+}dq'
\frac{1}{\omega_{\bf p}} \nonumber\\
\times&\left(1-\frac{p^2+k^2-q^2}{2pk}\frac{p^2+k'^2-q'^2}{2pk'}\right) 
f_{l\phi}(k,q) f_{l\phi}(k',q') f_N^{eq}(\omega_{\bf p})\nonumber\\
\times&\frac{\gamma\gamma'}{((\omega_{\bf p}-k-q)^2 +\gamma^2)
((\omega_{\bf p}-k'-q')^2 + \gamma^{'2})} \ ,
\end{align}
with the integration boundaries
\begin{align}
q_{\pm} = |p \pm k| \ , \quad q'_{\pm} = |p \pm k'| \ .
\end{align}
\begin{figure}
\begin{center}
\includegraphics[width=9cm]{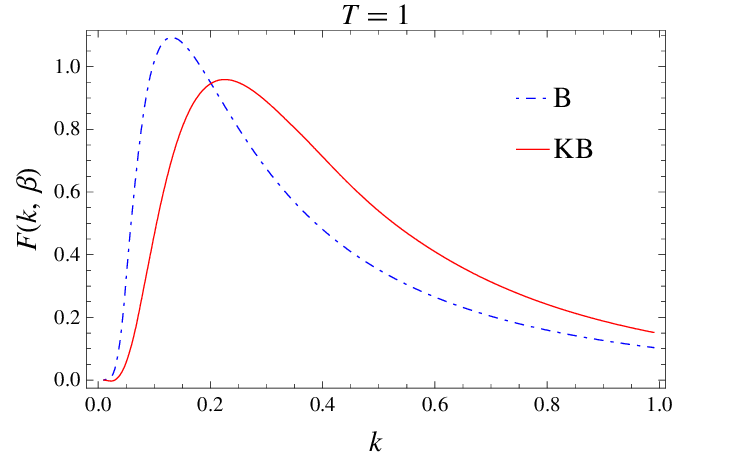}\\
\vspace{1cm}
\includegraphics[width=9cm]{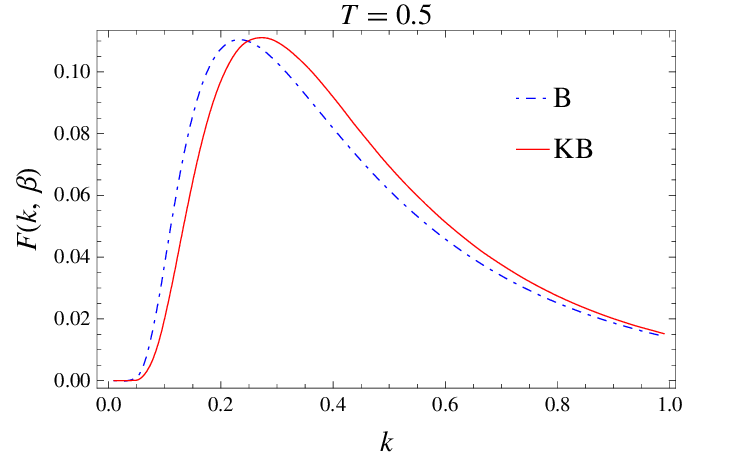}\\
\vspace{1cm}
\includegraphics[width=9cm]{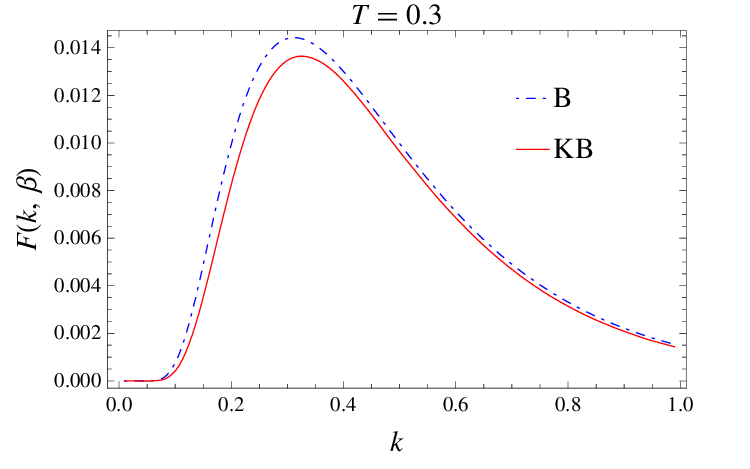}
\end{center}
\caption{Comparison of the lepton asymmetry distribution functions obtained
from Boltzmann equations (B, dot-dashed line) and Kadanoff-Baym equations
(KB, full line) for three different temperatures; temperature and momentum
are given in units of $M$.}\label{BKB} 
\end{figure}
\begin{figure}
\begin{center}
\includegraphics[width=9cm]{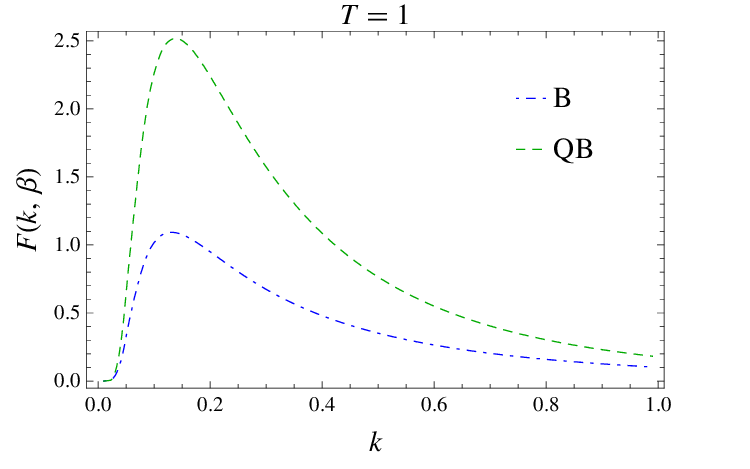}\\
\vspace{1cm}
\includegraphics[width=9cm]{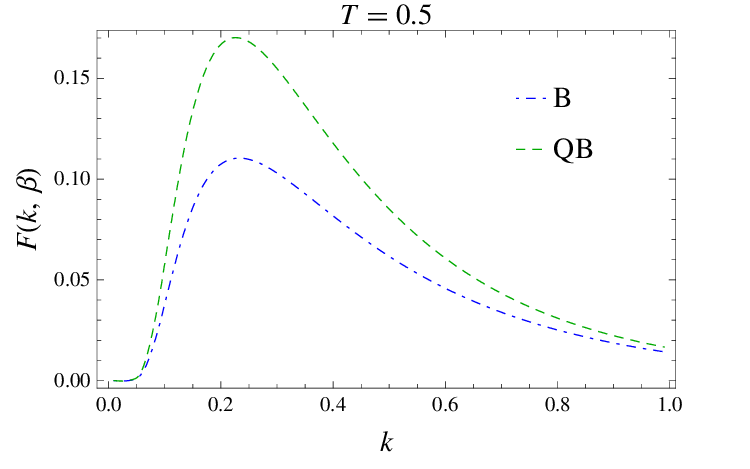}\\
\vspace{1cm}
\includegraphics[width=9cm]{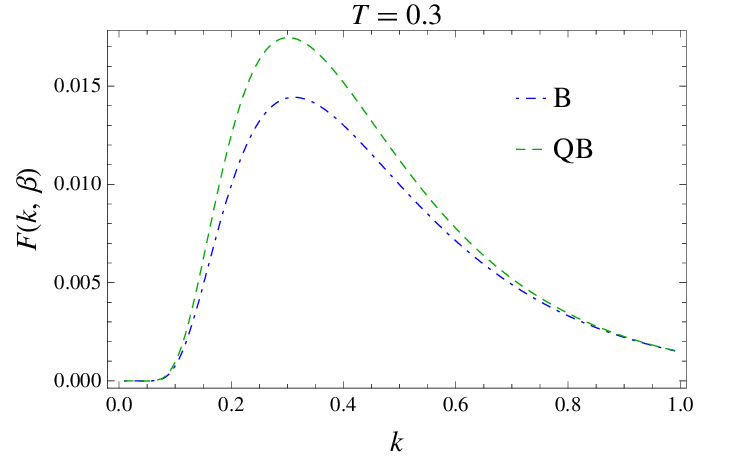}
\end{center}
\caption{Comparison of the lepton asymmetry distribution functions obtained
from Boltzmann equations (B, dot-dashed line) and quantum Boltzmann equations
(QB, dashed line) for three different temperatures; temperature and momentum
are given in units of $M$.}\label{BQB} 
\end{figure}
\begin{figure}
\begin{center}
\includegraphics[width=10cm]{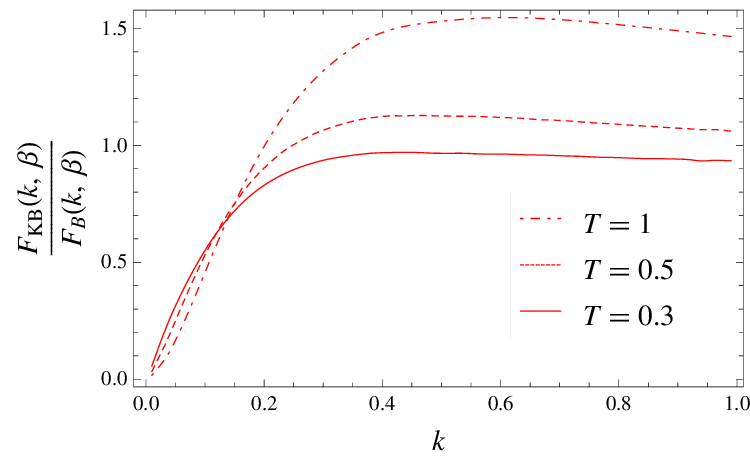}\\
\vspace{1cm}
\includegraphics[width=10cm]{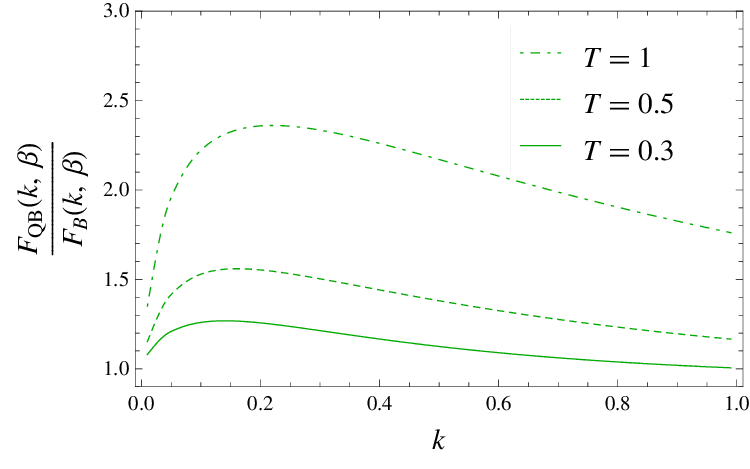}
\end{center}
\caption{Ratio of Kadanoff-Baym and Boltzmann lepton asymmetries (upper panel)
and ratio of quantum Boltzmann and Boltzmann lepton asymmetries (lower panel)
for three different temperatures; temperature and momentum are given in 
units of $M$.}\label{RsBKBQB}
\end{figure}
For the thermal widths we use the estimate 
$\gamma \simeq \gamma' \sim \frac{6g^2}{8\pi}T \sim 0.1~T$ (cf.~\cite{LeB}).
Note that the damping in a non-Abelian plasma is considerably stronger than
in an electromagnetic plasma at the same temperature.

It is instructive to compare the Boltzmann and Kadanoff-Baym results with
the prediction of quantum Boltzmann equations. As shown in \cite{Garny:2010nj,
Beneke:2010wd}, these equations lead to an additional statistical factor
compared to Boltzmann equations, which implies for the lepton asymmetry
\begin{align}
F_{\rm QB}(k,\beta) =
\frac{1}{k}&\int_{p_{\rm min}(k)}^{\infty}dp
\int_{k'_{\rm min}(p)}^{k'_{\rm max}(p)} k'dk' 
\frac{1}{\omega_{\bf p}} \\
\times&\left(1-\frac{2\omega_{\bf p} k-M^2}{2pk}
\frac{2\omega_{\bf p} k'-M^{2}}{2pk'}\right)
 f_{l\phi}(k,\omega_{\bf p}-k) f_{l\phi}(k',\omega_{\bf p}-k')
f_N^{eq}(\omega_{\bf p})\ . \nonumber
\end{align}
In \cite{Garny:2010nj,Beneke:2010wd}, this enhancement has been included in
an effective, temperature-dependent CP asymmetry.

In Fig.~\ref{BKB} Boltzmann and Kadanoff-Baym results for the lepton
asymmetry are compared. At momenta $k \sim 0.2$, where both distributions
peak, the differences are less than 20\%, at larger momenta they reach at
most 50\% (cf.~Fig.~\ref{RsBKBQB}). At temperatures $T \sim 0.3$, where 
leptogenesis takes place for typical neutrino parameters 
\cite{Buchmuller:2002rq,Buchmuller:2004nz}, differences are essentially
negligible. 

Boltzmann and quantum Boltzmann results for the lepton asymmetry are compared
in Fig.~\ref{BQB}. At momenta $k \sim 0.2$, where both distributions
are maximal, the differences can exceed 100\%, and they remain large also
at larger momenta (cf.~Fig.\ref{RsBKBQB}). An enhancement ${\cal O}(100\%)$
at $T \sim 1$ is qualitatively consistent with the enhancement found for
the temperature-dependent CP asymmetries in \cite{Garny:2010nj,Beneke:2010wd}.

The Kadanoff-Baym result strongly depends on the size of the thermal damping
rates. For $\gamma,\gamma' \rightarrow 0$, off-shell effects dissappear, and
the Kadanoff-Baym result approaches the quantum Boltzmann result. Numerically,
already for $\gamma \simeq \gamma' \sim 0.01~T$ the differences are negligible.
However, in a non-Abelian plasma, damping rates are large and, 
as a consequence,
they almost compensate the enhancement due to the additional statistical
factor contained in the quantum Boltzmann as well as the Kadanoff-Baym result.
We conclude that, according to our estimates, the conventional Boltzmann
equations provide rather accurate predictions for the lepton asymmetry.

\section{Summary and conclusions}
The goal of leptogenesis is the prediction of the cosmological baryon 
asymmetry, given neutrino masses an mixings. In a `theory of leptogenesis', 
it must be possible to quantify the theoretical error on this prediction.
This requires to go beyond Boltzmann as well as quantum Boltzmann equations,
such that the size of memory and off-shell effects can be systematically
computed. 

In the present paper we have shown how to calculate the lepton
asymmetry from first principles, i.e., in the framework of nonequilibrium
quantum field theory. Our calculation is entirely based on Green's functions,
and it therefore avoids all assumptions which are needed to arrive at
Boltzmann equations. 

Two key ingredients make the problem solvable. First, the thermal bath has
a large number of degrees of freedom, all standard model particles, 
compared to only one particle out of equilibrium, the heavy neutrino.
Hence, the backreaction of its equilibration on the temperature of the thermal 
bath can be neglected. Second, the heavy neutrino is only weakly coupled to 
the thermal bath and we can use perturbation theory in the corresponding 
Yukawa coupling $\lambda$. 

The weak coupling of the heavy neutrino to the bath allowed us to obtain
analytic expressions for the spectral function, which do not depend on initial 
conditions, and for the statistical propagator. In Section~4 we have discussed
two solutions of the Kadanoff-Baym equations, which correspond to thermal
and vacuum initial conditions. The statistical propagator which 
interpolates between vacuum at $t=0$ and thermal equilibrium at large times
can then be used in the computation of the lepton asymmetry.

Thermal leptogenesis has two vastly different scales, the width $\Gamma$ of 
the heavy neutrino on one side, and its mass $M$, temperature $T$ of the bath 
and thermal damping widths $\gamma$ on the other side,
\begin{align}
\Gamma \sim \lambda^2 M\ \ll\  \gamma \sim g^2 T\ <\ T\ \lsim M\ .\nonumber
\end{align}
Typical leptogenesis parameters (cf.~\cite{Buchmuller:2005eh}) are
$\Gamma \sim 10^{-7}~M$, $\gamma \sim 0.1~T$, $T \sim 0.3~M$, 
$M\sim 10^{10}~\mathrm{GeV}$. The existence of interactions in the plasma,
which are fast compared to the equilibration time $\tau_N =1/\Gamma$ of the 
heavy neutrino, is always implicitly assumed to justify the use of 
Boltzmann equations for the calculation of the asymmetry, but their effects
are usually not explicitly taken into account.

The main result of this paper is the computation of the lepton asymmetry in
Section~5, where the nonequilibrium propagators of the heavy neutrino and
free equilibrium propagators for massless lepton and Higgs fields are used.
Compared to Boltzmann and quantum Boltzmann equations, the crucial difference 
of the result (\ref{lkb1}) - (\ref{lkb6}) are the memory effects, oscillations 
with frequencies ${\cal O}(M)$, much faster than the heavy neutrino 
equilibration time $\tau_N =1/\Gamma$. These oscillations strongly suppress
the generated lepton asymmetry $L_{\bf k}(t,t)$ compared to the Boltzmann
result $f_L(t,k)$. In fact, as shown in appendix~C, the ratio 
$L_{\bf k}(t,t)/f_L(t,k)$ vanishes in the `zero-width' limit 
$\Gamma/M \rightarrow 0$, with $\tau = \Gamma t$ fixed.

This situation changes when the interactions, which in the Boltzmann approach
are assumed to establish kinetic equilibrium, are explicitly included 
in the calculation. 
Lepton and Higgs fields in the thermal bath then acquire large thermal
damping widths $\gamma \sim g^2 T$, which cut off the oscillations. As
a consequence, the predicted lepton asymmetry is similar to the quantum
Boltzmann result, except for off-shell effects which are now included.
For small damping widths, $\gamma \ll T$, the off-shell effects are 
negligible. They are large, however, in the standard model plasma. According
to our calculation, using $\gamma \sim 0.1~T$, the damping effects 
essentially compensate the enhancement due to the additional statistical factor
of the quantum Boltzmann equations. We conclude that, after all corrections
are taken into account, the conventional Boltzmann equations again provide
rather accurate predictions for the lepton asymmetry. Note that the classical
Boltzmann behaviour emerges at large times, $t \gsim 1/\Gamma > 1/\gamma$,
while at early times all terms are of similar magnitude, and all quantum
effects have to be kept.

As already emphasized in \cite{Anisimov:2010aq}, it is of crucial importance
to include gauge interactions in the Kadanoff-Baym approach to make further
progress towards a `theory of leptogenesis'. It remains to be seen whether
the qualitative effects of thermal damping, as discussed in this paper, will
then be confirmed or whether new surprises are encountered.

\section*{Acknowlegements}
We would like to thank D.~B\"odeker, O.~Philipsen, M.~Shaposhnikov 
and C.~Weniger for 
helpful discussions, and J.~Schmidt for sharing his expertise. This work
was supported by the German Science Foundation (DFG) within the Collaborative
Research Center 676 ``Particles, Strings and the Early Universe''
and by the Swiss National Science Foundation.

\clearpage

\begin{appendix}

\section{Thermal propagators}
In the following we list all propagators, which are needed in the calculation
described in Section~5, as functions of relative time $y=t_1-t_2$ and total
time $t=(t_1+t_2)/2$.

\begin{itemize}

\item Free massive scalar ($\omega_{\textbf{q}} = \sqrt{m^2+\textbf{q}^2}$)
\begin{align}
\Delta^{-}_{\textbf{q}}(y) &= \frac{1}{\omega_{\textbf{q}}}
\sin(\omega_{\textbf{q}}y)\ , \label{D-}\\
\Delta^{+}_{\textbf{q}}(y) &= \frac{1}{2\omega_{\textbf{q}}}
\coth\left(\frac{\beta\omega_{ \textbf{q}}}{2}\right)
\cos(\omega_{\textbf{q}}y)\ ,\label{D+}\\
\Delta^{11}_{\textbf{q}}(y) &= \frac{1}{2\omega_{\textbf{q}}}
\left(\coth\left(\frac{\beta\omega_{\textbf{q}}}{2}\right)
\cos(\omega_{ \textbf{q}}y)-i\sin(\omega_{\textbf{q}}|y|)\right) \\
&= \Delta^{+}_{\textbf{q}}(y) - 
\frac{i}{2}{\rm sign}(y)\Delta^{-}_{\textbf{q}}(y)\ , \nonumber \\
\Delta^{22}_{ \textbf{q}}(y) &= \frac{1}{2\omega_{\textbf{q}}}
\left(\coth\left(\frac{\beta\omega_{\textbf{q}}}{2}\right)
\cos(\omega_{ \textbf{q}}y)+i\sin(\omega_{\textbf{q}}|y|)\right) \\
&= \Delta^{+}_{\textbf{q}}(y) +
\frac{i}{2}{\rm sign}(y)\Delta^{-}_{\textbf{q}}(y)\ ,\nonumber \\
\Delta^{>}_{ \textbf{q}}(y) &= \frac{1}{2\omega_{\textbf{q}}}
\left(\coth\left(\frac{\beta\omega_{\textbf{q}}}{2}\right)
\cos(\omega_{ \textbf{q}}y)-i\sin(\omega_{ \textbf{q}}y)\right)\ ,\\
\Delta^{<}_{ \textbf{q}}(y) &= \frac{1}{2\omega_{\textbf{q}}}
\left(\coth\left(\frac{\beta\omega_{\textbf{q}}}{2}\right)
\cos(\omega_{ \textbf{q}}y)+i\sin(\omega_{ \textbf{q}}y)\right)\ .
\end{align}

\item Free massive Dirac fermion ($\omega_{\textbf{k}} = 
\sqrt{m^2+\textbf{k}^2}$)
\begin{align}
S^{-}_{\textbf{k}}(y) &= i\gamma_0\cos(\omega_{\textbf{k}}y) +
\frac{m-{\bf k}\pmb{\gamma}}{\omega_{\textbf{k}}}
\sin(\omega_{\textbf{k}}y)\ , \label{S-}\\
S^{+}_{\textbf{k}}(y) &= -\frac{1}{2}\tanh\left(\frac{\beta\omega_{\textbf{k}}}
{2}\right)\left(i\gamma_0\sin(\omega_{\textbf{k}}y) -
\frac{m-{\bf k}\pmb{\gamma}}{\omega_{\textbf{k}}}
\cos(\omega_{\textbf{k}}y)\right)\ ,\label{S+}\\ 
S^{11}_{\textbf{k}}(y) &= \frac{\gamma_0}{2}\left(
\cos(\omega_{\textbf{k}}y)\textrm{sign}(y)
-i\tanh\left(\frac{\beta\omega_{\textbf{k}}}{2}\right)
\sin(\omega_{\textbf{k}}y)\right)\nonumber\\
&\quad +\frac{m-{\bf k}\pmb{\gamma}}{2\omega_{\textbf{k}}}\left(\tanh\left(
\frac{\beta\omega_{\textbf{k}}}{2}\right)\cos(\omega_{\textbf{k}}y)
-i\sin(\omega_{\textbf{k}}|y|)\right) \\
&=S^{+}_{\textbf{k}}(y) - \frac{i}{2}{\rm sign}(y)S^{-}_{\textbf{k}}(y)\ ,
\nonumber \\
S^{22}_{\textbf{k}}(y) &= \frac{\gamma_0}{2}\left(
-\cos(\omega_{\textbf{k}}y)\textrm{sign}(y)
-i\tanh\left(\frac{\beta\omega_{\textbf{k}}}{2}\right)
\sin(\omega_{\textbf{k}}y)\right)\nonumber\\
&\quad +\frac{m-{\bf k}\pmb{\gamma}}{2\omega_{\textbf{k}}}\left(\tanh\left(
\frac{\beta\omega_{\textbf{k}}}{2}\right)\cos(\omega_{\textbf{k}}y)
+i\sin(\omega_{\textbf{k}}|y|)\right) \\
&=S^{+}_{\textbf{k}}(y) + \frac{i}{2}{\rm sign}(y)S^{-}_{\textbf{k}}(y)\ ,
\nonumber \\
S^{>}_{ \textbf{k}}(y) &= \frac{\gamma_0}{2}\left(\cos(\omega_{
\textbf{k}}y)-i\tanh\left(\frac{\beta\omega_{ \textbf{k}}}{2}\right)
\sin(\omega_{\textbf{k}}y)\right)\nonumber\\
&+\frac{m-{\bf k}\pmb{\gamma}}{2\omega_{\textbf{k}}}\left(\tanh\left(
\frac{\beta\omega_{\textbf{k}}}{2}\right)\cos(\omega_{\textbf{k}}y)
-i\sin(\omega_{ \textbf{k}}y)\right)\ ,  \\
S^{<}_{\textbf{k}}(y) &= \frac{\gamma_0}{2}\left(-\cos(\omega_{\textbf{k}}y)
-i\tanh\left(\frac{\beta\omega_{ \textbf{k}}}{2}\right)
\sin(\omega_{ \textbf{k}}y)\right) \nonumber\\
&+\frac{m-{\bf k}\pmb{\gamma}}{2\omega_{\textbf{k}}}\left(\tanh\left(
\frac{\beta\omega_{ \textbf{k}}}{2}\right)\cos(\omega_{\textbf{k}}y)
+i\sin(\omega_{ \textbf{k}}y)\right)\ .
\end{align}
The propagators for a massless left-handed fermion are obtained by the
substitutions $\omega_{\textbf{k}} \rightarrow k=|\textbf{k}|$, 
$S^{...}_{ \textbf{k}} \rightarrow P_L S^{...}_{ \textbf{k}}$, where
$P_L = (1-\gamma_5)/2$. 

\item Free massive Majorana fermion ($\omega_{\textbf{p}} = 
\sqrt{M^2+\textbf{p}^2}$)
\begin{align}
G^{-}_{\textbf{p}}(y) &= \Big(i\gamma_0\cos(\omega_{ \textbf{p}}y)+
\frac{M-{\bf p}\gamma}{\omega_{\textbf{p}}}
\sin(\omega_{\textbf{p}}y)\Big)C^{-1}\ , \\
G^{+}_{\textbf{p}}(y) &= -\frac{1}{2}\tanh\left(
\frac{\beta\omega_{\textbf{p}}}{2}\right)
\left(i\gamma_0\sin(\omega_{\textbf{p}}y)
-\frac{M-{\bf p}\pmb{\gamma}}{\omega_{\textbf{p}}}
\cos(\omega_{\textbf{p}}y)\right)C^{-1}\ ,\\
G^{11}_{ \textbf{p}}(y) &= \left[\frac{\gamma_0}{2}\left(
\cos(\omega_{\textbf{p}}y)\textrm{sign}(y)-i\tanh\left(
\frac{\beta\omega_{\textbf{p}}}{2}\right)\sin(\omega_{\textbf{p}}y)\right)
\right.\nonumber\\
&\left.+\frac{M-{\bf p}\pmb{\gamma}}{2\omega_{\textbf{p}}}\left(\tanh\left(
\frac{\beta\omega_{\textbf{p}}}{2}\right)\cos(\omega_{\textbf{p}}y)
-i\sin(\omega_{\textbf{p}}|y|)\right)\right]C^{-1}\ ,\\
G^{22}_{\textbf{p}}(y)&= \left[\frac{\gamma_0}{2}\left(
-\cos(\omega_{\textbf{p}}y)\textrm{sign}(y)-i\tanh\left(
\frac{\beta\omega_{\textbf{p}}}{2}\right)\sin(\omega_{\textbf{p}}y)\right)
\right.\nonumber\\
&\left.+\frac{M-{\bf p}\pmb{\gamma}}{2\omega_{\textbf{p}}}\left(\tanh\left(
\frac{\beta\omega_{\textbf{p}}}{2}\right)\cos(\omega_{\textbf{p}}y)
+i\sin(\omega_{\textbf{p}}|y|)\right)\right]C^{-1}\ ,\\
G^{>}_{ \textbf{p}}(y)&= \left[\frac{\gamma_0}{2}\left(\cos(\omega_{
\textbf{p}}y)-i\tanh\left(\frac{\beta\omega_{ \textbf{p}}}{2}\right)
\sin(\omega_{\textbf{p}}y)\right)\right.\nonumber\\
&\left.+\frac{M-{\bf p}\pmb{\gamma}}{2\omega_{\textbf{p}}}\left(
\tanh\left(\frac{\beta\omega_{\textbf{p}}}{2}\right)\cos(\omega_{
\textbf{p}}y)-i\sin(\omega_{\textbf{p}}y)\right)\right]C^{-1}\ , \\
G^{<}_{\textbf{p}}(y) &= \left[\frac{\gamma_0}{2}\left(
-\cos(\omega_{\textbf{p}}y)-i\tanh\left(
\frac{\beta\omega_{\textbf{p}}}{2}\right)
\sin(\omega_{ \textbf{p}}y)\right)\right.\nonumber\\
&\left.+\frac{M-{\bf p}\pmb{\gamma}}{2\omega_{ \textbf{p}}}
\left(\tanh\left(\frac{\beta\omega_{ \textbf{p}}}{2}\right)
\cos(\omega_{\textbf{p}}y)+i\sin(\omega_{\textbf{p}}y)\right)\right]C^{-1}\ . 
\end{align}\\

\item Nonequilibrium massive Majorana fermion (interpolation between
vacuum at $t=y=0$ and thermal equilibrium at $t=\infty$, and memory integral)
\begin{align}
G^-_{\textbf{p}}(y) &= \left(i\gamma_0\cos(\omega_{\textbf{p}}y)+
\frac{M-{\bf p}\pmb{\gamma}}{\omega_{\textbf{p}}}
\sin(\omega_{ \textbf{p}}y)\right)e^{-\Gamma_{{\bf p}}|y|/2}C^{-1}\ , 
\label{Gne1}\\
G^{+}_{\textbf{p}}(t,y) &= -\left(i\gamma_0\sin(\omega_{\textbf{p}}y)
-\frac{M-{\bf p}\pmb{\gamma}}{\omega_{\textbf{p}}}
\cos(\omega_{\textbf{p}}y)\right) \nonumber\\
&\quad\times\left(\frac{1}{2}\tanh\left(
\frac{\beta\omega_{\textbf{p}}}{2}\right)
e^{-\Gamma_{{\bf p}}|y|/2}+f_N^{eq}(\omega_{\bf p})e^{-\Gamma_{{\bf p}} t}
\right)C^{-1}\ , \label{Gne2}\\
G^{11}_{\textbf{p}}(t,y) &= G^{+}_{\textbf{p}}(t,y)
- \frac{i}{2}{\rm sign}(y)G^{-}_{\textbf{p}}(y) \ ,\label{Gne3}\\
G^{22}_{\textbf{p}}(t,y) &= G^{+}_{\textbf{p}}(t,y)
+ \frac{i}{2}{\rm sign}(y)G^{-}_{\textbf{p}}(y) \ ,\label{Gne4}\\
G^{>}_{\textbf{p}}(t,y) &= G^{+}_{\textbf{p}}(t,y)
- \frac{i}{2}G^{-}_{\textbf{p}}(y) \ ,\label{Gne5}\\
G^{<}_{\textbf{p}}(t,y) &= G^{+}_{\textbf{p}}(t,y)
+ \frac{i}{2} G^{+}_{\textbf{p}}(y) \ ,\label{Gne6}\\
G^{+}_{\textbf{p},\mathrm{mem}}(t,y) &= -\frac{1}{2}\tanh\left(
\frac{\beta\omega_{\textbf{p}}}{2}\right)\left(i\gamma_0
\sin(\omega_{\textbf{p}}y)-\frac{M-{\bf p}\pmb{\gamma}}{\omega_{\textbf{p}}}
\cos(\omega_{\textbf{p}}y)\right) \nonumber\\
&\quad\quad \times\left(e^{-\Gamma_{{\bf p}}|y|/2}
-e^{-\Gamma_{{\bf p}} t}\right)C^{-1}\ .
\end{align}

\end{itemize}

\section{Feynman rules}
For completeness, we list in the following the Feynman rules for the
Standard Model Lagrangian with right-handed neutrinos given in 
Eq.~(\ref{LRN2}); $\alpha,\beta$ are spinor indices and
$a,b,\ldots$ are SU(2) indices.\\

\noindent\textbullet\;  Majorana neutrino\\
\begin{figure}[h]
\vspace*{-0.3cm}
\hspace*{0.5cm}
\begin{minipage}[c]{5 cm}
\psfrag{x2}{$x_{2,\beta}$}
\psfrag{x1}{$x_{1,\alpha}$}
\psfrag{N}{$N$}
\includegraphics[width=5 cm]{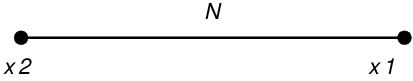}
\end{minipage}\hspace*{1cm}
\begin{minipage}[c]{5 cm}
$G_{\alpha\beta}(x_1,x_2)$\
\end{minipage}
\end{figure}\\
%
\noindent\textbullet\; Lepton doublet\\
\begin{figure}[h]
\vspace*{-0.3cm}
\hspace*{0.5cm}
\begin{minipage}[c]{5 cm}
\psfrag{N}{$l$}
\psfrag{x1}{$x_{1,\alpha,a,i}$}
\psfrag{x2}{$x_{2,\beta,b,j}$}
\includegraphics[width=5 cm]{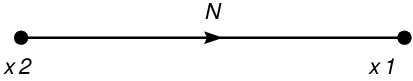}
\end{minipage}\hspace*{1cm}
\begin{minipage}[c]{5 cm}
$\delta_{ij}\delta_{ab}S_{\alpha\beta}(x_1,x_2)$
\end{minipage}
\end{figure}
\newpage
\noindent\textbullet\; Higgs doublet\\
\begin{figure}[h]
\vspace*{-0.3cm}
\hspace*{0.5cm}
\begin{minipage}[c]{5 cm}
\psfrag{N}{$\phi$}
\psfrag{x1}{$x_{1,a}$}
\psfrag{x2}{$x_{2,b}$}
\includegraphics[width=5 cm]{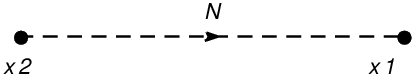}
\end{minipage}\hspace*{1cm}
\begin{minipage}[c]{5 cm}
$\delta_{ab}\Delta(x_1,x_2)$
\end{minipage}
\end{figure}\\

\noindent\textbullet\; Vertices
\vspace*{0.5cm}
\begin{figure}[h!]
\vspace*{-0.5cm}
\hspace*{0.5cm}
\begin{minipage}[c]{4 cm}
\psfrag{n}{$N$}
\psfrag{be}{$\beta$}
\psfrag{iaa}{$\hspace{-0.3cm}i,\alpha,a$}
\psfrag{l}{$l$}
\psfrag{b}{$b$}
\psfrag{p}{$\phi$}
\includegraphics[width=4 cm]{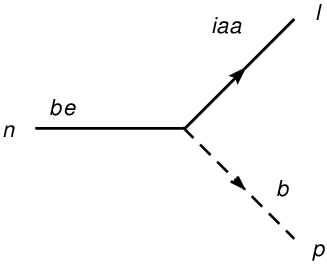}
\end{minipage}\hspace*{1cm}
\begin{minipage}[c]{4 cm}
$i\lambda^*_{i1}\epsilon_{ab}(P_R)_{\alpha\beta}$
\end{minipage}\\
%
\vspace*{0.5cm}
\hspace*{0.5cm}
\begin{minipage}[c]{4 cm}
\psfrag{n}{$N$}
\psfrag{be}{$\beta$}
\psfrag{iaa}{$\hspace{-0.3cm}i,\alpha,a$}
\psfrag{l}{$l$}
\psfrag{b}{$b$}
\psfrag{p}{$\phi$}
\includegraphics[width=4 cm]{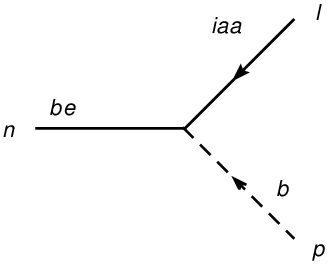}
\end{minipage}\hspace*{1cm}
\begin{minipage}[c]{4 cm}
$i\lambda_{i1}(CP_L)_{\beta\alpha}\epsilon_{ab}$
\end{minipage}\\
%
\vspace*{0.5cm}
\hspace*{0.5cm}
\begin{minipage}[c]{4 cm}
\psfrag{c}{$c$}
\psfrag{iaa}{$i,\alpha,a$}
\psfrag{l}{$l$}
\psfrag{ibb}{$j,\beta,b$}
\psfrag{p}{$\phi$}
\psfrag{d}{$d$}
\includegraphics[width=4 cm]{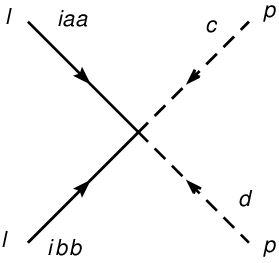}
\end{minipage}\hspace*{1cm}
\begin{minipage}[c]{5 cm}
$i\eta_{ij}(\epsilon_{ac}\epsilon_{bd}
+\epsilon_{ad}\epsilon_{bc})(CP_L)_{\alpha\beta}$
\end{minipage}\\
%
\vspace*{0.5cm}
\hspace*{0.5cm}
\begin{minipage}[c]{4cm}
\psfrag{c}{$c$}
\psfrag{iaa}{$i,\alpha,a$}
\psfrag{l}{$l$}
\psfrag{ibb}{$j,\beta,b$}
\psfrag{p}{$\phi$}
\psfrag{d}{$d$}
\includegraphics[width=4cm]{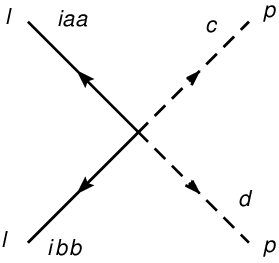}
\end{minipage}\hspace*{1cm}
\begin{minipage}[c]{5 cm}
$i\eta_{ij}^*(\epsilon_{ac}\epsilon_{bd}
+\epsilon_{ad}\epsilon_{bc})(P_RC)_{\alpha\beta}$
\end{minipage}
\end{figure}

\newpage
\section{Zero-width limit}
In this section we consider the Kadanoff-Baym result
for the lepton asymmetry normalised to the Boltzmann result,  
$L_{\bf k}(t,t)/f_L(t,k)$, in the zero-width limit as defined in 
Eq.~(\ref{zero}), i.e.,
\begin{align}
\frac{\Gamma}{M} \rightarrow 0\ ,\quad \tau=\Gamma t\;\; \mathrm{fixed}\ .
\nonumber
\end{align}
To this end we have to evaluate the corresponding momentum integral 
(\ref{lkb2}) in this limit.

\subsection{Boltzmann equation}

Consider first the Boltzmann result for the lepton asymmetry given in 
Eq.~(\ref{soll2}),
\begin{align}
f_{Li}(t,k) = - \epsilon_{ii}\frac{16\pi}{k}
&\int_{\bf q,p,q',k'} k\cdot k'\ (2\pi)^4\delta^4(k+q-p)
(2\pi)^4\delta^4(k'+q'-p) \nonumber\\
&\times f_{l\phi}(k,q)
f_N^{eq}(\omega_{\bf p})\frac{1}{\Gamma}\left(1-e^{-\Gamma t}\right)\ .
\end{align}
The integration over ${\bf q}$ and ${\bf q'}$ can be performed using the 
$\delta$-functions, which leads to
\begin{align}
f_{Li}(t,k) = - \frac{\epsilon_{ii}}{16\pi^3}
&\int d^3p\int d^3k'\ \frac{k\cdot k'}{kk'}\frac{1}{\omega_{\bf p} q q'}\
\delta(k+q-\omega_{\bf p})\delta(k'+q'-\omega_{\bf p}) \nonumber\\
&\times f_{l\phi}(k,q)f_N^{eq}(\omega_{\bf p})
\frac{1}{\Gamma}\left(1-e^{-\Gamma t}\right)\ ,
\end{align}
where $q=|{\bf q}|$ and $q'=|{\bf q}'|$. The product of 4-vectors,
$k\cdot k'=kk'(1-\hat{\bf k}\cdot\hat{\bf k}')$, depends on the angles between 
the different momenta. It is convenient to define the angles with respect to 
the momentum ${\bf p}$: $\theta=\angle({\bf k}, {\bf p})$, 
$\theta'=\angle({\bf k}', {\bf p})$ and 
$\varphi'=\angle({\bf k}_{\perp}, {\bf k}'_{\perp})$; here ${\bf k}_{\perp}$ 
and ${\bf k}'_{\perp}$ are perpendicular to the vector ${\bf p}$, i.e.,
${\bf k}={\bf k}_{\parallel}+{\bf k}_{\perp}$ and 
${\bf k}'={\bf k}'_{\parallel}+{\bf k}'_{\perp}$. In terms of these angles
the unit vectors $\hat{\bf k}$ and $\hat{\bf k}'$ are given by 
(see Fig.~\ref{angles})
\begin{figure}[htbp]\label{angles}
\begin{centering}
\includegraphics[width=0.6\textwidth]{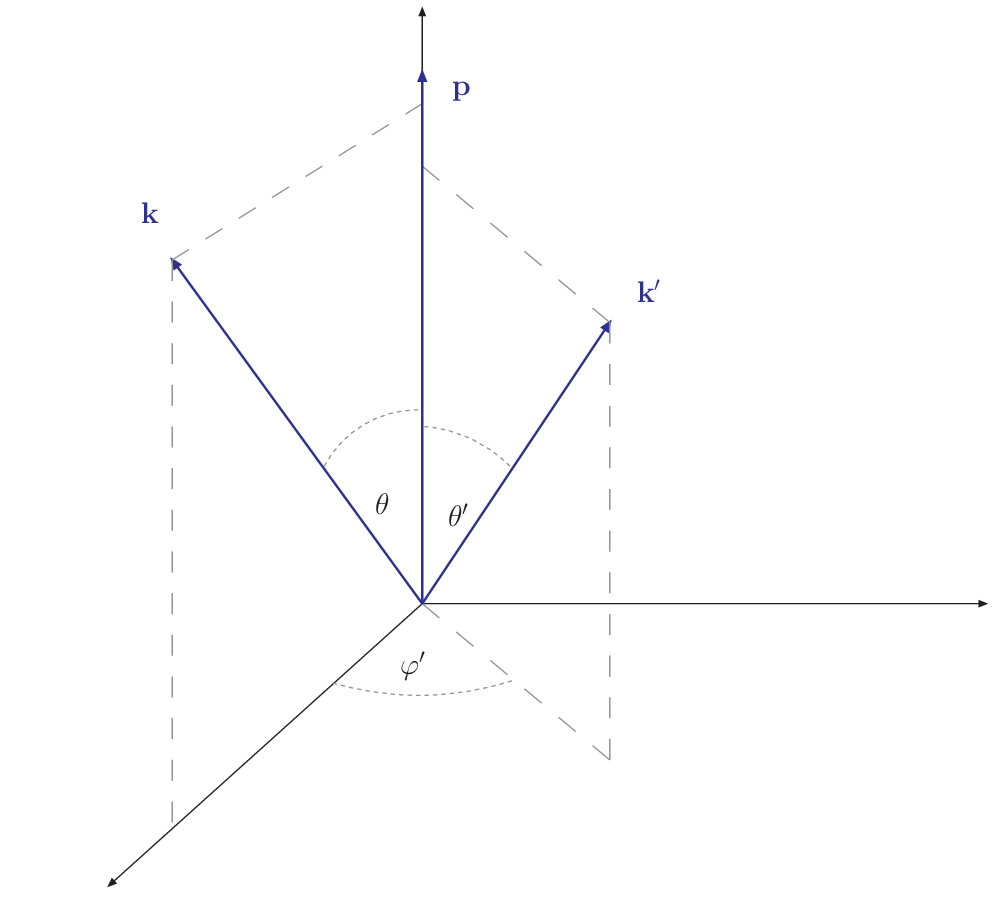}
\caption{Integration angles}
\label{angles}
\end{centering}
\label{angles}
\end{figure}
\begin{eqnarray}\label{versors}
\hat{\bf k}=\left(\begin{array}{c}\cos\theta \\\sin\theta \\0\end{array}
\right)\ ,\qquad
\hat{\bf k}'=\left(\begin{array}{c}\cos\theta' \\\sin\theta'\cos\varphi' \\
\sin\theta'\sin\varphi'\end{array}\right)\ ,
\end{eqnarray}
with $\hat{\bf k}\cdot\hat{\bf k}'=\cos\theta\cos\theta'
+\sin\theta\sin\theta'\cos\varphi'$.  We then obtain
\begin{align}
f_{Li}(t,{\bf k}) = - \frac{\epsilon_{ii}}{16\pi^3}
&\int d^3p\int_0^{\infty} k'^2dk'\int_{-1}^{1}d\cos\theta'
\int_{0}^{2\pi}d\varphi'\ \frac{1}{\omega_{\bf p} q q'}\nonumber\\ 
&\times (1-\cos\theta\cos\theta'-\sin\theta\sin\theta'\cos\varphi') \\
&\times \delta(k+q-\omega_{\bf p})\delta(k'+q'-\omega_{\bf p})
f_{l\phi}(k,q)f_N^{eq}(\omega_{\bf p})
\frac{1}{\Gamma}\left(1-e^{-\Gamma t}\right)\ . \nonumber
\end{align}
Momentum conservation relates the energies $q$ and $q'$ to $p$, $k$, $k'$
and the angles $\theta$ and $\theta'$,
\begin{align}
q&=|{\bf p}-{\bf k}|=(p^2+k^2-2pk\cos\theta)^{1/2}\;,\label{fofo1}\\
q'&=|{\bf p}-{\bf k}'|=(p^2+k'^2-2pk\cos\theta')^{1/2}\;.\label{fofo2}
\end{align}

We can now make use of rotational invariance of the distribution function,
\begin{align}
f_{Li}(t,k)=\frac{1}{4\pi}\int d\Omega_{\bf k}\ f_{Li}(t,k) \ .
\end{align}
Changing variables,
\begin{align}\label{varchange}
dq=-\frac{pk}{q}d\cos\theta\ , \quad dq'=-\frac{pk'}{q'}d\cos\theta' \ ,
\end{align}
one arrives at
\begin{align}
f_{Li}(t,k) = - &\frac{\epsilon_{ii}}{4\pi}\frac{1}{k}
\int dp\int_0^{\infty}k'dk'\int_{q_+}^{q_-} dq \int_{q_{+}'}^{q_-'} dq'\ 
\left(1-\frac{p^2+k^2-q^2}{2pk}\frac{p^2+k^{'2}-q^{'2}}{2pk'}\right) 
\nonumber\\
&\times\frac{1}{\omega_{\bf p}}\delta(k+q-\omega_{\bf p})
\delta(k'+q'-\omega_{\bf p})
f_{l\phi}(k,q)f_N^{eq}(\omega_{\bf p})
\frac{1}{\Gamma}\left(1-e^{-\Gamma t}\right)\ ,
\end{align}
where the limits of integration are given by the maximal and minimal value 
of $q$ and $q'$, respectively,
\begin{align}\label{qbound}
q_{\pm}=|k\pm p|\ , \qquad q'_{\pm}=|k'\pm p|\ .
\end{align}

Consider now the argument of one $\delta$-function, 
$\Omega_1=\omega_{\bf p}-k-q$, 
with $\Omega_1^{\rm min}=\omega_{\bf p}-k-q_+$ and 
$\Omega_1^{\rm max}=\omega_{\bf p}-k-q_{-}$ (cf.~Eq.(\ref{I})). Obviously, the 
conditions $\Omega_1^{\rm min}<0$ and $\Omega_1^{\rm max}>0$ limit the 
integration range in $p$ for given momentum $k$,
\begin{equation}\label{pbound}
p>\frac{|M^2-4k^2|}{4k} \equiv p_{\rm min}(k)  \ .
\end{equation}
Similarly, the constraint $p>(M^2-4k^{'2})/(4k')$ restricts the integration
range in $k'$ for given $p$,
\begin{align}\label{kbound}
k'>\frac{\omega_{\bf p}-p}{2} \equiv k'_{\rm min}(p)\ , \qquad 
k'<\frac{\omega_{\bf p}+p}{2} \equiv k'_{\rm max}(p)\ .
\end{align}
Changing again variables from $q$ and $q'$ to $\Omega_1$ and $\Omega_3$, 
respectively, and using
\begin{align}
\frac{\partial(p,k',\Omega_1,\Omega_3)}{\partial(p,k',q,q')} = 1\ ,
\end{align}
the integral can now be written as
\begin{align}
f_{Li}(t,k) = - \frac{\epsilon_{ii}}{4\pi}\frac{1}{k}
&\int_{p_{\rm min}(k)}^{\infty} dp\int_{k'_{\rm min}(p)}^{k'_{\rm max}(p)} dk' 
\int_{\Omega_1^{\rm min}}^{\Omega_1^{\rm max}} d\Omega_1 
\int_{\Omega_3^{\rm min}}^{\Omega_3^{\rm max}} d\Omega_3 
\nonumber \\ 
&\times\frac{1}{\omega_{\bf p}}\delta(\Omega_1)\delta(\Omega_3)
\left(1-\frac{p^2+k^2-q^2}{2pk}\frac{p^2+k^{'2}-q^{'2}}{2pk'}\right) 
\nonumber\\ 
&\times f_{l\phi}(k,q)f_N^{eq}(\omega_{\bf p})
\frac{1}{\Gamma}\left(1-e^{-\Gamma t}\right)\ .
\end{align}
The limits of integration have been chosen such that they contain the points
$\Omega_1=0$ and $\Omega_3=0$, which correspond to energy conservation, 
$q=\omega_{\bf p}-k$ and 
$q'=\omega_{\bf p}-k'$, respectively. Hence, the integration on $\Omega_1$ and 
$\Omega_3$ can
trivially be carried out, and we obtain the final result
\begin{align}
f_{Li}(t,k) = - \frac{\epsilon_{ii}}{4\pi}\frac{1}{k}
&\int_{p_{\rm min}(k)}^{\infty}dp\int_{k'_{\rm min}(p)}^{k'_{\rm max}(p)}dk'\ 
\frac{1}{\omega_{\bf p}}\left(1-\frac{2\omega_{\bf p} k-M^2}{2pk}
\frac{2\omega_{\bf p} k'-M^2}{2pk'}\right) \nonumber\\
&\times f_{l\phi}(k,\omega_{\bf p}-k)f_N^{eq}(\omega_{\bf p})
\frac{1}{\Gamma}\left(1-e^{-\Gamma t}\right)\ . \label{intboltz}
\end{align}

\subsection{Kadanoff-Baym equation}

We are now ready to evaluate the leading contribution of the Kadanoff-Baym 
result for the lepton asymmetry. It is given by Eq.~(\ref{lkb3}) with
$\alpha=\beta=1$, and it can be written in the form 
\begin{align}\label{kbint}
L_{{\bf k}ii}(t,t) =  -\epsilon_{ii}\ &8\pi 
\int_{\bf q,q'} \frac{k\cdot k'}{kk'\omega_{\bf p}}\ 
f_{l\phi}(k,q) f_{l\phi}(k',q') f_N^{eq}(\omega_{\bf p})\nonumber\\
&\times\frac{\frac{1}{2}\Gamma}{((\omega_{\bf p}-k-q)^2 +\frac{\Gamma^2}{4})
((\omega_{\bf p}-k'-q')^2 + \frac{\Gamma^2}{4})} \nonumber\\
&\times\bigg[\left(e^{-\frac{\Gamma t}{2}}
-\cos((\omega_{\bf p}-k-q)t)\right)\left(e^{-\frac{\Gamma t}{2}}
-\cos((\omega_{\bf p}-k'-q')t)\right) \nonumber\\
&\quad -\sin((\omega_{\bf p}-k-q)t)\sin((\omega_{\bf p}-k'-q')t)\bigg]\ .
\end{align}
We first change variables, ${\bf (q,q') \rightarrow (p,k')}$, with
${\bf p = q + k = q' + k'}$, and use rotational invariance,
\begin{align}
L_{{\bf k}ii}(t,t) = \frac{1}{4\pi}\int d\Omega_{\bf k}\ 
L_{{\bf k}ii}(t,t) \ .
\end{align}
Choosing again angles  
according to Fig.~\ref{angles}, the integral (\ref{kbint}) becomes
\begin{align}
L_{{\bf k}ii}(t,t) &\propto \int d\Omega_{\bf k}\int d\Omega_{{\bf k}'}
\int_0^{2\pi} d\varphi'\ 
\frac{k\cdot k'}{kk'}\ F(\theta,\theta',\cdots) \nonumber \\ 
&=  \int_{-1}^1 d\cos\theta\int_{-1}^1 d\cos\theta'\int_0^{2\pi} d\varphi'\
(1-\cos\theta\cos\theta' \nonumber\\      
&\hspace{4.5cm} -\sin\theta\sin\theta'\cos\varphi')\ F(\theta,\theta',\cdots) 
\nonumber \\
&= (2\pi)^2 \int_{-1}^1 d\cos\theta\int_{-1}^1 d\cos\theta'\ 
(1-\cos\theta\cos\theta')F(\theta,\theta',\cdots)\ ,
\end{align}
where we have used that the function $F(\theta,\theta',\cdots)$ does not
depend on the angle $\varphi'$.
As in the previous section, we now change the integration variables
from $(\theta,\theta')$ to $(q,q')$, and using Eq.~(\ref{varchange})
we obtain
\begin{align}\label{kbint2}
L_{{\bf k}ii}(t,k) =  &-\frac{\epsilon_{ii}}{8\pi^3}\frac{1}{k} 
\int_{p_{\rm min}(k)}^{\infty}dp \int_{k'_{\rm min}(p)}^{k'_{\rm max}(p)}k'dk'
\int_{q_-}^{q_+}dq\int_{q'_-}^{q'_+}dq'
\frac{1}{\omega_{\bf p}}\nonumber\\
&\times\left(1-\frac{p^2+k^2-q^2}{2pk}\frac{p^2+k'^2-q'^2}{2pk'}\right) 
f_{l\phi}(k,q) f_{l\phi}(k',q') f_N^{eq}(\omega_{\bf p})\nonumber\\
&\times\frac{\frac{1}{2}\Gamma}{((\omega_{\bf p}-k-q)^2 +\frac{\Gamma^2}{4})
((\omega_{\bf p}-k'-q')^2 + \frac{\Gamma^2}{4})} \nonumber\\
&\times\bigg[\left(e^{-{\Gamma t\over 2}}-\cos((\omega_{\bf p}-k-q)t)\right)
\left(e^{-{\Gamma t\over 2}}-\cos((\omega_{\bf p}-k'-q')t)\right)\nonumber\\
&\qquad -\sin((\omega_{\bf p}-k-q)t)\sin((\omega_{\bf p}-k'-q')t)\bigg]\ ,
\end{align}
where the limits of integration are given in 
Eqs.~(\ref{qbound}) - (\ref{kbound}). We have restricted the integration
over $p$ and $k$ to the range for which the intervals $[q_-,q_+]$ and
$[q_-',q_+']$ contain points satisfying $\omega_{\bf p}-k-q = 0$ and
$\omega_{\bf p}-k'-q' = 0$, respectively. This finite part of the
integral could then be $\mathcal{O}(1/\Gamma)$, which is required to match
the Boltzmann result for the lepton asymmetry. The remaining part is
$\mathcal{O}(1)$ and therefore suppressed compared to the Boltzmann result.

Remarkably, the integral (\ref{kbint2}) is a sum of terms each of which 
factorizes into a product where one factor depends on
$q$ but not on $q'$, whereas the other factor depends on $q'$ but not on $q$.
Hence one obtains
\begin{align}
L_{{\bf k}ii}(t,t) \propto \int_{p_{\rm min}(k)}^{\infty}dp
\int_{k'_{\rm min}(p)}^{k'_{\rm max}(p)}k'dk'\
\sum_{i}\mathcal{P}_i(q_-,q_+)\mathcal{Q}_i(q_-',q_+')\ ,
\end{align}
where we have dropped the dependence of the factors $\mathcal{P}_i$ and 
$\mathcal{Q}_i$ on $k$, $p$ and $k'$ for simplicity. Because of the
factorization, we can now perform the integrations on $q$ and $q'$ separately. 

Naively, one may think that in the zero-width limit $\Gamma/M\rightarrow0$ 
the cosine terms can be set to one.  But for large time $t$, they oscillate 
fast, which leads to a different result. Consider the following contribution 
to the integral (\ref{kbint2}),
\begin{align}\label{P}
\mathcal{P}(q_{-},q_+)= -\int_{q_-}^{q_+}dq 
\frac{F(q)}{(\omega_{\bf p}-k-q)^2 + \frac{\Gamma^2}{4}}
\cos((\omega_{\bf p}-k-q)t)\ ,
\end{align}
where $F(q)$ has no poles. Changing the integration variable from $q$ to
$z=2\Omega_1/\Gamma$, with $\Omega_1=\omega_{\bf p}-k-q$, one obtains
\begin{align}
\mathcal{P}(z_{\rm min},z_{\rm max}) = \frac{i}{2\Gamma}
&\int_{z_{\rm min}}^{z_{\rm max}}dz\ 
F_i\left(\omega_{\bf p}-k-\frac{\Gamma}{2}z\right) \nonumber\\
&\times \left(\frac{1}{z-i}-\frac{1}{z+1}\right)
\left(e^{iz\frac{\Gamma}{2}t}+e^{-iz\frac{\Gamma}{2}t}\right)\ ,
\end{align}
where $z_{\rm min}=2\Omega_1^{\rm min}/\Gamma$ and 
$z_{\rm max}=2\Omega_1^{\rm max}/\Gamma$,
with $z_{\rm min} < 0$ and $z_{\rm max} > 0$. In the limit 
$\Gamma/M \rightarrow0$ with $\tau=\Gamma t$ fixed, the integration limits 
approach $z_{\rm min} \rightarrow -\infty$ and 
$z_{\rm max} \rightarrow +\infty$, respectively. The integral is now easily
evaluated by means of the residue theorem leading to the result 
\begin{align}\label{primadon1}
\Gamma\mathcal{P}(z_{\rm min},z_{\rm max})\Big|_{\Gamma\rightarrow0}
&= -\pi \left(F\left(\omega_{\bf p}-k-i\frac{\Gamma}{2}\right)
e^{-\frac{\tau}{2}} + F\left(\omega_{\bf p}-k+i\frac{\Gamma}{2}\right)
e^{-\frac{\tau}{2}}\right)\bigg|_{\Gamma\rightarrow0} \nonumber\\
&= -2\pi F(\omega_{\bf p}-k)e^{-\frac{\tau}{2}}\ .
\end{align}

In Eq.~(\ref{kbint2}) the term $\mathcal{P}$ appears together with a second
term,
\begin{align}\label{P'}
\mathcal{P'}(q_{-},q_+)= \int_{q_-}^{q_+}dq 
\frac{F(q)}{(\omega_{\bf p}-k-q)^2 + \frac{\Gamma^2}{4}}\ 
e^{-\frac{\tau}{2}} \ ,
\end{align}
which can be evaluated in the same way as $\mathcal{P}$ in the zero-width
limit, yielding
\begin{align}\label{primadon2}
\Gamma\mathcal{P}'(z_{\rm min},z_{\rm max})\Big|_{\Gamma\rightarrow0}
&= \pi \left(F\left(\omega_{\bf p}-k-i\frac{\Gamma}{2}\right)
+ F\left(\omega_{\bf p}-k+i\frac{\Gamma}{2}\right)
\right)\bigg|_{\Gamma\rightarrow 0}\ e^{-\frac{\tau}{2}} \nonumber\\
&= 2\pi F(\omega_{\bf p}-k)\ e^{-\frac{\tau}{2}}\ .
\end{align}
Clearly, the two terms $\mathcal{P}$ and $\mathcal{P}'$ add up to zero.
The same result is obtained for the second factor $\mathcal{Q}$ after the $q'$ 
integration, as well as for the product of two sinus functions.

We conclude that the integral (\ref{kbint}) does not contain a contribution
$\mathcal{O}(1/\Gamma)$. Hence, the ratio of Kadanoff-Baym result and 
Boltzmann result, $L_{\bf k}(t,t)/f_L(t,k)$, approaches zero in the limit
$\Gamma/M \rightarrow 0$, $\tau=\Gamma t\ \mathrm{fixed}$.

\section{Equilibrium contribution}

In Section~5 we argued that the equilibrium part of the heavy neutrino
propagator does not contribute to the lepton asymmetry. In this section we 
verify this claim.

The heavy neutrino propagator has an equilibrium and a nonequilibrium part,
\begin{align}
G_{\bf p}(t_1,t_3) = G^{\rm eq}_{\bf p}(t_1-t_3)+\tilde{G}_{\bf p}(t_1,t_3)\ ,
\end{align}
whose main difference lies in the time dependence,
\begin{align}\label{timedifference}
G^{\rm eq}_{\bf p}(t_1-t_3) \propto e^{-\frac{\Gamma}{2}|t_1-t_3|}\ , \quad
\tilde{G}_{\bf p}(t_1,t_3) \propto e^{-\frac{\Gamma}{2}(t_1+t_3)} \ .
\end{align}
The computation of the lepton asymmetry in Section~5 was based on the
nonequilbrium part, and it involved the time integral $\mathcal{I}$ 
(cf.~Eq.~(\ref{I})). Because of the different time dependence given in 
Eq.~(\ref{timedifference}), the contribution
of the equilibrium part to the asymmetry involves instead the integral 
\begin{align}\label{J}
\mathcal{J}(t) = \int_0^{t}dt_1\int_0^t dt_2\int_0^{t_2}dt_3\ 
e^{-i\Omega_1 t_1+i\Omega_2t_2+i\Omega_3t_3}e^{-\frac{\Gamma}{2}|t_1-t_3|}\ ,
\end{align}
which differs from $\mathcal{I}$ only with respect to the damping factor.
$\Omega_1$, $\Omega_2$ and $\Omega_3$ are different linear combinations of
energies, which satisfy $\Omega_1 = \Omega_2 +\Omega_3$.

In order to evaluate the integral $\mathcal{J}$, we have to split the
time integration,
\begin{align}
\mathcal{J}(t)=\int_0^{t}dt_1\bigg[&\int_0^{t_1} dt_2\int_0^{t_2}dt_3\ 
e^{-i\Omega_1 t_1+i\Omega_2t_2+i\Omega_3t_3}
e^{-\frac{\Gamma}{2}(t_1-t_3)} \nonumber\\
+&\int_{t_1}^{t} dt_2 \bigg(\int_0^{t_1}dt_3\ 
e^{-i\Omega_1 t_1+i\Omega_2t_2+i\Omega_3t_3}
e^{-\frac{\Gamma}{2}(t_1-t_3)}\nonumber\\
&\phantom{\int_{t_1}^{t_2} dt_2}+\int_{t_1}^{t_2}dt_3\ 
e^{-i\Omega_1 t_1+i\Omega_2t_2+i\Omega_3t_3}
e^{-\frac{\Gamma}{2}(t_3-t_1)}\bigg)\bigg]\  .
\end{align}
Note the change of sign in the damping factor of the last two terms. As in
Section~5, it is convenient to use the variables 
$\bar{\Omega}_1=\Omega_1-\frac{i}{2}\Gamma$ and 
$\bar{\Omega}_3=\Omega_3-\frac{i}{2}\Gamma$, for which the integral 
simplifies to
\begin{align}
\mathcal{J}(t) = \int_0^{t}dt_1\bigg[&\int_0^{t_1} dt_2\int_0^{t_2}dt_3\ 
e^{-i\bar{\Omega}_1 t_1+i\Omega_2t_2+i\bar{\Omega}_3t_3} \\
+&\int_{t_1}^{t} dt_2\bigg(\int_0^{t_1}dt_3 
e^{-i\bar{\Omega}_1 t_1+i\Omega_2t_2+i\bar{\Omega}_3t_3}
+ \int_{t_1}^{t_2}dt_3\ 
e^{-i\bar{\Omega}^*_1 t_1+i\Omega_2t_2+i\bar{\Omega}^*_3t_3}\bigg)\bigg]\ .
\nonumber
\end{align}
Performing the $t_3$ integral
and using the relation $\Omega_1=\Omega_2+\Omega_3$, we obtain
\begin{align}
\mathcal{J}(t) = \int_0^{t}dt_1\bigg[&\int_0^{t_1}dt_2\ 
e^{-i\bar{\Omega}_1 t_1}\frac{1}{i\bar{\Omega}_3}
\left(e^{i\bar{\Omega}_1t_2}-e^{i\Omega_2t_2}\right) \nonumber\\
+&\int_{t_1}^{t}dt_2\bigg(e^{-i\bar{\Omega}_1 t_1}\frac{1}{i\bar{\Omega}_3}
\left(e^{i\bar{\Omega}_3t_1}-1\right) 
e^{i\Omega_2t_2}\nonumber\\
&\phantom{\int_{t_1}^{t_2} dt_2}
+e^{-i\bar{\Omega}^*_1 t_1}\frac{1}{i\bar{\Omega}^*_3}
\left(e^{i\bar{\Omega}^*_1t_2}-e^{i\bar{\Omega}^*_3t_1}
e^{i\Omega_2t_2}\right)\bigg)\bigg]\ .
\end{align}
%
It is now straightforward to carry out the integrations over $t_1$ and $t_2$,
which leads to
\begin{align}\label{ReJ}
\mathcal{J}(t) + \mathcal{J}^*(t) =
&\frac{2}{(\Omega_1^2+\frac{\Gamma^2}{4})(\Omega_3^2+\frac{\Gamma^2}{4})
(\Omega_1-\Omega_3)}\ \times \\
&\bigg[\Gamma\left(\Omega_1+\Omega_3\right)
\left(\cos((\Omega_1-\Omega_3)t)-1 + 
(\cos(\Omega_1t)-\cos(\Omega_3t))e^{-\frac{\Gamma t}{2}}\right) \nonumber\\
&+\left(2\Omega_1\Omega_3-\frac{\Gamma^2}{2}\right)
\left(\sin((\Omega_1-\Omega_3)t) -
(\sin(\Omega_1t)-\sin(\Omega_3t))e^{-\frac{\Gamma t}{2}}\right)\bigg]\ .
\nonumber
\end{align}
Note that the expression has no pole at $\Omega_1=\Omega_3$.

As in appendix~C we now have to evaluate the momentum integral
\begin{align}
\mathcal{S} = \int_{\Omega_1^{\rm min}}^{\Omega_1^{\rm max}}d\Omega_{1}
\int_{\Omega_3^{\rm min}}^{\Omega_3^{\rm max}}d\Omega_{3}\ 
(\mathcal{J} + \mathcal{J}^*)\ ,
\end{align}
with the integration limits given below Eq.~(\ref{qbound}). 
To perform the zero-width limit, we again introduce the variables 
$z_{1,3}=2\Omega_{1,3}/\Gamma$.
For $\Gamma\rightarrow 0$, the limits of integration $z^{\rm min}_{1,3}$ 
and $z^{\rm max}_{1,3}$ approach $-\infty$ and $+\infty$, respectively.
The $z_3$-integration can now be carried out by means of the residue
theorem. The integrand of the remaining $z_1$-integration has a double pole.
The integration can again be performed using the residue theorem, and we
find that $\Gamma\mathcal{S}$ approaches zero in the limit
$\Gamma/M \rightarrow 0$, $\tau=\Gamma t\ \mathrm{fixed}$. Hence, the 
equilibrium part of the heavy neutrino propagator does not contribute at
leading order in $\Gamma/M$.




\newpage
\thispagestyle{empty}
\section{Erratum}

1. Starting from Eq.~(2.11), $f_{Li}$ has to be replaced by $f_{Li}/2$;
starting from Eq.~(5.38), $L_{{\bf k}ii}$ has to be replaced by 
$L_{{\bf k}ii}/2$. In Eq.~(3.25) $P_L$ has to be replaced by $P_L\gamma^0$;
in Eq.~(5.33) $M$ has to be dropped; in the trace of Eq.~(5.34) a factor
$P_L$ has to be included; 
in Eqs.~(5.19), (5.20) and (5.45) the sign on the r.h.s.
has to be reversed. In Eqs.~(C.14) and (C.15) $dk'$ has to be replaced by 
$k'dk'$; in the first line of Eq.~(C.18) the 
$\varphi'$-integral has to be dropped and in the second line a factor $2\pi$
has to be included.

2. The statement after Eq.~(5.43), that $\mathcal{O}(t)$ does not contribute
to the asymmetry to leading order in $\Gamma$, is not correct. The leading 
contribution is given by
\begin{align}
L'_{{\bf k}ii}(t,t) =  -\epsilon_{ii}\ 16\pi 
&\int_{\bf q,q'} \frac{k\cdot k'}{kk'\omega_{\bf p}}\ f_N^{eq}(\omega_{\bf p}) 
f_{l\phi}(k,q) f_{l\phi}(k',q')
\frac{1}{(\Omega_1^2+\frac{\Gamma^2}{4})
(\Omega_3^2 +\frac{\Gamma^2}{4})} \nonumber \\ 
&\times\frac{2\Omega_1\Omega_3 +\frac{\Gamma^2}{2}}{\Omega_2}
\left(\sin(\Omega_2 t) - e^{-\frac{\Gamma t}{2}}\left(\sin(\Omega_1 t)
-\sin(\Omega_3 t)\right)\right) \ .
\end{align}
Changing integration variables, as
described in detail in Appendix~C (cf. Eq.~(C.19) and the following
discussion), one obtains
\begin{align}
L'_{{\bf k}ii}(t,t) =  &-\frac{\epsilon_{ii}}{2\pi^3\Gamma}\frac{1}{k} 
\int_{p_{\rm min}(k)}^{\infty}dp \int_{k'_{\rm min}(p)}^{k'_{\rm max}(p)}k'dk'
\int_{z_1^{\rm min}}^{z_1^{\rm max}}dz_1
\int_{z_3^{\rm min}}^{z_3^{\rm max}}dz_3
\frac{1}{\omega_{\bf p}}\nonumber\\
&\times\left(1-\frac{p^2+k^2-q^2}{2pk}\frac{p^2+k'^2-q'^2}{2pk'}\right) 
f_{l\phi}(k,q) f_{l\phi}(k',q') f_N^{eq}(\omega_{\bf p})\\
&\times\frac{1}{\left(z_1^2+1\right)\left(z_3^2+1\right)}
\frac{z_1 z_3 +1}{z_1 - z_3}
\left(\sin((z_1 - z_3)\tfrac{\tau}{2}) - 
e^{-\tfrac{\tau}{2}}\left(\sin(z_1\tfrac{\tau}{2})
-\sin(z_3\tfrac{\tau}{2})\right)\right) , \nonumber
\end{align}
where $\tau=\Gamma t$, $q=\omega_{\bf p} - k - \Gamma z_1/2$,
$q'=\omega_{\bf p} - k' - \Gamma z_3/2$ and
$z_{1,3}^{\rm min}=2\Omega_{1,3}^{\rm min}/\Gamma$,  
$z_{1,3}^{\rm max}=2\Omega_{1,3}^{\rm max}/\Gamma$,
with $z_{1,3}^{\rm min} < 0$ and $z_{1,3}^{\rm max} > 0$. In the limit 
$\Gamma/M \rightarrow0$ with $\tau=\Gamma t$ fixed, the integration limits 
approach $z_{1,3}^{\rm min} \rightarrow -\infty$ and 
$z_{1,3}^{\rm max} \rightarrow +\infty$, respectively. The integral can now 
be evaluated by means of Cauchy's theorem yielding the result 
\begin{align}
L'_{{\bf k}ii}(t,t) \rightarrow  
&-\frac{\epsilon_{ii}}{2\pi}\frac{1}{k} 
\int_{p_{\rm min}(k)}^{\infty}dp \int_{k'_{\rm min}(p)}^{k'_{\rm max}(p)}k'dk'
\frac{1}{\omega_{\bf p}} 
\left(1-\frac{2\omega_{\bf p} k-M^2}{2pk}
\frac{2\omega_{\bf p} k'-M^2}{2pk'}\right) \nonumber\\
&\times f_{l\phi}(k,\omega_{\bf p}-k)f_{l\phi}(k,\omega_{\bf p}-k')
f_N^{eq}(\omega_{\bf p})\frac{1-e^{-\tau}}{\Gamma}\ ,
\end{align}
which, up to the statistical factor, agrees with the Boltzmann result (C.15).
Hence, contrary to statements in the paper, the Boltzmann result is also
obtained for vanishing thermal damping widths in the limit 
$\Gamma \rightarrow 0$.\\

\noindent
We thank Mathias Garny for pointing out to us that  $\mathcal{O}(t)$ yields
the Boltzmann result in the zero-width limit.

\end{appendix}

\end{document}